\documentclass{egpubl}
\usepackage{egsr18-e}

\WsPaper         % uncomment for final version of EG Workshop contribution
 \electronicVersion % can be used both for the printed and electronic version

\ifpdf \usepackage[pdftex]{graphicx} \pdfcompresslevel=9

\else \usepackage[dvips]{graphicx} \fi

\PrintedOrElectronic

\usepackage{t1enc,dfadobe}

\usepackage{egweblnk}
\usepackage{cite}
\usepackage{hyperref}
\usepackage{appendix}
\usepackage{amsmath}
\usepackage{amsfonts}
\usepackage{subcaption} % subfigure
\usepackage{overpic}
\usepackage{contour}
\usepackage{xspace}
\usepackage[normalem]{ulem} % for strikeout \sout

\usepackage{wrapfig}

\usepackage[ruled,linesnumbered,vlined,algo2e]{algorithm2e}
\usepackage{scalerel} % very small subscripts

\newcommand{\ud}{\,\mathrm{d}} % for integrals
\newcommand{\SHBR}{Y_{\mathbb{R}}} % for integrals
\newcommand{\SHBC}{Y_{\mathbb{C}}} % for integrals
\newcommand{\iu}{{i\mkern1mu}}
\newcommand{\icaption}[1]{\caption{#1}}
\newcommand{\rev}[1]{\textcolor{black}{#1}}

\makeatletter
\def\mathcolor#1#{\@mathcolor{#1}}
\def\@mathcolor#1#2#3{%
  \protect\leavevmode
  \begingroup\color#1{#2}#3\endgroup
}
\makeatother

\definecolor{orange}{rgb}{1,0.5,0}
\definecolor{darkgreen}{rgb}{0,0.5,0}
\definecolor{red}{rgb}{1,0.0,0}

\usepackage{titletoc}

\newcommand{\nocontentsline}[3]{}
\newcommand{\tocless}[2]{\bgroup\let\addcontentsline=\nocontentsline#1{#2}\egroup}

\usepackage{todonotes}

\title%
{
$P_N$-Method for Multiple Scattering in Participating Media
}

\author[D. Koerner \& J. Portsmouth \& W. Jakob]
{
David Koerner$^{1}$
\qquad Jamie Portsmouth$^{2}$
\qquad Wenzel Jakob$^{3}$
\\
$^1$University of Stuttgart
\qquad $^2$Solid Angle
\qquad $^3$\'{E}cole polytechnique f\'{e}d\'{e}rale de Lausanne (EPFL)
}

\begin{document}
% uncomment for using teaser
\teaser
{
\centering
\begin{subfigure}{0.24\linewidth}
%\missingfigure{}
\includegraphics[width=\linewidth]{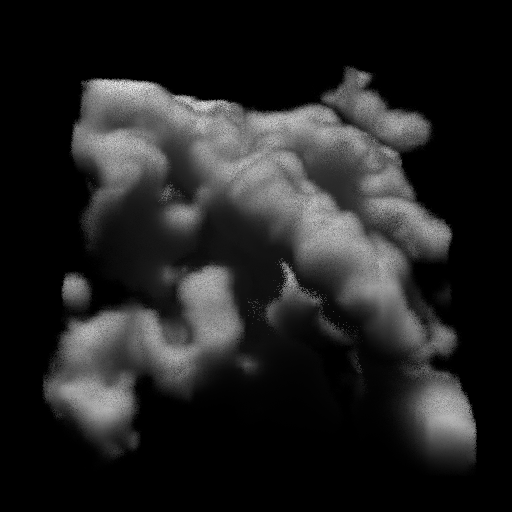}
\vspace{-0.17in}
\caption{Classical diffusion}
\label{fig:teaser_cda}
\end{subfigure}%
\hspace{0.002\linewidth}
\begin{subfigure}{0.24\linewidth}
\includegraphics[width=\linewidth]{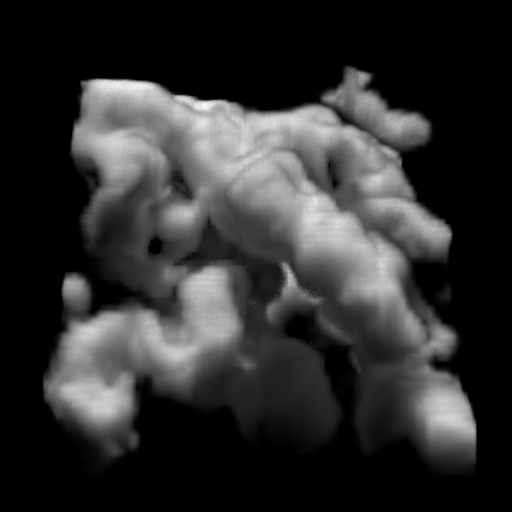}
\vspace{-0.17in}
\caption{$P_5$ (ours)}
\label{fig:nebulae_ours}
\label{fig:teaser_pn}
\end{subfigure}%
\hspace{0.002\linewidth}
\begin{subfigure}{0.24\linewidth}
%\missingfigure{}
\includegraphics[width=\linewidth]{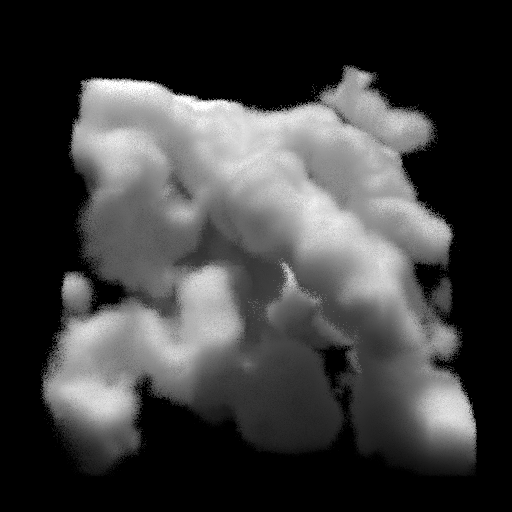}
\vspace{-0.17in}
\caption{Flux-limited diffusion}
\label{fig:teaser_fld}
\end{subfigure}%
\hspace{0.002\linewidth}
\begin{subfigure}{0.24\linewidth}
%\missingfigure{}
\includegraphics[width=\linewidth]{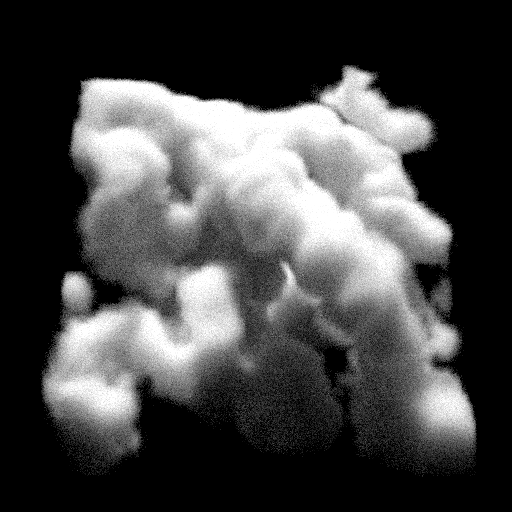}
\vspace{-0.17in}
\caption{Brute force path tracing}
\label{fig:teaser_gt}
\end{subfigure}%
\vspace{-0.1in}
\icaption
{
Flux-limited diffusion (\subref{fig:teaser_fld}) is an extension to the classical diffusion approximation (\subref{fig:teaser_cda}). It improves accuracy, but is based on ad-hoc assumptions about volumetric transport. We add the $P_N$-method (\subref{fig:teaser_pn}), which allows solution of the RTE with configurable accuracy, to the toolbox of methods in volume rendering and investigate its benefits and trade-offs against flux-limited diffusion.
}
\label{fig:teaser}
}

\maketitle

%-------------------------------------------------------------------------

\begin{abstract}
Rendering highly scattering participating media using brute force path tracing is a challenge. The diffusion approximation reduces the problem to solving a simple linear partial differential equation. Flux-limited diffusion introduces non-linearities to improve the accuracy of the solution, especially in low optical depth media, but introduces several ad-hoc assumptions. Both methods are based on a spherical harmonics expansion of the radiance field that is truncated after the first order. In this paper, we investigate the open question of whether going to higher spherical harmonic orders provides a viable improvement to these two approaches. Increasing the order introduces a set of complex coupled partial differential equations (the \emph{$P_N$-equations}), whose growing number make them difficult to work with at higher orders. We thus \rev{use} a computer algebra framework for representing and manipulating the underlying mathematical equations, and use it to derive the real-valued $P_N$-equations for arbitrary orders. We further present a staggered-grid $P_N$-solver and generate its stencil code directly from the expression tree of the $P_N$-equations. Finally, we discuss how our method compares to prior work for various standard problems.
%\begin{classification} % according to http://www.acm.org/class/1998/
%\CCScat{Computer Graphics}{I.3.7}{Three-Dimensional Graphics and Realism}{Raytracing}
%\end{classification}
\end{abstract}

%-------------------------------------------------------------------------

\tocless\section{Introduction}
%\vspace{0.1in}

Simulating light transport in participating media remains a challenging problem for image synthesis in computer graphics. Due to their ability to produce unbiased results and conceptual simplicity, Monte Carlo based techniques have become the standard approach~\cite{Novak18}. The main downside of these methods are their computational demands when rendering media with strong scattering or anisotropy.

Deterministic methods have enjoyed less popularity, because they suffer from discretization artifacts, produce biased results, cannot be coupled easily with surface rendering problems and are trickier to implement. However, their appeal lies in the fact that they produce a global solution across the whole domain and have better performance for certain problems~\cite{Brunner02}.

The work on path-guiding techniques from recent years~\cite{Muller17} has shown how approximate representations of the steady-state transport in a scene can be used to accelerate Monte Carlo integration techniques, such as path tracing. Instead of generating these approximate representations using Monte Carlo methods, deterministic methods may offer a viable alternative. Hybrid methods could combine the performance benefits of deterministic methods with accurate and unbiased Monte Carlo results.
Deterministic methods also lend themselves to applications where fast approximate solutions are preferable over correct, but slowly converging results. 

For these reasons, we suggest it is important for volume-rendering researchers to study deterministic methods and have a solid understanding of their characteristics and performance traits for typical rendering problems.

The $P_N$-method is a deterministic method of solving the radiative transfer equation~(RTE) which is used in other fields such as medical imaging and nuclear sciences, but has not found use in computer graphics thus far. The purpose and main contribution of our paper is to gain a solid understanding of its foundations and present a method for using it in the context of rendering. In particular, we present these theoretical and practical contributions:
\begin{itemize}
	\item We derive and present the time-independent real-valued $P_N$-equations and write them down in a very concise and compact form which we have not found anywhere else in the literature.
	\item We introduce a staggered-grid solver, for which we generate stencil code automatically from a computer algebra representation of the $P_N$-equations. This allows us to deal with the increasingly complex equations which the $P_N$-method produces for higher order. It further allows our solver to be used for any (potentially coupled) partial differential equations, which result in a system of linear equations after discretization.
	\item Finally, we compare the $P_N$-method for higher orders against flux-limited diffusion and ground truth Monte Carlo integration.
\end{itemize}

In the next section, we will discuss related work and its relation to our contribution. In Section~\ref{sec:discretized_rte} we revisit the deterministic approach to light transport simulation in participating media and outline the discretization using spherical harmonics. In Section~\ref{sec:car} we introduce our computer algebra representation, which we required to derive the real-valued $P_N$-equations, presented in Section~\ref{sec:real_valued_pn_eq}. This representation is also a key component of our solver, which we present in Section~\ref{sec:pnsolver}. Section~\ref{sec:rendering} discusses application of the solution in the context of rendering. We compare our $P_N$-solver against flux-limited diffusion for a set of standard problems in Section~\ref{sec:results}. Finally, Section~\ref{sec:conclusion} concludes with a summary and review of future work.

%-------------------------------------------------------------------------

\tocless\section{Previous work}

Light transport in participating media is governed by the RTE, first studied in the context of astrophysics by Chandrasekhar~\cite{Chandrasekhar60} and later introduced to computer graphics by Kajiya~\cite{Kajiya86}. In computer graphics today, this equation is typically solved using Monte Carlo methods~\cite{Novak18}. However in strongly scattering or highly anisotropic media these methods can become prohibitively expensive, for example in the case of a high albedo medium such as milk where tracing paths with a huge number of scattering events is necessary.

%Consider a high albedo medium like milk, where tracing paths with many scattering events is necessary.
In contrast to path-tracing, the $P_N$-method~\cite{Brunner02} gives a solution by solving a system of linear equations. It is derived by discretizing the angular variable of the radiative transfer equation into spherical harmonics (SH). This gives rise to a set of coupled, complex-valued partial differential equations, called the $P_N$-equations. The subscript $N$ refers to the spherical harmonics truncation order.

\begin{figure}[t]
\centering
\begin{subfigure}{0.45\columnwidth}
%\missingfigure{test}
\includegraphics[width=\columnwidth]{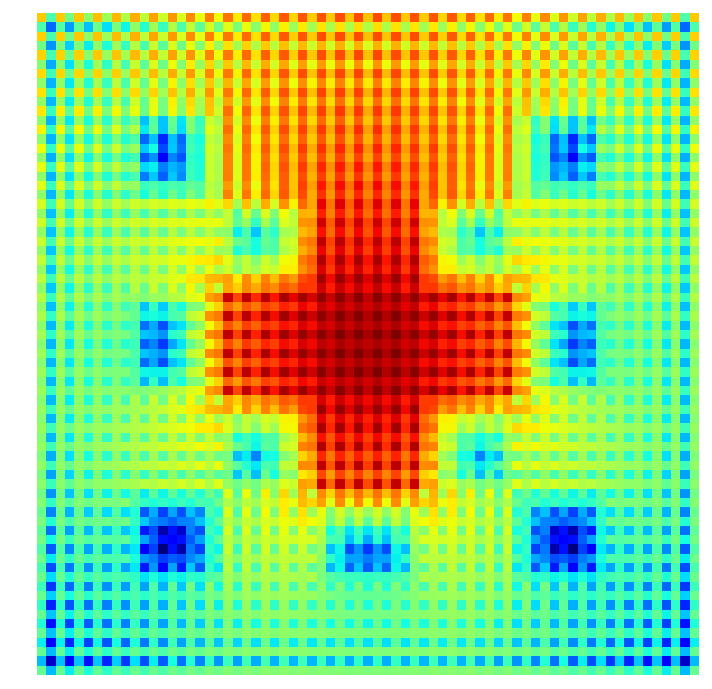}
\end{subfigure}%
\hspace{0.05\columnwidth}
\begin{subfigure}{0.45\columnwidth}
%\missingfigure{test2}
\includegraphics[width=\columnwidth]{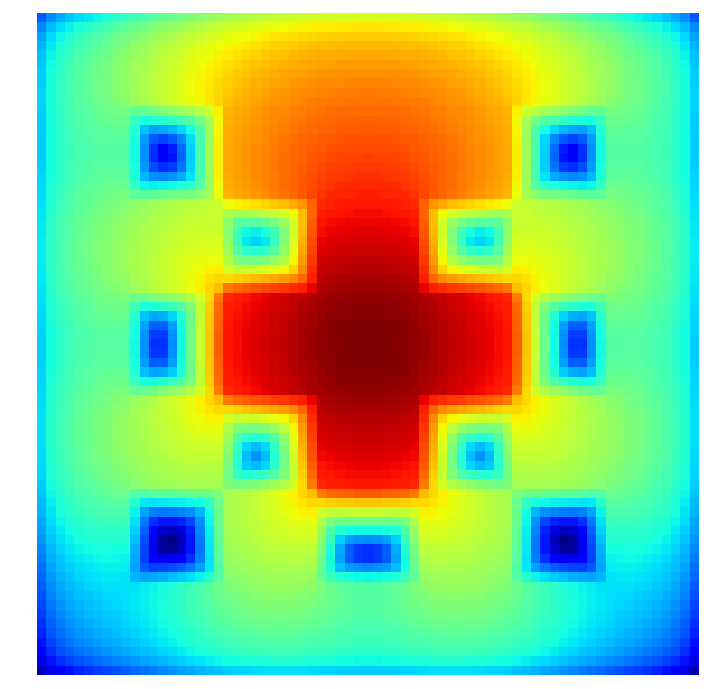}
\end{subfigure}%
%\vspace{-0.1in}
\icaption{Solving the 2D checkerboard problem using naive collocated grids produces oscillating artifacts (left). Our solver uses staggered grids and produces artifact free results (right). \vspace{-0.1in}}
\label{fig:artifacts}
\end{figure}

The $P_N$-method has a long history in other fields and was never applied in graphics. Kajiya~\cite{Kajiya84} explained the theory, but did not give any details on implementation or how to solve it. In fact, as Max~\cite{Max95} pointed out, it is not clear if Kajiya succeeded at all at applying the method, as all of the results in his paper were produced with a simpler method. This is further strengthened by the fact that a straightforward finite difference discretization of the $P_N$-equations produces unusable results, due to oscillation artifacts in the solution ~\cite{Seibold14}. We use a staggered-grid solver, motivated by the solver of Seibold et al.~\cite{Seibold14}, that produces artifact-free solutions (see Figure~\ref{fig:artifacts}).

Related to the $P_N$-method, the classical diffusion approximation (CDA) is a deterministic method which arrives at a solution by solving a system of linear equations. It corresponds to the $P_N$-equations when $N=1$ (truncation after the first SH order), which can be collapsed to a simple diffusion equation, giving the method its name. CDA has a long history in other domains, such as astrophysics and nuclear physics~\cite{Brunner02}, and was introduced to graphics by Stam \cite{Stam95}.

CDA suffers from severe energy loss close to regions with strong density gradients~\cite{Koerner14}. The problem can be addressed by a modification known as the Variable Eddington factor (VEF) method~\cite{Brunner02}, which non-linearly adjusts the diffusion coefficient to improve the solution near density gradients and low-density regions. Flux-limited diffusion, developed in the context of astrophysics by Levermore et al.~\cite{Levermore81} and later introduced to graphics by Koerner et al.~\cite{Koerner14}, is the most prominent example.
VEF is based on modifying the closure of the moment hierarchy in a way which more accurately than CDA models the transition from the diffusive regime (in which photons effectively undergo a random walk) to the transport regime (in which photons travel along straight lines) as the medium opacity decreases. VEF methods produce better results than CDA for isotropic media, but they do not provide a satisfactory treatment of anisotropic media.

Although the $P_N$-method should provide an increasingly accurate solution to the full RTE including anisotropy as the truncation moment order is increased, at the expense of more computation time, it is still an open and unresolved question how high a truncation order is needed for a particular problem in order to produce significantly better results than first order non-linear diffusion methods (FLD and VEF). This question has also been raised in other domains~\cite{Olson00}.

%\vspace{1in}

%\begin{itemize}
  %\item Alternative deterministic methods
  %\begin{itemize}
  %  \item Heuristics \cite{Kaplanyan10} \cite{Elek14}
  %  \item Lattice Boltzmann Methods \cite{Geist04}
  %  \item Discrete Ordinates \cite{Languenou95}
  %\end{itemize}
  %However, among all the deterministic methods, the diffusion approximation has been the most popular due to its intuition and simplicity.

%\end{itemize}

\newpage 

%-------------------------------------------------------------------------
\tocless\section{Discretized Radiative Transfer Equation \label{sec:discretized_rte}}

Our method derives from the radiative transfer equation (RTE), which expresses the change of the radiance field $L$, with respect to an infinitesimal change of position in direction $\omega$ at point $\vec{x}$:
\begin{align}
%\label{eq:rte}
\left(\nabla\cdot\omega\right)L\left(\vec{x}, \omega \right)
=&
-\sigma_t\left(\vec{x}\right) L\left(\vec{x}, \omega \right)\nonumber\\
&
+\sigma_s\left(\vec{x}\right) \int_{\Omega}
{
p\left(\omega'\cdot\omega\right)L\left(\vec{x}, \omega' \right)\ud\omega'
}\nonumber\\
&
+Q\left(\vec{x}, \omega\right)\nonumber
\  .
\label{eq:rte}
\end{align}

The left hand side (LHS) is the transport term, and we refer to the terms on the right hand side (RHS) as collision, scattering, and source term, respectively. The symbols $\sigma_t$, $\sigma_s$, $p$, and $Q$ refer to the extinction- and scattering coefficient, phase function and emission.

The RTE is often given in operator notation, where transport, collision, and scattering are expressed as operators $\mathcal{T}$, $\mathcal{C}$ and $\mathcal{S}$, which are applied to the radiance field $L$:
\begin{align}
\mathcal{T}\left(L\right) = -\mathcal{C}\left(L\right) + \mathcal{S}\left(L\right) + Q
\ .
\end{align}

Deterministic methods are derived by discretizing the angular and spatial domain. This gives rise to a linear system of equations, which can be solved using standard methods. For the $P_N$-method, the angular variable is first discretized, using a truncated spherical harmonics expansion. This results in the $P_N$-equations, a system of coupled PDEs that still depend on a continuous spatial variable.

The number of equations grows with the truncation order $N$. This is why the discretization is laborious and difficult to do without errors if done by hand. We therefore use a computer algebra representation to automate this process. After giving an outline of the general discretization in this section, we will present our computer algebra representation in the next section. The $P_N$-equations that result from our automated discretization are given in Section~\ref{sec:real_valued_pn_eq}.

Since the radiance field $L$ is real, we use the real-valued SH basis functions $\SHBR^{l,m}$, which are defined in terms of the complex-valued SH basis functions $\SHBC^{l,m}$ as follows~\cite{Jarosz09}:
\begin{align}
\SHBR^{l,m}=
\left\{
\begin{array}{lr}
\frac{\iu}{\sqrt{2}}\left(\SHBC^{l,m}-\left(-1\right)^m\SHBC^{l,-m}\right), & \text{for } m < 0\\
\SHBC^{l,m}, & \text{for } m = 0\\
\frac{1}{\sqrt{2}}\left(\SHBC^{l,-m}-\left(-1\right)^m\SHBC^{l,m}\right) & \text{for } m > 0
\end{array}
\right.
\label{eq:sh_real_basis}
\end{align}

We express the projection into the spherical harmonics basis functions with a projection operator $\mathcal{P}$:
\begin{align}
\mathcal{P}^{l, m}(f) = \int_{\Omega}f(\vec{x}, \omega) \SHBR^{l,m}(\omega)\,\mathrm{d}\omega = f^{l,m}\left(\vec{x}\right)
\ .
\nonumber
\end{align}

The $P_N$-equations are derived by first expressing all direction-dependent parameters in spherical harmonics. The radiance field $L$ in Equation~\ref{eq:rte} is therefore replaced by its SH reconstruction $\widehat{L}$, introducing an error due to truncation at order $N$:
\begin{align}
\widehat{L}\left(\vec{x}, \omega\right) =
\sum_{l=0}^{N}
{
\sum_{m=-l}^{l}
{
L^{l,m}\left(\vec{x}\right)\SHBR^{l,m}\left(\omega\right)
}
}
\approx
L\left(\vec{x}, \omega\right)
\ .
\nonumber
\end{align}
%\vspace{0.5in}

After substitution, all angular parameters are expressed in terms of spherical harmonics, but they still depend on the continuous angular variable $\omega$. As a next step, we project each term of the RTE into spherical harmonics, using the projection operator $\mathcal{P}$. This produces a single equation for each $l,m$-pair. The $P_N$-equations therefore can be written as:
\begin{align}
\mathcal{P}^{l,m}\mathcal{T}\left(\widehat{L}\right)
=
-\mathcal{P}^{l,m}\mathcal{C}\left(\widehat{L}\right)
+\mathcal{P}^{l,m}\mathcal{S}\left(\widehat{L}\right)
+\mathcal{P}^{l,m}\left(Q\right)
\ .
\label{eq:pn_operator_notation}
\end{align}

Once the $P_N$-equations have been found, the spatial variable $\vec{x}$ is discretized using a finite difference (FD) voxel grid (using central differences for differential operators).

\begin{figure}[t]
\centering
%\missingfigure{test}
\includegraphics[width=\columnwidth]{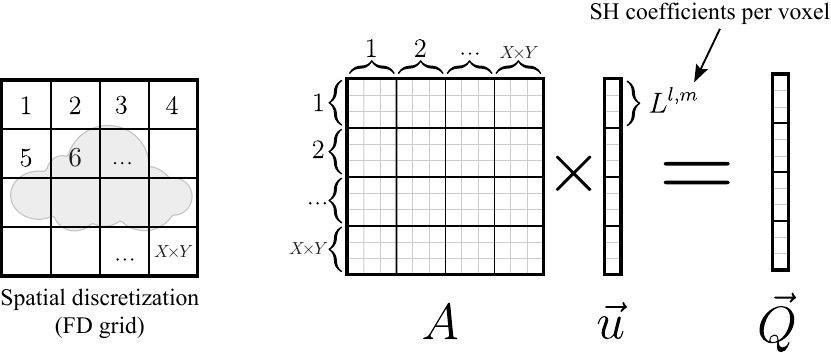}
%\missingfigure{figures fig matrix layout}
\vspace{-0.3in}
\icaption{Structure of coefficient matrix $A$ and solution vector $\vec{u}$ after discretization of the $P_N$-equations on a finite difference grid.}
\label{fig:matrix_layout}
\end{figure}

Following this discretization, the radiance field $L$, is represented as a set of SH coefficients per voxel. Flattening these over all voxels into a single vector, gives the solution vector $\vec{u}$. The RHS vector $\vec{Q}$ is produced similarly. The projected operators can be expressed as linear transformations, which can be collapsed into a single coefficient matrix $A$ (see Figure~\ref{fig:matrix_layout}):
\begin{align}
(T+C-S)\vec{u} = A\vec{u} = \vec{Q}
\ .
\end{align}

$T$, $C$, $S$ are matrices, which result from the discretized transport, collision and scattering operators respectively.

\tocless\section{Computer Algebra Representation \label{sec:car}}

\begin{figure}[h]
%\vspace{1in}
\centering
\includegraphics[width=0.95\columnwidth]{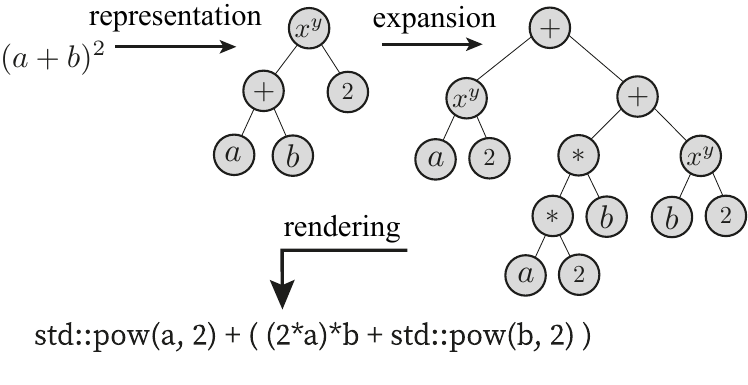}
%\missingfigure{figures fig car}
\icaption{\rev{A} computer algebra framework allows us to represent equations as mathematical expressions trees. It further provides a set of functions for manipulating the tree according to valid mathematical operations, such as the binomial expansion above. Frontends allow generation of source code from the expression tree.\vspace{-0.2in} \label{fig:car}}
\end{figure}

So far, we have only given the $P_N$-equations in high-level operator notation (Equation~\ref{eq:pn_operator_notation}). Carrying out the full derivation creates large, unwieldy equations and requires a string of expansions and applications of identities. These are challenging to manipulate if done by hand. We \rev{therefore used a computer algebra representation}, which allowed us to derive and discretize the $P_N$-equation in a semi-automatic fashion (Figure~\ref{fig:car})).

\rev{It represents the equations using a tree of mathematical expressions, which represent numbers, symbols and other expression types, such as integrals, derivatives, sums, products and functions. Further, manipulators can be executed on these expression trees to perform substitution, constant folding, reordering of nested integrals, application of identities and more complex operations. Finally, frontends allow rendering the expression tree into different forms, such as \LaTeX~and C++ source code. While we ended up implementing our own lightweight framework, off-the-shelf packages, such as SymPy (www.sympy.org), exist and would be equally suitable for our use case.}

\rev{Using the computer algebra representation, we perform the derivation steps required to arrive at the real-valued $P_N$-equations (the derivation steps shown in the Appendix were almost all rendered from the expression tree). More importantly, we use the representation to perform the discretization and generate the stencil code used by our solver. This is detailed in section~\ref{sec:solver_precomputation}}.

\tocless\section{Real-valued $P_N$-Equations \label{sec:real_valued_pn_eq}}

With the help of \rev{a} computer algebra representation framework, we are able to easily derive and work with the large and unwieldy $P_N$-equations. We present here the final real-valued $P_N$-equations, for a general $N$ and in three dimensions, in a very compact form which we have not found elsewhere in the literature. The derivation of the real-valued equations is rather long and takes many pages to describe in detail. We therefore give only the final result in this section, and present the full derivation for reference in Appendix~\ref{appendix}. 

Since the real-valued SH bases (Equation~\ref{eq:sh_real_basis}) have different definitions for $m<0$, $m=0$ or $m>0$,  we get different projections for $\mathcal{S}^{l,m}$, depending on the sign of $m$.
%\vspace{5in}
%\begin{align}
%&-\frac{1}{2}c^{\scaleto{l-1,m-1}{4pt}}
%\partial_y
%L^{\scaleto{l-1,-m+1}{4pt}}
%%\\
%+\frac{1}{2}d^{\scaleto{l+1,m-1}{4pt}}
%\partial_y
%L^{\scaleto{l+1,-m+1}{4pt}}
%%\\
%-\frac{1}{2}\beta^{\scaleto{m}{4pt}}e^{\scaleto{l-1,m+1}{4pt}}
%\partial_y
%L^{\scaleto{l-1,-m-1}{4pt}}
%\nonumber
%\\&
%+\frac{1}{2}\beta^{\scaleto{m}{4pt}}f^{\scaleto{l+1,m+1}{4pt}}
%\partial_y
%L^{\scaleto{l+1,-m-1}{4pt}}
%%\\
%+\frac{1}{2}\delta_{\scaleto{m\neq -1}{4pt}}c^{\scaleto{l-1,m-1}{4pt}}
%\partial_x
%L^{\scaleto{l-1,m-1}{4pt}}
%\nonumber
%\\&
%-\frac{1}{2}\delta_{\scaleto{m\neq -1}{4pt}}e^{{l-1,m+1}}
%\partial_x
%L^{\scaleto{l-1,m+1}{4pt}}
%%\\
%+\frac{1}{2}f^{\scaleto{l+1,m+1}{4pt}}
%\partial_x
%L^{\scaleto{l+1,m+1}{4pt}}
%%\\
%-\frac{1}{2}d^{\scaleto{l+1,m-1}{4pt}}
%\partial_x
%L^{\scaleto{l+1,m-1}{4pt}}
%\nonumber
%\\&
%+a^{\scaleto{l-1,m}{4pt}}
%\partial_z
%L^{\scaleto{l-1,m}{4pt}}
%%\\
%+b^{\scaleto{l+1,m}{4pt}}
%\partial_z
%L^{\scaleto{l+1,m}{4pt}}
%%\\
%+\sigma_t L^{\scaleto{l,m}{4pt}}
%%\\
%-\sigma_s\lambda_{\scaleto{l}{4pt}}p^{\scaleto{l,0}{4pt}}L^{\scaleto{l,m}{4pt}}
%%\\
%= Q^{\scaleto{l,m}{4pt}}
%\label{eq:rpn_m_<_z}
%\end{align}

For $m=0$ we have
\begin{align}
&
\frac{1}{\sqrt{2}}c^{\scaleto{l-1,-1}{4pt}}\partial_x L^{\scaleto{l-1,1}{4pt}}
-\frac{1}{\sqrt{2}}d^{\scaleto{l+1,-1}{4pt}}\partial_x L^{\scaleto{l+1,1}{4pt}}
%\\&
\frac{1}{\sqrt{2}}c^{\scaleto{l-1,-1}{4pt}}\partial_y L^{\scaleto{l-1,-1}{4pt}}
\nonumber
\\&
-\frac{1}{\sqrt{2}}d^{\scaleto{l+1,-1}{4pt}}\partial_y L^{\scaleto{l+1,-1}{4pt}}
%\\&
+a^{\scaleto{l-1,0}{4pt}}\partial_z L^{\scaleto{l-1,0}{4pt}}
+b^{\scaleto{l+1,0}{4pt}}\partial_z L^{\scaleto{l+1,0}{4pt}}
\nonumber
\\&
+\sigma_t L^{\scaleto{l,m}{4pt}}
-\sigma_s\lambda_{\scaleto{l}{4pt}}p^{\scaleto{l,0}{4pt}}L^{\scaleto{l,m}{4pt}}
= Q^{\scaleto{l,m}{4pt}}\quad .
\label{eq:rpn_m_=_z}
\end{align}
%and for $m>0$:
%\begin{align}
%&
%\frac{1}{2}c^{\scaleto{l-1,-m-1}{4pt}}\partial_x L^{\scaleto{l-1,m+1}{4pt}}
%%\\&
%-\frac{1}{2}d^{\scaleto{l+1,-m-1}{4pt}}\partial_x L^{\scaleto{l+1,m+1}{4pt}}
%%\\&
%-\frac{1}{2}\beta^{\scaleto{m}{4pt}}e^{\scaleto{l-1,m-1}{4pt}}\partial_x L^{\scaleto{l-1,m-1}{4pt}}
%\nonumber
%\\&
%+\frac{1}{2}\beta^{\scaleto{m}{4pt}}f^{l+1,-m+1}\partial_x L^{\scaleto{l+1,m-1}{4pt}}
%%\\&
%+\frac{1}{2}c^{\scaleto{l-1,-m-1}{4pt}}\partial_y L^{\scaleto{l-1,-m-1}{4pt}}
%\nonumber
%\\&
%-\frac{1}{2}d^{\scaleto{l+1,-m-1}{4pt}}\partial_y L^{\scaleto{l+1,-m-1}{4pt}}
%%\nonumber
%%\\&
%+\delta_{\scaleto{m\neq 1}{4pt}}\frac{1}{2}e^{\scaleto{l-1,-m+1}{4pt}}%\partial_y L^{\scaleto{l-1,-m+1}{4pt}}
%\nonumber
%\\&
%-\delta_{\scaleto{m\neq 1}{4pt}}\frac{1}{2}f^{\scaleto{l+1,-m+1}{4pt}}%\partial_y L^{\scaleto{l+1,-m+1}{4pt}}
%%\\&
%+a^{\scaleto{l-1,-m}{4pt}}\partial_z L^{\scaleto{l-1,m}{4pt}}
%%\nonumber
%%\\&
%+b^{\scaleto{l+1,-m}{4pt}}\partial_z L^{\scaleto{l+1,m}{4pt}}
%\nonumber
%\\&
%+\sigma_t L^{\scaleto{l,m}{4pt}}
%-\sigma_s\lambda_{\scaleto{l}{4pt}}p^{\scaleto{l,0}{4pt}}L^{\scaleto{l,m}{4pt}}
%= Q^{\scaleto{l,m}{4pt}}
%%\nonumber
%\label{eq:rpn_m_>_z}
%\end{align}

For $m<0$ (upper sign) and $m>0$ (lower sign) we have
\begin{align}
&
\frac{1}{2}c^{\scaleto{l-1,\pm m-1}{4pt}}\partial_x L^{\scaleto{l-1,m\mp 1}{4pt}}
%\\&
-\frac{1}{2}d^{\scaleto{l+1,\pm m-1}{4pt}}\partial_x L^{\scaleto{l+1,m\mp 1}{4pt}}
%\\&
-\frac{1}{2}\beta_x^{\scaleto{m}{4pt}}e^{\scaleto{l-1,m\pm 1}{4pt}}\partial_x L^{\scaleto{l-1,m\pm 1}{4pt}}
\nonumber
\\&
+\frac{1}{2}\beta_x^{\scaleto{m}{4pt}}f^{l+1,\pm m+1}\partial_x L^{\scaleto{l+1,m\pm 1}{4pt}}
%\\&
\mp \frac{1}{2}c^{\scaleto{l-1,\pm m-1}{4pt}}\partial_y L^{\scaleto{l-1,-m\pm 1}{4pt}}
\nonumber
\\&
\pm \frac{1}{2}d^{\scaleto{l+1,\pm m-1}{4pt}}\partial_y L^{\scaleto{l+1,-m \pm 1}{4pt}}
%\nonumber
%\\&
\mp \beta_y^{\scaleto{m}{4pt}}\frac{1}{2}e^{\scaleto{l-1,\pm m+1}{4pt}}\partial_y L^{\scaleto{l-1,-m\mp 1}{4pt}}
\nonumber
\\&
\pm \beta_y^{\scaleto{m}{4pt}}\frac{1}{2}f^{\scaleto{l+1,\pm m+1}{4pt}}\partial_y L^{\scaleto{l+1,-m\mp 1}{4pt}}
%\\&
+a^{\scaleto{l-1,\pm m}{4pt}}\partial_z L^{\scaleto{l-1,\mp m}{4pt}}
%\nonumber
%\\&
+b^{\scaleto{l+1,\pm m}{4pt}}\partial_z L^{\scaleto{l+1,\mp m}{4pt}}
\nonumber
\\&
+\sigma_t L^{\scaleto{l,m}{4pt}}
-\sigma_s\lambda_{\scaleto{l}{4pt}}p^{\scaleto{l,0}{4pt}}L^{\scaleto{l,m}{4pt}}
= Q^{\scaleto{l,m}{4pt}}
\quad ,
%\nonumber
\label{eq:rpn_m_>_z}
\end{align}

with
\begin{align*}
\label{eq:real_sh_basis}
\beta_x^{m}=
\left\{
\begin{array}{ll}
0, & \text{for } m = -1\\
\frac{2}{\sqrt{2}}, & \text{for } m \neq 1\\
1, & \text{otherwise }
\end{array}
\right.
,\quad
\beta_y^{m}=
\left\{
\begin{array}{ll}
\frac{2}{\sqrt{2}}, & \text{for } m = -1\\
0, & \text{for } m \neq 1\\
1, & \text{otherwise }
\end{array}
\right.
\end{align*}

and
\begin{align*}
&
a^{\scaleto{l,m}{4pt}}= \sqrt{\frac{\left(l-m+1\right)\left(l+m+1\right)}{\left(2l+1\right)\left(2l-1\right)}} \qquad
b^{\scaleto{l,m}{4pt}}= \sqrt{\frac{\left(l-m\right)\left(l+m\right)}{\left(2l+1\right)\left(2l-1\right)}}
\\&
c^{\scaleto{l,m}{4pt}}= \sqrt{\frac{\left(l+m+1\right)\left(l+m+2\right)}{\left(2l+3\right)\left(2l+1\right)}} \qquad
d^{\scaleto{l,m}{4pt}}= \sqrt{\frac{\left(l-m\right)\left(l-m-1\right)}{\left(2l+1\right)\left(2l-1\right)}}
\\&
e^{\scaleto{l,m}{4pt}}= \sqrt{\frac{\left(l-m+1\right)\left(l-m+2\right)}{\left(2l+3\right)\left(2l+1\right)}} \qquad
f^{\scaleto{l,m}{4pt}}= \sqrt{\frac{\left(l+m\right)\left(l+m-1\right)}{\left(2l+1\right)\left(2l-1\right)}}
\end{align*}
\begin{align*}
\lambda_l=\sqrt{\frac{4\pi}{2l+1}}
\quad .
\end{align*}

In the next section we will present the solver that we use to solve the $P_N$-equations.
\vspace{1in}

%-------------------------------------------------------------------------
\begin{figure*}[t!]
\centering
%\missingfigure{$P_N$-solver overview: generate stencil code $\rightarrow$ build system $\rightarrow$ solve $\rightarrow$ render}
\includegraphics[width=\textwidth]{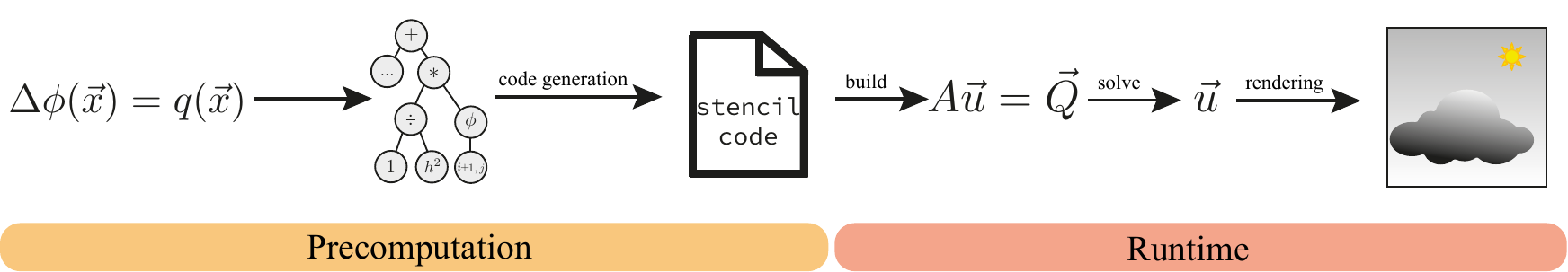}
%\missingfigure{figures fig pipeline}
\vspace{-0.2in}
\icaption{Overview of our $P_N$-solver. After generating the stencil source code from the expression trees representing the $P_N$-equations, the linear system $A\vec{u}=\vec{Q}$ is built using RTE parameter fields and additional user input, such as grid resolution and type of boundary conditions. The resulting system is solved for $\vec{u}$, which is then used in our rendering application.}
\label{fig:pnsolver}
\end{figure*}

\vspace{-0.75in}

\tocless\section{$P_N$-Solver \label{sec:pnsolver}}

The truncation order $N$ is the key input parameter to the solver. With higher values, manual implementation of the solver from the equations would be arduous, error-prone and time-consuming. We therefore decided to make use of the computer algebra representation and designed our solver around it.

The solver consists of two components. The first is a precomputation (Section~\ref{sec:solver_precomputation}), which is executed once for every single value of $N$. This step runs a partial evaluation on the mathematical expression tree and applies the spatial discretization in a reference space we call stencil space. The precomputation step automatically generates source code from the resulting expression tree.

The generated stencil code is compiled with the runtime component (Section~\ref{sec:solver_runtime}) of our solver. This component receives the actual problem as input, including grid resolution and RTE parameter fields. It then builds the linear system and solves it using standard methods. An overview of the solver is given in Figure~\ref{fig:pnsolver}.

\tocless\subsection{Precomputation \label{sec:solver_precomputation}}

The result of the precomputation is a stencil, which can be used during runtime to build the linear system for a given problem. The stencil is a pattern of indices into the solution vector, along with values. It expresses how the sum of the weighted solution vector components relate to the RHS for a given unknown in the system, and therefore contains most information required to fill the system matrix $A$ and RHS-vector $\vec{Q}$ row-by-row. Note that while the coefficients may change, the sparsity pattern of the stencil is identical for different rows.

\vspace{0.8in}

\begin{figure}[h]
\centering
%\missingfigure{test}
\includegraphics[width=0.7\columnwidth]{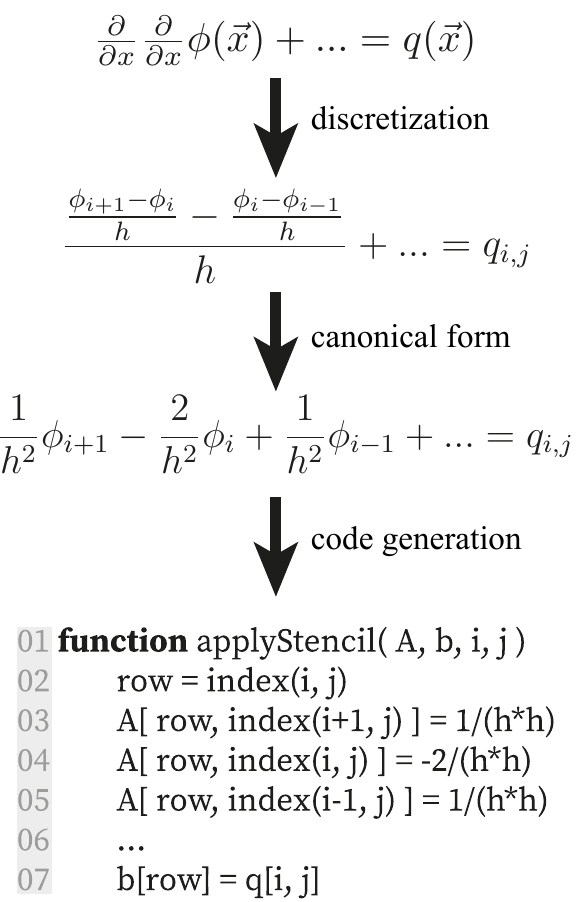}
%\missingfigure{figures fig stencil pipeline}
%\vspace{-0.2in}
\icaption{Creating the stencil code requires several steps, usually done by hand. We express the given problem in a computer algebra representation and use it to fully automate the process.}
\label{fig:stencile_pipeline}
\end{figure}

Stencil generation entails discretizing a PDE at a hypothetical center voxel $(i, j, k)$ (assumed to mean the voxel center most of the time). Finite differences create weighted references to other voxels (e.g. $i+1, j, k$). After bringing the discretized equation into canonical form (a weighted sum of unknowns), one can write the stencil by reading off the weights and offsets (Figure~\ref{fig:stencile_pipeline}). Voxel $(i, j, k)$ will only be known during runtime, when the stencil is executed for a particular unknown (row). Then the offsets can be used to find the component index into the solution-vector, and weights can be evaluated for concrete world space position. We refer to the space with the hypothetical voxel $(i, j, k)$ at the center as stencil space.

\vspace{0.1in}

\vspace{0.2in}

\begin{wrapfigure}{r}{0.6\columnwidth}
%\begin{center}
\hspace{-.2in}
\includegraphics[width=0.6\columnwidth]{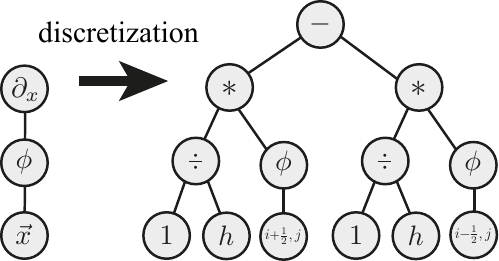}
%\missingfigure{figures fig discretization}
%\end{center}
\end{wrapfigure}
The spatial discretization is done by parsing the expression tree from the root. The discrete substitute for the continuous position variable $\vec{x}$ is initialized with $ijk$. Differential operator nodes are replaced by a subtree, which expresses the finite difference approximation (including voxel-size factor $h$). The subtree of the differential operator node is duplicated for different offsets to the discrete variable $(i, j, k)$. Since this offset only applies to the subtree, a stack is maintained by the parser to manage scope. Whenever the parser encounters the continuous variable $\vec{x}$ in the expression tree, its node in the tree is replaced by the discrete substitute, currently on top of the stack. Nested differential operators yield higher order finite difference stencils as expected.

\begin{wrapfigure}{r}{0.6\columnwidth}
%\begin{center}
\hspace{-.2in}
\includegraphics[width=0.6\columnwidth]{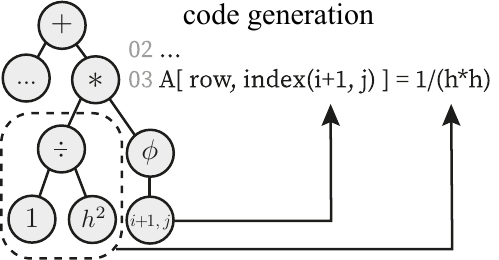}
%\missingfigure{figures fig codegen}
%\end{center}
\end{wrapfigure}
Factorization into canonical form is done as a manipulation pass on the mathematical expression tree. The result allows us to implement the code generation in a straightforward fashion. For each term, the $ijk$-offset is retrieved from the unknown. The factor-expression, which is multipled with the unknown, is extracted from the tree and used to generate source code for evaluating the factor-expression during runtime (including calls for evaluating RTE-parameter fields).

\begin{figure}[h]

%\vspace{1in}
\centering
\includegraphics[width=0.8\columnwidth]{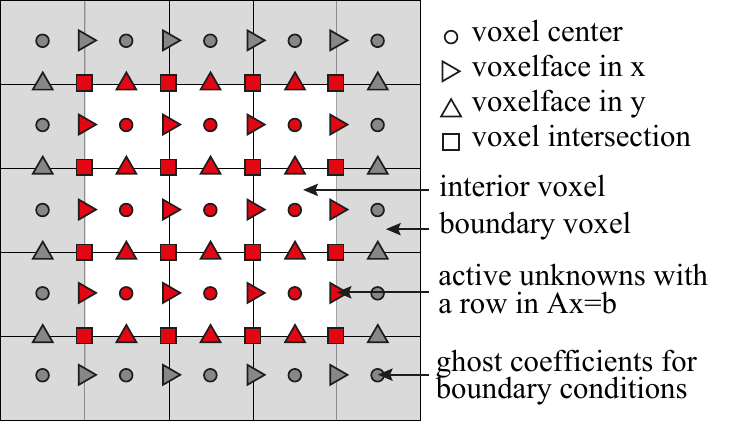}
%\missingfigure{figures fig staggered grids}
%\vspace{-0.2in}
\icaption{\emph{Staggering} distributes the coefficients of the solution vector $\vec{u}$ onto four disjoint grids, indicated by the symbols, in a way which guarantees second order accuracy~\cite{Seibold14}.\vspace{-.2in}}
\label{fig:staggeredgrid}
\end{figure}

\begin{wrapfigure}{r}{0.6\columnwidth}
%\begin{center}
\hspace{-.1in}
\includegraphics[width=0.6\columnwidth]{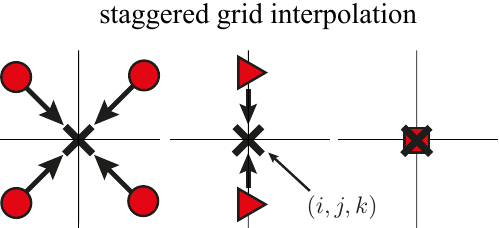}
%\missingfigure{figures fig staggered interpolation}
\vspace{-.25in}
%\end{center}
\end{wrapfigure}Our solver supports placement of coefficients at arbitrary staggered grid locations (see Figure~\ref{fig:staggeredgrid}). This means that during the discretization step, the discrete location $(i, j, k)$ (at which the unknown is meant to be evaluated) might not coincide with the location of the unknown. To solve this, depending on how the two are located relative to each other, the parser returns an expression which interpolates the coefficients at $(i, j, k)$ from their defined locations. If those happen to coincide, the coefficient itself is returned. This is also done for RTE parameters, such as $\sigma_t$ or $p^{l,m}$, which are always located at the voxel center. We use the staggering scheme introduced by Seibold et al.~\cite{Seibold14} for their solver \textsf{StaRMAP}, which they proved ensures second order accuracy, i.e. a truncation error of $O(h^2)$, and prevents the growth of non-physical oscillations in the solution (see their paper for the full details of the staggered coefficient placement).

%\subsection{Stencil Code Generation}

% After applying the discretization step to the expression tree of the $P_N$-equations, it is used to generate the stencil code. In numerical analysis, a stencil is an arrangement of voxels and weights, that relate values at different locations to each other and form the basis for propagating rows in the system matrix $A$ and RHS vector $\vec{Q}$ with values. The name comes from the fact, that the geometric structure and weights of the configuration do not change, when applied to different voxels. The same is true for the $P_N$-equations and we use this fact to generate a single function, which propagates rows in $A$ and $\vec{Q}$ for a given voxel.

%The $P_N$-equations express, how each coefficient of a voxel depends on other coefficients of the same or adjacent voxels. The unknowns in the terms give information about the coefficient index and voxel offset, and therefore identify a column offset in the matrix $A$. The factors to these coefficients may contain evaluations of RTE parameters, such as $\sigma_t$. Therefore, these factors can not be determined during stencil generation, but are rendered into code expressions, which are executed as part of the stencil function during runtime. Because the stencil code has been generated in stencil space relative to the voxel at $(0,0,0)$, we can run the same stencil code for every voxel, by simply applying an offset accordingly.

%\subsection{System Building and Solving}
\tocless\subsection{Runtime}
\label{sec:solver_runtime}

The stencil code is generated once for every value of $N$ and compiled with the runtime component of our solver. The runtime executes the stencil for every voxel to populate the system matrix $A$ and RHS vector $\vec{Q}$ with numerical values.

The number of rows is determined by the number of voxels times the number of coefficients per voxel (see Figure~\ref{fig:matrix_layout}) and can therefore become very large for high resolution and high truncation order. The matrix $A$ is square and extremely sparse, due to the local structure of the finite differences discretization. Unfortunately, it is non-symmetric due to the transport term and not diagonal dominant, which rules out many \rev{iterative} methods for solving linear systems. \rev{Iterative methods are useful, as they allow balancing accuracy against performance by tweaking the convergence threshold.} We address this by solving the normal form $A^TA\vec{u}=A^T\vec{Q}$ instead. This gives a symmetric and positive definite system matrix $A^TA$, albeit with a higher condition number. Investigation of other solution schemes (e.g. multigrid) would be an interesting avenue for future work. \rev{However, more importantly, in the presence of vacuum regions, the matrix $A$ becomes singular and the system cannot be solved at all. This requires the introduction of a minimum threshold for the extinction coefficient $\sigma_t$.}

Our solver supports both Neumann and Dirichlet boundary conditions (BC). They are handled transparently by the code which generates the stencil. Whenever the stencil accumulates coefficients into a boundary location, the framework either ignores the write operation (Dirichlet BC) or accumulates into the row and column in $A$ of the coefficient in the closest voxel inside the domain (Neumann BC). This is done by changing the index of the written component.

\tocless\section{Rendering \label{sec:rendering}}

We use an approach similar to Koerner et al.~\cite{Koerner14}, where we separate the radiance field into single scattered light $L_{ss}$ and multiple scattered light $L_{ms}$:
\begin{align}
L\left(\vec{x},\omega\right) = L_{ss}\left(\vec{x},\omega\right) + L_{ms}\left(\vec{x},\omega\right)
\ .
\end{align}
The single scattered light is folded into the emission term $Q$:
\begin{align}
Q(\vec{x}, \omega) = L_{ss}(\vec{x}, \omega) = \sigma_s\left(\vec{x}\right)\int_\Omega{ p\left(\omega'\rightarrow\omega\right) L_{u}\left(\vec{x}, \omega'\right)\ud\omega' }
\ .
\end{align}

This means that our solver will solve for the multiple scattered light $L_{ms}$. The quantity $L_u$ is the ``uncollided'' light, which was emitted from the light source and attenuated by the volume without undergoing any scattering event.
We compute it using a few light samples per voxel, which quickly converges to a useful result for Dirac delta light sources.

Running the solver gives solution vector $\vec{u}$. We then un-stagger the solution by interpolating all coefficients to voxel centers. The additional coefficients at boundary voxels are no longer needed. This operation is represented as a matrix that produces a three-dimensional voxel grid with SH coefficients for order $N$ at the center of each voxel.

For rendering, we use a simple forward path tracing approach, where we start tracing from the camera. At the first scattering event, we use next event estimation to account for $L_{ss}$. Then we sample a new direction according to the phase function. Instead of continuing tracing into the new direction, we evaluate the in-scattering integral using $\widehat{L}_{ms}$. The SH coefficients at $\vec{x}$ are found by trilinear interpolation from the voxel grid of SH coefficients.

%-------------------------------------------------------------------------
\tocless\section{Results \label{sec:results}}

In this section, we present results for a set of standard problems in order to validate and evaluate our method. We also compare against classical diffusion (CDA) and flux-limited diffusion (FLD).

Our computer algebra framework and the precomputation has been implemented in Python. The runtime component of our solver has been implemented in C++. We use a naive implementation of a CG solver, which has been modified such that we do not need to explicitly compute the normal form of the linear system to solve. We use the sparse matrix representation and sparse matrix vector product from the Eigen linear algebra library (eigen.tuxfamily.org).

The solver for classical diffusion is based on the diffusion equation, which is derived by collapsing the $P_1$-equations:
\begin{align}
\nabla\left(\frac{1}{3\sigma_t}\nabla L^{0,0}\right)  = -Q^{0,0}
\ .
\label{eq:cda}
\end{align}

Since our solver can work with any PDE which results in a linear system of equations, we put Equation~\ref{eq:cda} into our computer algebra representation and provide it as an input to our solver, which generates the correct stencil code automatically.

Since FLD is based on a non-linear diffusion equation, we were not able to use our system in the same way. Our implementation closely follows the implementation in~\cite{Koerner14} (though ours runs on CPU) and we refer to their paper for more details.

\tocless\subsection{2D checkerboard}

First we ran our solver on the 2D checkerboard, a very common test case in other fields. The problem has dimensions $7\times 7$ and is discretized with resolution $71\times 71$. Unit size blocks are filled with purely absorbing medium $\sigma_a=10$ in a checkerboard pattern. All other blocks are filled with a purely scattering medium with $\sigma_s=1$.

Solving the standard checkerboard problem allows us to validate our solver against the \textsf{StaRMAP} solver from Seibold et al.~\cite{Seibold14}, which solves for the time-dependent and complex-valued $P_N$-equations on a staggered grid, but in the 2D case only. The 2D case is derived by assuming that all SH coefficients, RTE parameters and boundary conditions are z-independent. This causes all SH coefficients and moment equations for which $l+m$ is odd to vanish. Due to the time-dependency, their approach is to do explicit incremental steps in time. We run their solver for many timesteps to get a result close to steady state.

\begin{figure}[!t]
\centering
\begin{subfigure}{0.47\columnwidth}
\includegraphics[width=\columnwidth]{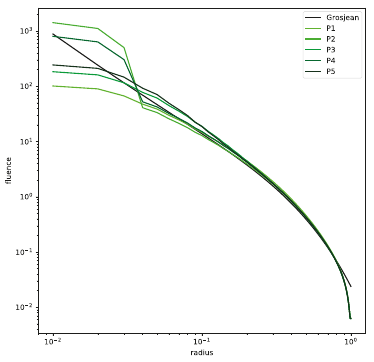}
%\missingfigure{figures fig pointsource pn}
\caption{$P_N$ vs. ground truth}
\label{fig:pointsource_pn}
\end{subfigure}%
\hspace{0.05\columnwidth}
\begin{subfigure}{0.47\columnwidth}
\includegraphics[width=\columnwidth]{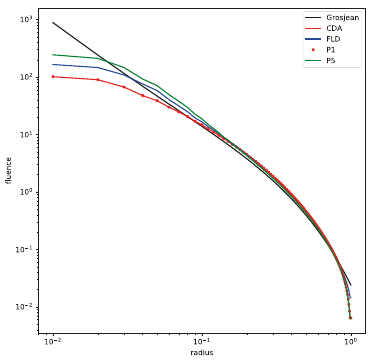}
%\missingfigure{figures fig pointsource p5}
\caption{$P_5$ vs. CDA and FLD}
\label{fig:pointsource_p5}
\end{subfigure}%
%\vspace{-0.1in}
\icaption{Lineplot through the 3D solution of our solver for the point source problem for various order $N$ (left). Solution for $P_5$ compared against classical diffusion, flux-limited diffusion and analytical solution (right). \vspace{-0.14in}}
\label{fig:pointsource}
\end{figure}

\begin{figure}[h]
\centering
\begin{subfigure}{0.49\columnwidth}
\includegraphics[width=\columnwidth]{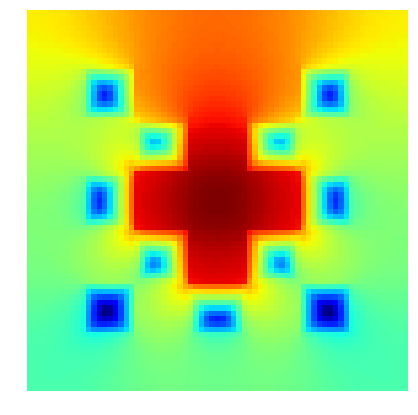}
\end{subfigure}%
\hspace{0.01\columnwidth}
\begin{subfigure}{0.49\columnwidth}
\includegraphics[width=\columnwidth]{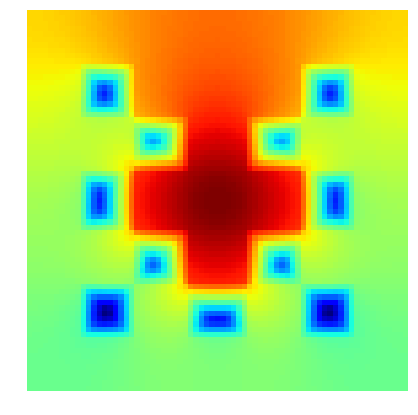}
\end{subfigure}%
\vspace{-0.1in}
\icaption{Comparison of the result (here the $L^{0,0}$ coefficient, or \emph{fluence}) for the checkerboard test using \textsf{StaRMAP}'s time-stepping solver~\cite{Seibold14} (left) against our steady-state solver (right) with $N=5$. Our results are in good agreement.}
\label{fig:vs_starmap}
\end{figure}

As can be seen in Figure~\ref{fig:vs_starmap}, the results from our solver are in good agreement with the results from Seibold et al.~\cite{Seibold14} and verify the correctness of our implementation. Converging to a residual of $10e^{-10}$ takes $0.27s$ for $P_1$ and $25s$ for $P_5$.

\tocless\subsection{Point source problem}

We also run our solver for the point source problem, a single point light in a homogeneous medium. This not only helps to validate our implementation for the 3D case, but also provides information on the accuracy of these methods. We use the Grosjean approximation, which was introduced by D'Eon et al.\cite{dEon11} as a very accurate approximation to the ground truth solution.

\begin{figure}[!t]
\centering
\includegraphics[width=0.7\columnwidth]{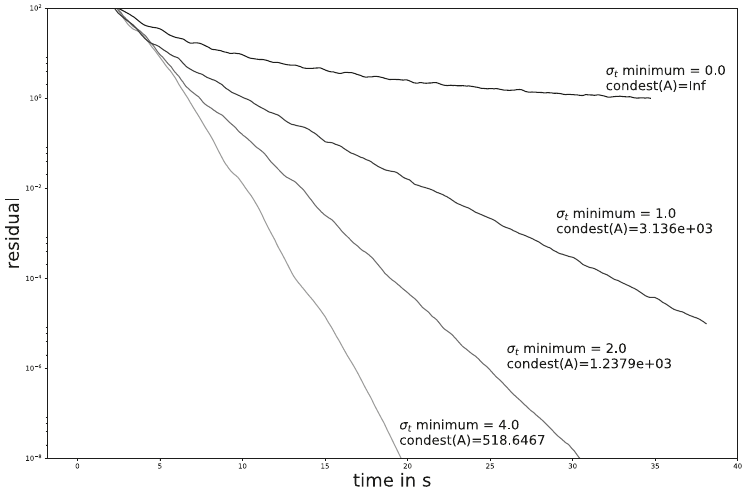}
%\missingfigure{figures fig nebulae p1 convergence}
%\vspace{-0.1in}
\icaption{Convergence behaviour of our solver with $N=1$ for the nebulae dataset and for varying minimum thresholds of the extinction coefficient $\sigma_t$. Threshold values and an estimate for the condition number of $A$ (MATLAB's \texttt{condest} function) are shown next to the plots. The convergence deteriorates as the threshold decreases. Once it reaches zero, the presence of pure vacuum makes the condition number infinite. \vspace{-0.2in}}
\label{fig:results_convergence}
\end{figure}

For our test case, we choose a FD resolution of $80\times80\times80$, an extinction coefficient $\sigma_t=8.0$ and albedo $\alpha=0.9$. We run the solver for different truncation values $N$. In Figure~\ref{fig:pointsource}~\subref{fig:pointsource_pn}, we see that the solution becomes increasingly accurate for higher truncation order. The ground truth is underestimated when $N$ is odd, and overestimated when $N$ is even. We further see in ~\ref{fig:pointsource}~\subref{fig:pointsource_p5} that the $P_1$ solution exactly matches the results from CDA, which confirms that the latter is only a collapsed version of the former. The time to solve is significant with $10m$ for $P_1$ and $45m$ with $P_5$. With these performance characteristics, our $P_N$-solver is clearly not competitive in comparison with our CDA and FLD solver, which are much faster.

\tocless\subsection{Nebulae}

Finally, we run our solver on a procedural cloud dataset to get an idea of its performance in more practical applications. Figure~\ref{fig:teaser}\subref{fig:nebulae_ours} shows the result of $P_5$ for a procedurally generated heterogeneous cloud with an isotropic phase function, with path-traced ground truth in Figure~\ref{fig:teaser}\subref{fig:teaser_gt}. We see that at order $N=5$, our method can capture indirect illumination similarly well as FLD (\ref{fig:teaser}\subref{fig:teaser_fld}) and is significantly better than CDA (\ref{fig:teaser}\subref{fig:teaser_cda}) as expected. The indirectly illuminated region at the bottom appears to be closer to the path-traced result as opposed to the solution from FLD which is very flat in that region. However, in many other areas, $P_5$ seems to still suffer a lot from the problem of energy loss near transitions from dense regions to vacuum. It appears that going higher order a few steps mitigates this only slowly at a high cost of compute and storage.
The main characteristic of the nebulae dataset is the presence of vacuum. We found that having vacuum regions in the dataset will cause the condition number to become infinite and the solver practically does not converge. We therefore introduced a minimum threshold for the extinction coefficient $\sigma_t$. Every voxel with an extinction coefficient smaller than the threshold is set to the threshold value. In Figure~\ref{fig:results_convergence} we show the effect of the minimum threshold on the convergence behaviour. As it increases, convergence improves.

%\begin{table}[h]
%\begin{center}
%\begin{tabular}{ c | c c c }
% & \textbf{2D Checkerboard} & \textbf{Point Source} & \textbf{Nebulae} \\
%\hline
%$P_1$ & 0.27s & 10m (64\times 64\times 64) & cell6 \\
%$P_2$ & 2.4s & cell8 & cell9 \\
%$P_3$ & 5.8s & cell8 & cell9 \\
%$P_4$ & 12.7s & cell8 & cell9 \\
%$P_5$ & 25s & ? & ?
%\end{tabular}
%\end{center}
%\icaption{Performance measurements for different datasets and varying truncation order $N$}
%\label{table2}
%\end{table}

%\begin{itemize}
%	\item mention performance.
%	\item run for various values of $N$
%	\item convergence behaviour in vacuum. Clarify
%	\item run for anisotropic problem
%\end{itemize}

%-------------------------------------------------------------------------
\newpage
\tocless\section{Conclusion \label{sec:conclusion}}

In this paper we introduced the $P_N$-method to the toolbox of deterministic methods for rendering participating media in computer graphics. We derived and presented the real-valued $P_N$-equations, along with a staggered grid solver with numerical stencils constructed automatically from the equations via a computer algebra system. We showed how to use the results in a rendering system and ran our solver for various standard problems for validation.

We originally set out to understand how non-linear diffusion methods compare to the $P_N$-method for increasing order. Our results indicate that although the lack of higher moments makes the FLD solution overly smooth, its energy conserving nature and comparably small resource footprint make it a better approach at present for most graphics applications compared to the $P_N$-method, which becomes increasingly costly for higher values of $N$.

The literature in other fields often states that the $P_N$ method---when solved in normal form as we do---is able to deal with vacuum regions. We found this misleading. The $P_N$-method in normal form indeed does not break down completely in the presence of vacuum as diffusion based methods do (due to $\sigma_t^{-1}$ in the diffusion coefficient). However, in the presence of vacuum, the condition number of the system matrix becomes infinite and does not converge either. Therefore $P_N$ based methods also require minimum thresholding of the extinction coefficient and offer no benefit for vacuum regions.

Much more work needs to be done in order to make the $P_N$-method competitive in performance to alternative solutions for volume rendering. We believe this can be made possible by employing a multigrid scheme for solving the linear system of equations. We implemented a multigrid solver, but did not find the expected performance improvements. This is possibly due to the coupling between coefficients within a voxel, which does not work well together with the smoothing step. We want to study this in the future.

Unique to our system is that it uses a computer algebra representation of the equation to solve as input. Discretization in angular and spatial domain is done using manipulation passes on the representation. The stencil code, which is used to populate the system of linear equations, is generated from the expression tree. This way, we can easily deal with coupled PDEs and avoid the time consuming and error prone process of writing stencil code by hand.

When researching the application of the $P_N$-method in other fields, we came across a rich variety of variations, such as simplified $P_N$ ($SP_N$)~\cite{Ryan10}, filtered $P_N$ ($FP_N$)~\cite{Radice13}, diffusion-correction $P_N$ ($DP_N$)~\cite{Schaefer11} and least-squares $P_N$ ($LSP_N$)~\cite{Hansen14}. These variations have been introduced to address certain problems of the standard $P_N$-method, such as ringing artifacts, dealing with vacuum regions and general convergence. Our solver can be applied to any (potentially coupled) PDE and therefore can generate stencil code for all these variations by simply expressing the respective PDEs in our computer algebra representation and providing this as an input to our solver. This would allow an exhaustive comparison of all these methods and we consider this as future work.

Finally, since the approach of our solver is very generic, we also would like to explore its application to other simulation problems in computer graphics.

% --------------------------------------------------------------------------------------------------

\section*{Acknowledgements}

We thank the anonymous reviewers for their valuable and encouraging comments and feedback. We also thank Martin Frank and Benjamin Seibold for the very valuable discussions and answers to our questions. We thank Bernd Eberhardt for feedback and support. This project has been partially funded by the MSC-BW project.

%-------------------------------------------------------------------------

\bibliographystyle{eg-alpha}

\tocless\bibliography{paper}

\newcommand{\etalchar}[1]{$^{#1}$}
\begin{thebibliography}{\uppercase{RARO13}}

\bibitem[Bru02]{Brunner02}
\textsc{{Brunner} T.~A.}:
\newblock {Forms of Approximate Radiation Transport}.
\newblock \emph{Tech. Rep. SAND2002-1778, Sandia National
Laboratories} (2002).

\bibitem[Cha60]{Chandrasekhar60}
\textsc{Chandrasekhar S.}:
\newblock \emph{Radiative Transfer}.
\newblock Dover Publications, 1960.

\bibitem[dI11]{dEon11}
\textsc{d'Eon E., Irving G.}:
\newblock A Quantized-Diffusion Model for Rendering Translucent Materials.
\newblock \emph{ACM TOG (Proc. of SIGGRAPH) 30}, 4 (July 2011), 56:1--56:14.

\bibitem[HPM{\etalchar{*}}14]{Hansen14}
\textsc{Hansen J., Peterson J., Morel J., Ragusa J., Wang Y.}:
\newblock A Least-Squares Transport Equation Compatible with Voids.
\newblock \emph{Journal of Computational and Theoretical Transport 43}, 1-7
  (2014), 374--401.

\bibitem[JCJ09]{Jarosz09}
\textsc{Jarosz, W., Carr, N.~A., Jensen, H.~W.}:
\newblock Importance Sampling Spherical Harmonics. 
\newblock \emph{Computer Graphics Forum (Proceedings of Eurographics)} (2009), 577--586.

\bibitem[Kaj86]{Kajiya86}
\textsc{Kajiya J.~T.}:
\newblock The rendering equation.
\newblock \emph{Computer Graphics (Proc. of SIGGRAPH)} (1986), 143--150.

\bibitem[KPS{\etalchar{*}}14]{Koerner14}
\textsc{Koerner D., Portsmouth J., Sadlo F., Ertl T., Eberhardt B.}:
\newblock Flux-limited Diffusion for Multiple Scattering in Participating
  media.
\newblock \emph{CoRR abs/1403.8105} (2014).

\bibitem[KVH84]{Kajiya84}
\textsc{Kajiya J.~T., Von~Herzen B.~P.}:
\newblock Ray tracing volume densities.
\newblock \emph{Computer Graphics (Proc. of SIGGRAPH) 18}, 3 (Jan. 1984),
  165--174.

\bibitem[LP81]{Levermore81}
\textsc{{Levermore} C.~D., {Pomraning} G.~C.}:
\newblock {A flux-limited diffusion theory}.
\newblock \emph{Astrophysical Journal 248} (1981), 321--334.

\bibitem[Max95]{Max95}
\textsc{Max N.}:
\newblock \emph{Efficient Light Propagation for Multiple Anisotropic Volume
  Scattering}.
\newblock Springer Berlin Heidelberg, Berlin, Heidelberg, 1995, pp.~87--104.

\bibitem[McC10]{Ryan10}
\textsc{McClarren R.~G.}:
\newblock Theoretical Aspects of the Simplified Pn Equations.
\newblock \emph{Transport Theory and Statistical Physics 39}, 2-4 (2010),
  73--109.

\bibitem[MGN17]{Muller17}
\textsc{Muller, T., Gross, M., Novak, J.}:
\newblock Practical Path Guiding for Efficient Light Transport Simulation.
\newblock \emph{Computer Graphics Forum} (2017), 36:91--36:100

\bibitem[NGHJ18]{Novak18}
\textsc{Novak, J., Georgiev, I., Hanika, J., Jarosz, W.}:
\newblock Monte Carlo Methods for Volumetric Light Transport Simulation. 
\newblock \emph{Computer Graphics Forum (Proceedings of Eurographics - State of the Art Reports)} (2018), 37(2).

\bibitem[OAH00]{Olson00}
\textsc{Olson G.~L., Auer L.~H., Hall M.~L.}:
\newblock Diffusion, P1, and other approximate forms of radiation transport.
\newblock \emph{Journal of Quantitative Spectroscopy and Radiative Transfer
  64}, 6 (2000), 619 -- 634.

\bibitem[RARO13]{Radice13}
\textsc{Radice D., Abdikamalov E., Rezzolla L., Ott C.~D.}:
\newblock A New Spherical Harmonics Scheme for Multi-Dimensional Radiation
  Transport I. Static Matter Configurations.
\newblock \emph{Journal of Computational Physics 242} (2013), 648 -- 669.

\bibitem[SF14]{Seibold14}
\textsc{Seibold B., Frank M.}:
\newblock StaRMAP---A second order staggered grid method for spherical
  harmonics moment equations of radiative transfer.
\newblock \emph{ACM Trans. Math. Softw. 41}, 1 (Oct. 2014), 4:1--4:28.

\bibitem[SFL11]{Schaefer11}
\textsc{Sch{\"a}fer M., Frank M., Levermore C.~D.}:
\newblock Diffusive Corrections to Pn Approximations.
\newblock \emph{Multiscale Modeling and Simulation 9} (2011), 1--28.

\bibitem[Sta95]{Stam95}
\textsc{Stam J.}:
\newblock Multiple scattering as a diffusion process.
\newblock In \emph{Proc. of Eurographics Workshop on Rendering Techniques}.
  Springer-Verlag, 1995, pp.~41--50.

\end{thebibliography}

%-------------------------------------------------------------------------

\newpage

\onecolumn

\renewcommand{\familydefault}{\sfdefault}

\begin{appendices}

%%%%%%%%%%%%%%%%%%%%%%%%%%%%%%%%%%%%%%%%%%%%%%%%%%%%%%%%%%%%%%%%%%%%%%%%%%%%%%%%%%%%%%%%%%%%%%%%%%%%%%%%%%%%%%%%%%%%%
\tocless\section{Full derivation of the $P_N$-equations \label{appendix}}
%%%%%%%%%%%%%%%%%%%%%%%%%%%%%%%%%%%%%%%%%%%%%%%%%%%%%%%%%%%%%%%%%%%%%%%%%%%%%%%%%%%%%%%%%%%%%%%%%%%%%%%%%%%%%%%%%%%%%

\end{appendices}

This appendix presents the detailed derivations of the complex-valued and real-valued $P_N$-equations. This was actually performed semi-automatically using our computer algebra representation of the equations, guaranteeing correctness, and here we just report the result of each stage in the derivation.

The starting point is the radiative transfer equation (RTE), which expresses the change of the radiance field $L$, with respect to an infinitesimal change of position into direction $\omega$ at point $\vec{x}$:
% RTE ------------
\begin{align}
%\label{eq:rte}
\left(\nabla\cdot\omega\right)L\left(\vec{x}, \omega \right)
=&
-\sigma_t\left(\vec{x}\right) L\left(\vec{x}, \omega \right)\nonumber\\
&
+\sigma_s\left(\vec{x}\right) \int_{\Omega}
{
p\left(\omega'\cdot\omega\right)L\left(\vec{x}, \omega' \right)\ud\omega'
} \ . \nonumber\\
&
+Q\left(\vec{x}, \omega\right)\nonumber
\end{align}
where the left hand side (LHS) is the transport term, and we refer to the terms on the right hand side (RHS) as collision, scattering, and source term, respectively. The symbols $\sigma_t$, $\sigma_s$, $p$, and $Q$ refer to the extinction coefficient, scattering coefficient, phase function and emission term.

The derivation of the $P_N$-equations is then done in two steps. First, the directional-dependent quantities are replaced by their SH-projected counterparts. For example the radiance field $L$ is replaced by its SH projection. This way the quantities are expressed in spherical harmonics, but still depend on direction $\omega$. In the second step, the RTE is projected into spherical harmonics, which is done by multiplying each term with the complex conjugate of the SH basis functions.

The SH basis functions are complex, which produces complex coefficients and complex $P_N$-equations. However, there are also real SH basis functions, which are defined in terms of the complex SH basis functions and which produce real coefficients and reconstructions. Since the radiance field $L$ is real, it is more convenient to work with the real SH basis functions.

In the next section, the complex $P_N$-equations are derived. In order to give the derivation a clearer structure, the two steps mentioned above are applied to each term individually in a separate subsection. The section concludes by putting all derived terms together. In Section~\ref{sec:real} the analogous derivation is followed to obtain the real $P_N$-equations which are used in the article.

\setcounter{tocdepth}{2}
\tableofcontents
\addtocontents{toc}{\protect\setcounter{tocdepth}{3}}

\section{Derivation of the complex-valued $P_N$-equations}

Deriving the $P_N$-equations consists of two main steps. First, all angular dependent quantities in the RTE are expressed in terms of spherical harmonics (SH) basis functions. After this, the RTE still depends on the angular variable. Therefore, the second step projects each term of the RTE by multiplying with the complex conjugate of the SH basis functions, followed by integration over solid angle to integrate out the angular variable. This gives an equation for each spherical harmonics coefficient.

Spherical Harmonics are a set of very popular and well known functions on the sphere. The complex-valued SH basis functions are given by
% complex valued SH ------------------
\begin{align}
\SHBC^{l,m}(\omega) = \SHBC^{l,m}(\theta, \phi)=
\begin{cases}
(-1)^m\sqrt{\frac{2l+1}{4\pi}\frac{(l-m)!}{(l+m)!}}e^{im\phi}P^{l,m}\left(\operatorname{cos}\left(\theta\right)\right), & \text{for $m\ge0$}\\
\left(-1\right)^m\overline{\SHBC^{l\left|m\right|}}(\theta, \phi), & \text{for $m<0$}
\end{cases}
\end{align}
where $P^{l,m}$ are the associated Legendre polynomials. The $\left(-1\right)^m$ factor is called the Condon-Shortley phase and is not part of the associated Legendre Polynomial (unlike some other definitions).

\subsection{Projecting Radiative Transfer Quantities}

% -----------------------------
\subsubsection{Radiance Field $L$ and Emission Field $Q$}
\label{sec:complex_proj_L}

Radiative transfer quantities, which depend on position $\vec{x}$ and angle $\omega$, are projected into spatially dependent SH coefficients for each SH basis function:
\begin{align}
L^{l,m}\left(\vec{x}\right)
&=
\int_\Omega{L\left(\vec{x}, \omega\right)\overline{\SHBC^{l,m}}\ud\omega} \ . \nonumber\\
Q^{l,m}\left(\vec{x}\right)
&=
\int_\Omega{Q\left(\vec{x}, \omega\right)\overline{\SHBC^{l,m}}\ud\omega} \ . \nonumber
\end{align}
The function is completely reconstructed by using all SH basis functions up to infinite order. The $P_N$-equations introduce a truncation error by only using SH basis functions up to order $N$ for the reconstruction $\hat{L}$ and $\hat{Q}$:
\begin{align}
\label{eq:sh_exp_L}
L\left(\vec{x}, \omega\right)
\approx
\hat{L}\left(\vec{x}, \omega\right) =
\sum_{l=0}^{N}
{
\sum_{m=-l}^{l}
{
L^{l,m}\left(\vec{x}\right)\SHBC^{l,m}\left(\omega\right)
}
}
=
\sum_{l,m}
{
L^{l,m}\left(\vec{x}\right)\SHBC^{l,m}\left(\omega\right)
} \ . \nonumber\\
Q\left(\vec{x}, \omega\right)
\approx
\hat{Q}\left(\vec{x}, \omega\right) =
\sum_{l=0}^{N}
{
\sum_{m=-l}^{l}
{
Q^{l,m}\left(\vec{x}\right)\SHBC^{l,m}\left(\omega\right)
}
}
=
\sum_{l,m}
{
Q^{l,m}\left(\vec{x}\right)\SHBC^{l,m}\left(\omega\right)
} \ .
\end{align}

% -------------------------------------
\subsubsection{Phase Function}
\label{sec:complex_proj_phase}

Throughout our article, we assume an isotropic phase function, which only depends on the angle between incident and outgoing vector $\omega_i$ and $\omega_o$ (note that in the graphics literature, these would often be called anisotropic). We will see later in section~\ref{sec:complex_scattering_term}, that this allows us to fix the outgoing vector $\omega_o$ at the pole axis $\vec{e}_3$ and compute the phase function SH coefficients by just varying the incident vector $\omega_i$.
%\operatorname{cos}\theta
\begin{align*}
p^{l,m}
=
\int_\Omega
{
p\left(\omega_i\cdot\vec{e}_3\right)
\SHBC^{l,m}\left(\omega_i\right)
\ud\omega_i
} \ .
\nonumber
\end{align*}

The expansion of the phase function can be further simplified because the phase function is rotationally symmetric around the pole axis $\vec{e}_3$. Consider the definition of the spherical harmonics basis function $\SHBC^{l,m}$:
\begin{align*}
\SHBC^{l,m}(\theta, \phi) = C^{l,m}e^{im\phi}P^{l,m}(\operatorname{cos}(\theta)) \ .
\end{align*}
Now we apply a rotation $R(\alpha)$ of $\alpha$ degrees around the pole axis. In spherical harmonics, this is expressed as:
\begin{align*}
\rho_{R(\alpha)}(\SHBC^{l,m}) = e^{-i m\alpha}\SHBC^{l,m} \ .
\end{align*}
If the phase function is rotationally symmetric around the pole axis, we have:
\begin{align*}
\rho_{R(\alpha)}(p) = p \ .
\end{align*}
and in spherical harmonics this would be:
\begin{align*}
\sum_{l,m}
{
e^{-i m\alpha}
p^{l,m}
\SHBC^{l,m} }\left(\omega_i\right)
=
\sum_{l,m}
{
p^{l,m}
\SHBC^{l,m}\left(\omega_i\right)
} \ .
\end{align*}
By equating coefficients we get:
\begin{align*}
p^{l,m} = p^{l,m}e^{-i m\alpha} \ .
\end{align*}

Since $e^{-i m\alpha}=1$ for all $\alpha$ only when $m=0$, we can conclude that $p^{l,m} = 0$ for all $m\ne0$. This means that for a function which is rotationally symmetric around the pole axis, only the $m=0$ coefficients will be valid. Therefore, our phase function reconstruction for a fixed outgoing vector ($\omega_o=\vec{e}_3$) only requires SH coefficients with $m=0$:
\begin{align}
\label{eq:sh_exp_phase}
p(\omega_i) =
\sum_l
{
p^{l0}
\SHBC^{l0}(\omega_i)
} \ .
\end{align}

\subsection{Projecting Terms of the RTE}

% ------------------------------------------------------------
\subsubsection{Transport Term}
\label{sec:complex_transport_term}

The transport term of the RTE is given as
\begin{align*}
(\omega\cdot\nabla)L(\vec{x}, \omega)
\end{align*}
Replacing $L$ with its expansion gives:
\begin{align*}
\left(\omega\cdot\nabla\right)
\left(
\sum_{l,m}
{
L^{l,m}\left(\vec{x}\right )
\SHBC^{l,m}\left(\omega\right)
}
\right)
\end{align*}

Next we multiply with $\overline{\SHBC^{l'm'}}(\omega)$ and integrate over solid angle:
\begin{align*}
\int_\Omega
{
\overline{Y^{l'm'}}(\omega)(\omega\cdot\nabla)
\sum_{l,m}
{
L^{l,m}\left(\vec{x}\right)
\SHBC^{l,m}\left(\omega\right)
}
}
\ud\omega
\end{align*}

We can pull the spatial derivative out of the integral to get:
\begin{align}
\nabla\cdot\int_\Omega
{
\omega\overline{\SHBC^{l'm'}}(\omega)
\sum_{l,m}
{
L^{l,m}\left(\vec{x}\right)
\SHBC^{l,m}\left(\omega\right)
}
\ud\omega
}
\label{eq:complex_transport_a}
\end{align}

We apply the following recursive relation for the spherical harmonics basis functions:
\begin{align}
\omega\;\overline{\SHBC^{l,m}} = \frac{1}{2}
\begin{pmatrix}
\ c^{l-1, m-1}\overline{\SHBC^{l-1,m-1}} - d^{l+1, m-1}\overline{\SHBC^{l+1,m-1}} - e^{l-1, m+1}\overline{\SHBC^{l-1,m+1}} + f^{l+1, m+1}\overline{\SHBC^{l+1,m+1}}\\
i\left(-c^{l-1, m-1}\overline{\SHBC^{l-1,m-1}} + d^{l+1, m-1}\overline{\SHBC^{l+1,m-1}} - e^{l-1, m+1}\overline{\SHBC^{l-1,m+1}} + f^{l+1, m+1}\overline{\SHBC^{l+1,m+1}}\right) \\
2\left(a^{l-1, m}\overline{\SHBC^{l-1,m}}+b^{l+1, m}\overline{\SHBC^{l+1,m}}\right)
\end{pmatrix}
\label{eq:recursive_relation}
\end{align}

with
\begin{align*}
a^{l,m}= \sqrt{\frac{\left(l-m+1\right)\left(l+m+1\right)}{\left(2l+1\right)\left(2l-1\right)}} \, , \qquad
b^{l,m}= \sqrt{\frac{\left(l-m\right)\left(l+m\right)}{\left(2l+1\right)\left(2l-1\right)}} \, , \qquad
c^{l,m}= \sqrt{\frac{\left(l+m+1\right)\left(l+m+2\right)}{\left(2l+3\right)\left(2l+1\right)}}\\
d^{l,m}= \sqrt{\frac{\left(l-m\right)\left(l-m-1\right)}{\left(2l+1\right)\left(2l-1\right)}} \, , \qquad
e^{l,m}= \sqrt{\frac{\left(l-m+1\right)\left(l-m+2\right)}{\left(2l+3\right)\left(2l+1\right)}} \, , \qquad
f^{l,m}= \sqrt{\frac{\left(l+m\right)\left(l+m-1\right)}{\left(2l+1\right)\left(2l-1\right)}}
\end{align*}

Note that the signs for the $x$- and $y$- component depend on the handedness of the coordinate system in which the SH basis functions are defined. Using this in Equation~\ref{eq:complex_transport_a} gives
\begin{align*}
\begin{pmatrix}
\frac{1}{2}\partial_x\\
\frac{i}{2}\partial_y\\
\partial_z
\end{pmatrix}
\cdot
\int_\Omega
&
\begin{pmatrix}
\ c^{l'-1, m'-1}\overline{\SHBC^{l'-1,m'-1}} - d^{l'+1, m'-1}\overline{\SHBC^{l'+1,m'-1}} - e^{l'-1, m'+1}\overline{\SHBC^{l'-1,m'+1}} + f^{l'+1, m'+1}\overline{\SHBC^{l'+1,m'+1}}\\
-c^{l'-1, m'-1}\overline{\SHBC^{l'-1,m'-1}} + d^{l'+1, m'-1}\overline{\SHBC^{l'+1,m'-1}} - e^{l'-1, m'+1}\overline{\SHBC^{l'-1,m'+1}} + f^{l'+1, m'+1}\overline{\SHBC^{l'+1,m'+1}} \\
a^{l'-1, m'}\overline{\SHBC^{l'-1,m'}}+b^{l'+1, m'}\overline{\SHBC^{l'+1,m'}}
\end{pmatrix}
&\sum_{l,m}{
L^{l,m}\left(\vec{x}\right )\SHBC^{l,m}\left(\omega\right)
}
\ud\omega
\end{align*}

Integrating the vector term over solid angle can be expressed as separate solid angle integrals over each component. These integrals over a sum of terms are split into separate integrals. We arrive at:
\begin{align*}
\begin{pmatrix}
\frac{1}{2}\partial_x\\
\frac{i}{2}\partial_y\\
\partial_z
\end{pmatrix}
\cdot
\begin{pmatrix}
\ c^{l'-1, m'-1}\sum_{l,m}{L^{l,m}\left(\vec{x}\right )\int_\Omega{\overline{\SHBC^{l'-1,m'-1}}\left(\omega\right)\SHBC^{l,m}\left(\omega\right)\ud\omega}} \quad - \quad ...\\
-c^{l'-1, m'-1}\sum_{l,m}{L^{l,m}\left(\vec{x}\right )\int_\Omega{\overline{\SHBC^{l'-1,m'-1}}\left(\omega\right)\SHBC^{l,m}\left(\omega\right)\ud\omega}} \quad + \quad ... \\
a^{l'-1, m'}\sum_{l,m}{L^{l,m}\left(\vec{x}\right )\int_\Omega{\overline{\SHBC^{l'-1,m'}}\left(\omega\right)\SHBC^{l,m}\left(\omega\right)\ud\omega}} \quad + \quad ...
\end{pmatrix}
\end{align*}

Applying the orthogonality property to the solid angle integrals will will select specific $l,m$ in each term:
\begin{align*}
\begin{pmatrix}
\frac{1}{2}\partial_x\\
\frac{i}{2}\partial_y\\
\partial_z
\end{pmatrix}
\cdot
\begin{pmatrix}
\ c^{l-1, m-1}L^{l-1,m-1} - d^{l+1, m-1}L^{l+1,m-1} - e^{l-1, m+1}L^{l-1,m+1} + f^{l+1, m+1}L^{l+1,m+1}\\
-c^{l-1, m-1}L^{l-1,m-1} + d^{l+1, m-1}L^{l+1,m-1} - e^{l-1, m+1}L^{l-1,m+1} + f^{l+1, m+1}L^{l+1,m+1} \\
a^{l-1, m}L^{l-1,m}+b^{l+1, m}L^{l+1,m}
\end{pmatrix}
\end{align*}

Which gives the final moment equation for the transport term:
\begin{align*}
=
&\frac{1}{2}\partial_x\left(c^{l-1, m-1}L^{l-1,m-1} - d^{l+1, m-1}L^{l+1,m-1} - e^{l-1, m+1}L^{l-1,m+1} + f^{l+1, m+1}L^{l+1,m+1}\right) + \\
&\frac{i}{2}\partial_y\left( -c^{l-1, m-1}L^{l-1,m-1} + d^{l+1, m-1}L^{l+1,m-1} - e^{l-1, m+1}L^{l-1,m+1} + f^{l+1, m+1}L^{l+1,m+1} \right) + \\
&\partial_z\left( a^{l-1, m}L^{l-1,m}+b^{l+1, m}L^{l+1,m} \right)
\\
=
&
\frac{1}{2}c^{l-1, m-1}\partial_x L^{l-1,m-1} - \frac{1}{2}d^{l+1, m-1}\partial_x L^{l+1,m-1} - \frac{1}{2}e^{l-1, m+1}\partial_x L^{l-1,m+1} + \frac{1}{2}f^{l+1, m+1}\partial_x L^{l+1,m+1}+
\\
&-\frac{i}{2}c^{l-1, m-1}\partial_y L^{l-1,m-1} + \frac{i}{2}d^{l+1, m-1}\partial_y L^{l+1,m-1} - \frac{i}{2}e^{l-1, m+1}\partial_y L^{l-1,m+1} + \frac{i}{2}f^{l+1, m+1}\partial_y L^{l+1,m+1}+
\\
&
a^{l-1, m}\partial_z L^{l-1,m}+b^{l+1, m}\partial_z L^{l+1,m} \ .
\end{align*}

% ------------------------------------------------------------
\subsubsection{Collision Term}

The collision term of the RTE is given as:
\begin{align*}
-\sigma_t\left(\vec{x}\right)L\left(\vec{x}, \omega\right)
\end{align*}
We first replace the radiance field $L$ with its spherical harmonics expansion:
\begin{align*}
-\sigma_t\left(\vec{x}\right)
\sum_{l,m}
{
L^{l,m}\left(\vec{x}\right )\SHBC^{l,m}\left(\omega\right)
}
\end{align*}
Multiplying with $\overline{\SHBC^{l'm'}}$ and integrating over solid angle gives, after pulling some factors out of the integral:
\begin{align*}
&-\sigma_t\left(\vec{x}\right)\sum_{l,m}{L^{l,m}\left(\vec{x}\right )\int_\Omega\overline{\SHBC^{l'm'}}\left(\omega\right)\SHBC^{l,m}\left(\omega\right)\ud\omega}\\
&= -\sigma_t\left(\vec{x}\right)\sum_{l,m}{L^{l,m}\left(\vec{x}\right )\delta_{ll'}\delta_{mm'}}\\
&= -\sigma_t\left(\vec{x}\right)L^{l,m}\left(\vec{x}\right ) \ .
\end{align*}

% ------------------------------------------------------------
\subsubsection{Scattering Term}
\label{sec:complex_scattering_term}

The scattering term in the RTE is given as:
\begin{align*}
\sigma_s(\vec{x})\int_{\Omega}p(\vec{x}, \omega'\cdot\omega)L(\vec{x}, \omega')\ud\omega'
\end{align*}

The phase function used in isotropic scattering medium only depends on the angle between incident and outgoing direction and therefore is rotationally symmetric around the pole-defining axis. This property allows us to define a rotation $R(\omega)$, which rotates the phase function such that the pole axis aligns with direction vector $\omega$. The rotated phase function is defined as:
\begin{align*}
\rho_{R(\omega)}(p)
\end{align*}
where $\rho$ is the rotation operator, which can be implemented by applying the inverse rotation $R(\omega)^{-1}$ to the arguments of $p$. With this rotated phase function, we now can express the integral of the scattering operator as a convolution denoted with the symbol $\circ$:
\begin{align}
\int_{\Omega}p(\vec{x}, \omega'\cdot\omega)L(\vec{x}, \omega')\ud\omega'
&=
L\circ \rho_{R(\omega)}(p) \nonumber\\
&=
\int_{\Omega'}{L(\vec{x}, \omega')\rho_{R(\omega)}(p)(\omega')\ud\omega'} \nonumber\\
&= \langle L,  \rho_{R(\omega)}(p)\rangle \ .
\label{eq:complex_scatt_conv}
\end{align}
As we evaluate the inner product integral of the convolution, the phase function rotates along with the argument $\omega$.

We now use the spherical harmonics expansions of $L$ (Equation~\ref{eq:sh_exp_L}) and $p$ (Equation~\ref{eq:sh_exp_phase}) in the definition for the inner product of our convolution (Equation~\ref{eq:complex_scatt_conv}):
\begin{align*}
\langle L,  \rho_{R(\omega)}(p)\rangle = \left < \sum_{l,m}{L^{l,m}(\vec{x}) \SHBC^{l,m}}, \rho_{R(\omega)}\left ( \sum_{l}{p^{l0}\SHBC^{l0}} \right )\right> \ .
\end{align*}
Due to linearity of the inner product operator, we can pull out the non-angular dependent parts of the expansions:
\begin{align*}
\langle L,  \rho_{R(\omega)}(p)\rangle
&=
\sum_{l,m}
{
L^{l,m}(\vec{x})
\left<
\SHBC^{l,m},
\rho_{R(\omega)}
\left(\sum_l{p^{l0} \SHBC^{l0}}\right)
\right>
} \ .
\end{align*}
and further:
\begin{align*}
\langle L,  \rho_{R(\omega)}(p)\rangle
&=
\sum_{l'}
{
\sum_{l,m}
{
p^{l'0}L^{l,m}(\vec{x})
\left<\SHBC^{l,m}, \rho_{R(\omega)}\left( \SHBC^{l'0} \right)\right>
}
} \ .
\end{align*}

The rotation $\rho_{R(\omega)}$ of a function with frequency $l$ gives a function of frequency $l$ again. In addition the spherical harmonics basis functions $\SHBC^{l,m}$ are orthogonal. We therefore have:
\begin{align*}
\left<
\SHBC^{l,m}, \rho_{R(\omega)}\left(\SHBC^{l'm'}\right)
\right> = 0       \qquad    \text{for all}\ \ l\ne l'
\end{align*}
which further simplifies our inner product integral to:
\begin{align*}
\langle L,  \rho_{R(\omega)}(p)\rangle
&=
\sum_{l,m}
{
p^{l0}L^{l,m}(\vec{x})
\left<
\SHBC^{l,m}, \rho_{R(\omega)}\left(\SHBC^{l0} \right )
\right>
} \ .
\end{align*}

What remains to be resolved is the inner product. We use the fact that the spherical harmonics basis functions $ \SHBC^{l,m}$ are eigenfunctions of the inner product integral operator in the equation above:
\begin{align*}
\left<
\SHBC^{l,m}, \rho_{R(\omega)}\left ( \SHBC^{l0} \right )\right> = \lambda_l \SHBC^{l,m}
\end{align*}
with
\begin{align*}
\lambda_l=\sqrt{\frac{4\pi}{2l+1}} \ .
\end{align*}
Replacing the inner product gives:
\begin{align*}
\langle L,  \rho_{R(\omega)}(p)\rangle
&=
\sum_{l,m}
{
\lambda_l
p^{l0}L^{l,m}(\vec{x})
\SHBC^{l,m}
} \ .
\end{align*}

This allows us to express the scattering term using SH expansions of phase function $p$ and radiance field $L$:
\begin{align*}
\sigma_s(\vec{x})\int_{\Omega}p(\vec{x}, \omega'\cdot\omega)L(\vec{x}, \omega')\ud\omega'
&=
\sigma_s(\vec{x})\langle L,  \rho_{R(\omega)}(p)\rangle
\\
&=
\sigma_s(\vec{x})
\sum_{l,m}
{
\lambda_l
p^{l0}L^{l,m}(\vec{x})
\SHBC^{l,m}
} \ .
\end{align*}

However, we haven't done a spherical harmonics expansion of the scattering term itself. It is still a scalar function which depends on direction $\omega$. We thus project the scattering term into spherical harmonics by multiplying with $\overline {\SHBC^{l'm'}}$ and integrating over solid angle $\omega$. We further pull out all factors which do not depend on $\omega$, and apply the SH orthogonality property to arrive at the scattering term of the complex-valued $P_N$-equations:
\begin{align}
&
\int_{\Omega}
{
\overline{\SHBC^{l'm'}}(\omega)
\sigma_s(\vec{x})
\sum_{l,m}
{
\lambda_l
p^{l0}L^{l,m}(\vec{x})
\SHBC^{l,m}\left(\omega\right)
}
\ud\omega
}
\nonumber\\
=&
\lambda_l
\sigma_s(\vec{x})
p^{l0}L^{l,m}(\vec{x})
\sum_{l,m}
{
\int_{\Omega}
{
\overline{\SHBC^{l'm'}}(\omega)
\SHBC^{l,m}\left(\omega\right)
\ud\omega
}
}
\nonumber\\
=&
\lambda_l
\sigma_s(\vec{x})
p^{l0}L^{l,m}(\vec{x})
\sum_{l,m}
{
\delta_{ll'}\delta_{mm'}
}
\nonumber\\
=&
\lambda_l
\sigma_s(\vec{x})
p^{l0}L^{l,m}(\vec{x}) \ .
\end{align}

% --------------------------------------------------------------
\subsubsection{Emission Term}

The emission term of the RTE is given as:
\begin{align}
Q\left(\vec{x}, \omega\right)
\end{align}
The derivation of the SH projected term is equivalent to the derivation of the projected collision term. Replacing the emission field with its SH projection and multiplying the term with the conjugate complex of $\SHBC$ results, after applying the orthogonality property, in:
\begin{align}
Q^{l,m}\left(\vec{x}, \omega\right)
\end{align}

% --------------------------------------------------------------
\subsection{Final Equation}

We arrive at the complex-valued $P_N$-equations after putting all the projected terms together:
\begin{align*}
\frac{1}{2}c^{l-1, m-1}\partial_x L^{l-1,m-1} - \frac{1}{2}d^{l+1, m-1}\partial_x L^{l+1,m-1} - \frac{1}{2}e^{l-1, m+1}\partial_x L^{l-1,m+1} + \frac{1}{2}f^{l+1, m+1}\partial_x L^{l+1,m+1}+
\\
-\frac{i}{2}c^{l-1, m-1}\partial_y L^{l-1,m-1} + \frac{i}{2}d^{l+1, m-1}\partial_y L^{l+1,m-1} - \frac{i}{2}e^{l-1, m+1}\partial_y L^{l-1,m+1} + \frac{i}{2}f^{l+1, m+1}\partial_y L^{l+1,m+1} +
\\
a^{l-1, m}\partial_z L^{l-1,m}+b^{l+1, m}\partial_z L^{l+1,m}
=
-\sigma_t\left(\vec{x}\right)L^{l,m}\left(\vec{x}\right )
+
\lambda_l
\sigma_s(\vec{x})
p^{l0}L^{l,m}(\vec{x}) + Q^{l,m}\left(\vec{x}, \omega\right) \ .
\end{align*}

\section{Derivation of the real-valued $P_N$-equations \label{sec:real}}

The real-valued $P_N$-equations are derived similar to their complex-valued counterpart, except that the real-valued SH basis functions $\SHBR$ are used instead of the complex-valued SH basis functions. The real-valued SH basis functions are defined in terms of complex-valued SH basis functions as follows:
% real valued SH ------------------
\begin{align}
\label{eq:real_sh_basis}
\SHBR^{l,m}=
\left\{
\begin{array}{lr}
\frac{\iu}{\sqrt{2}}\left(\SHBC^{l,m}-\left(-1\right)^m\SHBC^{l,-m}\right), & \text{for } m < 0\\
\SHBC^{l,m}, & \text{for } m = 0\\
\frac{1}{\sqrt{2}}\left(\SHBC^{l,-m}+\left(-1\right)^m\SHBC^{l,m}\right), & \text{for } m > 0
\end{array}
\right.
\end{align}
Note we use the subscript $\mathbb{R}$ and $\mathbb{C}$ do distinguish between real- and complex-valued SH basis functions respectively.

\subsection{Projecting Radiative Transfer Quantities}

% --------------------------------------------
\subsubsection{Radiance Field $L$ and Emission Field $Q$}
\label{sec:real_proj_L}

As with the complex-valued case, the angular dependent quantities are projected into SH coefficients. Here, those coefficients will be real-valued, since we use the real-valued SH basis.
\begin{align}
L^{l,m}\left(\vec{x}\right)
&=
\int_\Omega{L\left(\vec{x}, \omega\right)\SHBR^{l,m}\ud\omega} \ . \nonumber\\
Q^{l,m}\left(\vec{x}\right)
&=
\int_\Omega{Q\left(\vec{x}, \omega\right)\SHBR^{l,m}\ud\omega} \ . \nonumber
\end{align}

The reconstruction $\hat{L}$ and $\hat{Q}$, is found by a truncated linear combination of SH basis functions weighted by their respective coefficients:
\begin{align}
\label{eq:real_sh_exp_L}
\hat{L}\left(\vec{x}, \omega\right) =
\sum_{l,m}
{
L^{l,m}\left(\vec{x}\right)\SHBR^{l,m}\left(\omega\right)
} \ .
\\
\hat{Q}\left(\vec{x}, \omega\right) =
\sum_{l,m}
{
Q^{l,m}\left(\vec{x}\right)\SHBR^{l,m}\left(\omega\right)
} \ .
\end{align}
We later will have to apply identities and properties for the complex-valued SH basis functions and therefore need to expand the real-valued basis function in $\hat{L}$. The real-valued basis function is different depending on $m$ and therefore gives different expansions for the sign of $m$:
\begin{align}
\hat{L}\left(\vec{x}, \omega\right)
=&
\left\{
\begin{array}{lr}
\sum_{l,m}L^{l,m}\left(\vec{x}\right)\frac{\iu}{\sqrt{2}}\left(\SHBC^{l,m}-\left(-1\right)^m\SHBC^{l,-m}\right), & \text{for } m < 0\\
\sum_{l,m}L^{l,m}\left(\vec{x}\right)\SHBC^{l,m}, & \text{for } m = 0\\
\sum_{l,m}L^{l,m}\left(\vec{x}\right)\frac{1}{\sqrt{2}}\left(\SHBC^{l,-m}-\left(-1\right)^m\SHBC^{l,m}\right), & \text{for } m > 0
\end{array}
\right.
\nonumber\\
=&
\left\{
\begin{array}{lr}
\sum_{l}\sum_{m=-l}^{-1}L^{l,m}\left(\vec{x}\right)\frac{\iu}{\sqrt{2}}\left(\SHBC^{l,m}-\left(-1\right)^m\SHBC^{l,-m}\right), & \text{for } m < 0\\
L^{l,0}\left(\vec{x}\right)\SHBC^{l,0}, & \text{for } m = 0\\
\sum_{l}\sum_{m=1}^{l}L^{l,m}\left(\vec{x}\right)\frac{1}{\sqrt{2}}\left(\SHBC^{l,-m}-\left(-1\right)^m\SHBC^{l,m}\right), & \text{for } m > 0
\end{array}
\right.
\nonumber\\
=&
\sum_{l=0}^{N}
\left(
\sum_{m=-l}^{-1}
{
L^{{l,m}}\left (\vec{x} \right )\left(\frac{i}{\sqrt{2}}\SHBC^{l, m}(\omega )-\frac{i}{\sqrt{2}}\left({-1}\right)^{m}\SHBC^{l, -m}(\omega )\right)
}
\right.
\nonumber\\
&
+L^{l,0}\left (\vec{x} \right )\SHBC^{l, 0}(\omega )
\nonumber\\
&
+
\left.
\sum_{m=1}^{l}
{
L^{{l,m}}\left (\vec{x} \right )\left(\frac{1}{\sqrt{2}}\SHBC^{l, -m}(\omega )+\frac{1}{\sqrt{2}}\left({-1}\right)^{m}\SHBC^{l, m}(\omega )\right)
}
\right)
\nonumber\\=&
\frac{i}{\sqrt{2}}\left(\sum_{l=0}^{N}{\sum_{m=-l}^{-1}{L^{{l,m}}\left (\vec{x} \right )\SHBC^{l, m}(\omega )}}\right)
-\frac{i}{\sqrt{2}}\left(\sum_{l=0}^{N}{\sum_{m=-l}^{-1}{L^{{l,m}}\left (\vec{x} \right )\left({-1}\right)^{m}\SHBC^{l, -m}(\omega )}}\right)
\nonumber\\&
+\sum_{l=0}^{N}{L^{{l,0}}\left (\vec{x} \right )\SHBC^{l, 0}(\omega )}
\nonumber\\&
+\frac{1}{\sqrt{2}}\left(\sum_{l=0}^{N}{\sum_{m=1}^{l}{L^{{l,m}}\left (\vec{x} \right )\SHBC^{l, -m}(\omega )}}\right)
+\frac{1}{\sqrt{2}}\left(\sum_{l=0}^{N}{\sum_{m=1}^{l}{L^{{l,m}}\left (\vec{x} \right )\left({-1}\right)^{m}\SHBC^{l, m}(\omega )}}\right) \ .
\end{align}

% ----------------------------------------------------
\subsubsection{Phase Function}

The real-valued spherical harmonics expansion of the phase function follows the same derivation as the complex-valued expansion from section~\ref{sec:complex_proj_phase}. We first fix the outgoing direction vector $\omega_o$ to always align with the $z$-axis ($\omega_o=\vec{e}_3$). We compute the spherical harmonics projection over incident direction vector $\omega_i$, using the real-valued spherical harmonics basis functions:
\begin{align*}
p^{l,m}
=
\int_\Omega
{
p\left(\omega_i\cdot\vec{e}_3\right)
\SHBR^{l,m}\left(\omega_i\right)
\ud\omega_i
} \ .
\nonumber
\end{align*}

Phase functions which only depend on the angle between incident and outgoing vectors are rotationally symmetric around the pole axis. Like with the complex-values spherical harmonics basis functions, such a rotation $R$ of angle $\alpha$ around the pole axis is given by:
\begin{align*}
\rho_{R(\alpha)}(\SHBR^{l,m}) = e^{-i m\alpha}\SHBR^{l,m} \ .
\end{align*}
We formulate symmetry around the pole axis with the following constraint:
\begin{align*}
\sum_{l,m}
{
e^{-i m\alpha}
p^{l,m}
\SHBR^{l,m} }\left(\omega_i\right)
=
\sum_{l,m}
{
p^{l,m}
\SHBR^{l,m}\left(\omega_i\right)
} \ .
\end{align*}
By comparing coefficients we get
\begin{align*}
p^{l,m} = p^{l,m}e^{-i m\alpha} \ .
\end{align*}

From this we can infer that $p^{l,m} = 0$ for all $m\ne0$, if the phase function is rotationally symmetric around the pole axis. We therefore have the same property as we have with complex-valued expansions of functions, which are symmetric about the pole axis: only the $m=0$ coefficients are needed for reconstruction. We therefore have
\begin{align}
\label{eq:sh_exp_phase}
p(\omega_i) =
\sum_l
{
p^{l0}
\SHBR^{l0}(\omega_i)
} \ .
\end{align}

\subsection{Projecting Terms of the RTE}

% -----------------------------------------------------
\subsubsection{Transport Term}

The transport term of the RTE is given as
\begin{align}
(\omega\cdot\nabla)L(\vec{x}, \omega)
=
\omega_{x}\partial_xL\left (\vec{x} ,\omega \right )+\omega_{y}\partial_yL\left (\vec{x} ,\omega \right )+\omega_{z}\partial_zL\left (\vec{x} ,\omega \right ) \ .
\label{eq:real_transport_term}
\end{align}

To improve readability, we first project the term into SH by multiplying with the conjugate complex of the SH basis, and replace $L$ by its SH expansion afterwards. This order was reversed, when we derived the complex-valued $P_N$-equation in section~\ref{sec:complex_transport_term}.

We now multiply Equation~\ref{eq:real_transport_term} with the real-valued SH basis and integrate over solid angle. However, the SH basis is different for $m'<0$, $m'=0$ and $m'>0$, and therefore will give us different $P_N$-equations depending on $m'$. We will go through the derivation in detail for the $m'<0$ case and give the results for the other cases at the end Multiplying the expanded transport term with the SH basis for $m'<0$ and integrating over solid angle gives:
\begin{align*}
&\int{\left(\frac{-i}{\sqrt{2}}\overline{Y^{l', m'}}(\omega )-\frac{-i}{\sqrt{2}}\left({-1}\right)^{m'}\overline{Y^{l', -m'}}(\omega )\right)\left(\omega_{x}\partial_xL\left (\vec{x} ,\omega \right )+\omega_{y}\partial_yL\left (\vec{x} ,\omega \right )+\omega_{z}\partial_zL\left (\vec{x} ,\omega \right )\right)\ud\omega}
\\
=&
\int-\frac{i}{\sqrt{2}}\overline{Y^{l', m'}}(\omega )\omega_{x}\partial_xL\left (\vec{x} ,\omega \right )-\frac{i}{\sqrt{2}}\overline{Y^{l', m'}}(\omega )\omega_{y}\partial_yL\left (\vec{x} ,\omega \right )-\frac{i}{\sqrt{2}}\overline{Y^{l', m'}}(\omega )\omega_{z}\partial_zL\left (\vec{x} ,\omega \right )
\\&
+\frac{i}{\sqrt{2}}\left({-1}\right)^{m'}\overline{Y^{l', -m'}}(\omega )\omega_{x}\partial_xL\left (\vec{x} ,\omega \right )+\frac{i}{\sqrt{2}}\left({-1}\right)^{m'}\overline{Y^{l', -m'}}(\omega )\omega_{y}\partial_yL\left (\vec{x} ,\omega \right )
\\&
+\frac{i}{\sqrt{2}}\left({-1}\right)^{m'}\overline{Y^{l', -m'}}(\omega )\omega_{z}\partial_zL\left (\vec{x} ,\omega \right )\ud\omega \ .
\end{align*}

After expanding the integrand and splitting the integral, we apply the recursive relation from Equation~\ref{eq:recursive_relation} to get:
\begin{align*}
&
\frac{i}{\sqrt{2}}\frac{1}{2}c^{{l'-1,m'-1}}\int{\partial_xL\left (\vec{x} ,\omega \right )\overline{Y^{l'-1, m'-1}}(\omega )\ud\omega}
-\frac{i}{\sqrt{2}}\frac{1}{2}d^{{l'+1,m'-1}}\int{\partial_xL\left (\vec{x} ,\omega \right )\overline{Y^{l'+1, m'-1}}(\omega )\ud\omega}
\\&
-\frac{i}{\sqrt{2}}\frac{1}{2}e^{{l'-1,m'+1}}\int{\partial_xL\left (\vec{x} ,\omega \right )\overline{Y^{l'-1, m'+1}}(\omega )\ud\omega}
+\frac{i}{\sqrt{2}}\frac{1}{2}f^{{l'+1,m'+1}}\int{\partial_xL\left (\vec{x} ,\omega \right )\overline{Y^{l'+1, m'+1}}(\omega )\ud\omega}
\\&
-\frac{i}{\sqrt{2}}\frac{i}{2}c^{{l'-1,m'-1}}\int{\partial_yL\left (\vec{x} ,\omega \right )\overline{Y^{l'-1, m'-1}}(\omega )\ud\omega}
+\frac{i}{\sqrt{2}}\frac{i}{2}d^{{l'+1,m'-1}}\int{\partial_yL\left (\vec{x} ,\omega \right )\overline{Y^{l'+1, m'-1}}(\omega )\ud\omega}
\\&
-\frac{i}{\sqrt{2}}\frac{i}{2}e^{{l'-1,m'+1}}\int{\partial_yL\left (\vec{x} ,\omega \right )\overline{Y^{l'-1, m'+1}}(\omega )\ud\omega}
+\frac{i}{\sqrt{2}}\frac{i}{2}f^{{l'+1,m'+1}}\int{\partial_yL\left (\vec{x} ,\omega \right )\overline{Y^{l'+1, m'+1}}(\omega )\ud\omega}
\\&
-\frac{i}{\sqrt{2}}a^{{l'-1,m'}}\int{\partial_zL\left (\vec{x} ,\omega \right )\overline{Y^{l'-1, m'}}(\omega )\ud\omega}
-\frac{i}{\sqrt{2}}b^{{l'+1,m'}}\int{\partial_zL\left (\vec{x} ,\omega \right )\overline{Y^{l'+1, m'}}(\omega )\ud\omega}
\\&
-\frac{i}{\sqrt{2}}\left({-1}\right)^{m'}\frac{1}{2}c^{{l'-1,-m'-1}}\int{\partial_xL\left (\vec{x} ,\omega \right )\overline{Y^{l'-1, -m'-1}}(\omega )\ud\omega}
\\&
+\frac{i}{\sqrt{2}}\left({-1}\right)^{m'}\frac{1}{2}d^{{l'+1,-m'-1}}\int{\partial_xL\left (\vec{x} ,\omega \right )\overline{Y^{l'+1, -m'-1}}(\omega )\ud\omega}
\\&
+\frac{i}{\sqrt{2}}\left({-1}\right)^{m'}\frac{1}{2}e^{{l'-1,-m'+1}}\int{\partial_xL\left (\vec{x} ,\omega \right )\overline{Y^{l'-1, -m'+1}}(\omega )\ud\omega}
\\&
-\frac{i}{\sqrt{2}}\left({-1}\right)^{m'}\frac{1}{2}f^{{l'+1,-m'+1}}\int{\partial_xL\left (\vec{x} ,\omega \right )\overline{Y^{l'+1, -m'+1}}(\omega )\ud\omega}
\\&
+\frac{i}{\sqrt{2}}\left({-1}\right)^{m'}\frac{i}{2}c^{{l'-1,-m'-1}}\int{\partial_yL\left (\vec{x} ,\omega \right )\overline{Y^{l'-1, -m'-1}}(\omega )\ud\omega}
\\&
-\frac{i}{\sqrt{2}}\left({-1}\right)^{m'}\frac{i}{2}d^{{l'+1,-m'-1}}\int{\partial_yL\left (\vec{x} ,\omega \right )\overline{Y^{l'+1, -m'-1}}(\omega )\ud\omega}
\\&
+\frac{i}{\sqrt{2}}\left({-1}\right)^{m'}\frac{i}{2}e^{{l'-1,-m'+1}}\int{\partial_yL\left (\vec{x} ,\omega \right )\overline{Y^{l'-1, -m'+1}}(\omega )\ud\omega}
\\&
-\frac{i}{\sqrt{2}}\left({-1}\right)^{m'}\frac{i}{2}f^{{l'+1,-m'+1}}\int{\partial_yL\left (\vec{x} ,\omega \right )\overline{Y^{l'+1, -m'+1}}(\omega )\ud\omega}
\\&
+\frac{i}{\sqrt{2}}\left({-1}\right)^{m'}a^{{l'-1,-m'}}\int{\partial_zL\left (\vec{x} ,\omega \right )\overline{Y^{l'-1, -m'}}(\omega )\ud\omega}+\frac{i}{\sqrt{2}}\left({-1}\right)^{m'}b^{{l'+1,-m'}}\int{\partial_zL\left (\vec{x} ,\omega \right )\overline{Y^{l'+1, -m'}}(\omega )\ud\omega} \ .
\end{align*}

Before we further expand the radiance field $L$ into its SH expansion, we will simplify coefficients by using the following relations:
\begin{align}
a^{l,m} = a^{l,-m}, \qquad
b^{l,m} = b^{l,-m}, \qquad
c^{l,m} = e^{l,-m}, \qquad
d^{l,m} = f^{l,-m} \ .
\label{eq:recursion_identities}
\end{align}

This allows us to rewrite the equation above as:
\begin{align*}
-i\alpha_c\int{\partial_yL\left (\vec{x} ,\omega \right )\overline{Y^{l'-1, m'-1}}(\omega )\ud\omega}
%\\
+\left({-1}\right)^{m'}i&\alpha_c\int{\partial_yL\left (\vec{x} ,\omega \right )\overline{Y^{l'-1, -m'+1}}(\omega )\ud\omega}
\\
+i\alpha_d\int{\partial_yL\left (\vec{x} ,\omega \right )\overline{Y^{l'+1, m'-1}}(\omega )\ud\omega}
%\\
-\left({-1}\right)^{m'} i &\alpha_d\int{\partial_yL\left (\vec{x} ,\omega \right )\overline{Y^{l'+1, -m'+1}}(\omega )\ud\omega}
\\
-i\alpha_e \int{\partial_yL\left (\vec{x} ,\omega \right )\overline{Y^{l'-1, m'+1}}(\omega )\ud\omega}
%\\
+\left({-1}\right)^{m'}i &\alpha_e \int{\partial_yL\left (\vec{x} ,\omega \right )\overline{Y^{l'-1, -m'-1}}(\omega )\ud\omega}
\\
+i\alpha_f \int{\partial_yL\left (\vec{x} ,\omega \right )\overline{Y^{l'+1, m'+1}}(\omega )\ud\omega}
%\\
-\left({-1}\right)^{m'}i &\alpha_f \int{\partial_yL\left (\vec{x} ,\omega \right )\overline{Y^{l'+1, -m'-1}}(\omega )\ud\omega}
% ---------------------------------------
\\
+\alpha_c\int{\partial_xL\left (\vec{x} ,\omega \right )\overline{Y^{l'-1, m'-1}}(\omega )\ud\omega}
%\\
+\left({-1}\right)^{m'}&\alpha_c\int{\partial_xL\left (\vec{x} ,\omega \right )\overline{Y^{l'-1, -m'+1}}(\omega )\ud\omega}
% ---------------------------------------
\\
-\alpha_e\int{\partial_xL\left (\vec{x} ,\omega \right )\overline{Y^{l'-1, m'+1}}(\omega )\ud\omega}
%\\
-\left({-1}\right)^{m'}&\alpha_e\int{\partial_xL\left (\vec{x} ,\omega \right )\overline{Y^{l'-1, -m'-1}}(\omega )\ud\omega}
% ---------------------------------------
\\
+\alpha_f\int{\partial_xL\left (\vec{x} ,\omega \right )\overline{Y^{l'+1, m'+1}}(\omega )\ud\omega}
%\\
+\left({-1}\right)^{m'}&\alpha_f\int{\partial_xL\left (\vec{x} ,\omega \right )\overline{Y^{l'+1, -m'-1}}(\omega )\ud\omega}
\\
% ---------------------------------------------
-\alpha_d\int{\partial_xL\left (\vec{x} ,\omega \right )\overline{Y^{l'+1, m'-1}}(\omega )\ud\omega}
%\\
% ---------------------------------------
-\left({-1}\right)^{m'}&\alpha_d\int{\partial_xL\left (\vec{x} ,\omega \right )\overline{Y^{l'+1, -m'+1}}(\omega )\ud\omega}
\\
% ---------------------------------------
-\alpha_a\int{\partial_zL\left (\vec{x} ,\omega \right )\overline{Y^{l'-1, m'}}(\omega )\ud\omega}
%\\
+\left({-1}\right)^{m'}&\alpha_a\int{\partial_zL\left (\vec{x} ,\omega \right )\overline{Y^{l'-1, -m'}}(\omega )\ud\omega}
% ---------------------------------------
\\
-\alpha_b\int{\partial_zL\left (\vec{x} ,\omega \right )\overline{Y^{l'+1, m'}}(\omega )\ud\omega}
%\\
+\left({-1}\right)^{m'}&\alpha_b\int{\partial_zL\left (\vec{x} ,\omega \right )\overline{Y^{l'+1, -m'}}(\omega )\ud\omega}
\end{align*}
with
\begin{align*}
\alpha_c = \frac{i}{\sqrt{2}}\frac{1}{2}c^{{l'-1,m'-1}}
,\qquad
%\\
\alpha_e = \frac{i}{\sqrt{2}}\frac{1}{2}e^{{l'-1,m'+1}}
,\qquad
%\\
\alpha_d = \frac{i}{\sqrt{2}}\frac{1}{2}d^{{l'+1,m'-1}}
\\
\alpha_f = \frac{i}{\sqrt{2}}\frac{1}{2}f^{{l'+1,m'+1}}
,\qquad
%\\
\alpha_a = \frac{i}{\sqrt{2}}a^{{l'-1,m'}}
,\qquad
%\\
\alpha_b = \frac{i}{\sqrt{2}}b^{{l'+1,m'}} \ .
\end{align*}

In the next step, we substitute the radiance field function $L$ with its spherical harmonics expansion and arrive at the following expression after further expansions and transformations:
\begin{align}
-i\alpha_c\partial_y\sum_{l,m}L^{l,m}\left (\vec{x}\right)\int{Y_{\mathbb{R}}^{l,m}\overline{\SHBC^{l'-1, m'-1}}(\omega )\ud\omega}
%\label{eq:real_transport_expansion_unsimplified_term1}
%\\
+\left({-1}\right)^{m'}i&\alpha_c\partial_y\sum_{l,m}L^{l,m}\left (\vec{x}\right)\int{Y_{\mathbb{R}}^{l,m}\overline{\SHBC^{l'-1, -m'+1}}(\omega )\ud\omega}
%\label{eq:real_transport_expansion_unsimplified_term2}
\label{eq:real_transport_expansion_unsimplified_term1_term2}
\\
+i\alpha_d\partial_y\sum_{l,m}L^{l,m}\left (\vec{x}\right)\int{Y_{\mathbb{R}}^{l,m}\overline{\SHBC^{l'+1, m'-1}}(\omega )\ud\omega}
%\label{eq:real_transport_expansion_unsimplified_term3}
%\\
-\left({-1}\right)^{m'} i &\alpha_d\partial_y\sum_{l,m}L^{l,m}\left (\vec{x}\right)\int{Y_{\mathbb{R}}^{l,m}\overline{\SHBC^{l'+1, -m'+1}}(\omega )\ud\omega}
%\label{eq:real_transport_expansion_unsimplified_term4}
\label{eq:real_transport_expansion_unsimplified_term3_term4}
\\
-i\alpha_e \partial_y\sum_{l,m}L^{l,m}\left (\vec{x}\right)\int{Y_{\mathbb{R}}^{l,m}\overline{\SHBC^{l'-1, m'+1}}(\omega )\ud\omega}
%\label{eq:real_transport_expansion_unsimplified_term5}
%\\
+\left({-1}\right)^{m'}i &\alpha_e \partial_y\sum_{l,m}L^{l,m}\left (\vec{x}\right)\int{Y_{\mathbb{R}}^{l,m}\overline{\SHBC^{l'-1, -m'-1}}(\omega )\ud\omega}
%\label{eq:real_transport_expansion_unsimplified_term6}
\\
+i\alpha_f \partial_y\sum_{l,m}L^{l,m}\left (\vec{x}\right)\int{Y_{\mathbb{R}}^{l,m}\overline{\SHBC^{l'+1, m'+1}}(\omega )\ud\omega}
%\label{eq:real_transport_expansion_unsimplified_term7}
%\\
-\left({-1}\right)^{m'}i &\alpha_f \partial_y\sum_{l,m}L^{l,m}\left (\vec{x}\right)\int{Y_{\mathbb{R}}^{l,m}\overline{\SHBC^{l'+1, -m'-1}}(\omega )\ud\omega}
% ---------------------------------------
%\label{eq:real_transport_expansion_unsimplified_term8}
\\
+\alpha_c\partial_x\sum_{l,m}L^{l,m}\left (\vec{x}\right)\int{Y_{\mathbb{R}}^{l,m}\overline{\SHBC^{l'-1, m'-1}}(\omega )\ud\omega}
%\label{eq:real_transport_expansion_unsimplified_term9}
%\\
+\left({-1}\right)^{m'}&\alpha_c\partial_x\sum_{l,m}L^{l,m}\left (\vec{x}\right)\int{Y_{\mathbb{R}}^{l,m}\overline{\SHBC^{l'-1, -m'+1}}(\omega )\ud\omega}
% ---------------------------------------
%\label{eq:real_transport_expansion_unsimplified_term10}
\\
-\alpha_e\partial_x\sum_{l,m}L^{l,m}\left (\vec{x}\right)\int{Y_{\mathbb{R}}^{l,m}\overline{\SHBC^{l'-1, m'+1}}(\omega )\ud\omega}
%\label{eq:real_transport_expansion_unsimplified_term11}
%\\
-\left({-1}\right)^{m'}&\alpha_e\partial_x\sum_{l,m}L^{l,m}\left (\vec{x}\right)\int{Y_{\mathbb{R}}^{l,m}\overline{\SHBC^{l'-1, -m'-1}}(\omega )\ud\omega}
% ---------------------------------------
%\label{eq:real_transport_expansion_unsimplified_term12}
\\
+\alpha_f\partial_x\sum_{l,m}L^{l,m}\left (\vec{x}\right)\int{Y_{\mathbb{R}}^{l,m}\overline{\SHBC^{l'+1, m'+1}}(\omega )\ud\omega}
%\label{eq:real_transport_expansion_unsimplified_term13}
%\\
+\left({-1}\right)^{m'}&\alpha_f\partial_x\sum_{l,m}L^{l,m}\left (\vec{x}\right)\int{Y_{\mathbb{R}}^{l,m}\overline{\SHBC^{l'+1, -m'-1}}(\omega )\ud\omega}
%\label{eq:real_transport_expansion_unsimplified_term14}
\\
% ---------------------------------------
-\alpha_d\partial_x\sum_{l,m}L^{l,m}\left (\vec{x}\right)\int{Y_{\mathbb{R}}^{l,m}\overline{\SHBC^{l'+1, m'-1}}(\omega )\ud\omega}
%\label{eq:real_transport_expansion_unsimplified_term15}
%\\
% ---------------------------------------
-\left({-1}\right)^{m'}&\alpha_d\partial_x\sum_{l,m}L^{l,m}\left (\vec{x}\right)\int{Y_{\mathbb{R}}^{l,m}\overline{\SHBC^{l'+1, -m'+1}}(\omega )\ud\omega}
%\label{eq:real_transport_expansion_unsimplified_term16}
\\
% ---------------------------------------
-\alpha_a\partial_z\sum_{l,m}L^{l,m}\left (\vec{x}\right)\int{Y_{\mathbb{R}}^{l,m}\overline{\SHBC^{l'-1, m'}}(\omega )\ud\omega}
%\label{eq:real_transport_expansion_unsimplified_term17}
% ---------------------------------------
%\\
+\left({-1}\right)^{m'}&\alpha_a\partial_z\sum_{l,m}L^{l,m}\left (\vec{x}\right)\int{Y_{\mathbb{R}}^{l,m}\overline{\SHBC^{l'-1, -m'}}(\omega )\ud\omega}
% ---------------------------------------
%\label{eq:real_transport_expansion_unsimplified_term18}
\\
-\alpha_b\partial_z\sum_{l,m}L^{l,m}\left (\vec{x}\right)\int{Y_{\mathbb{R}}^{l,m}\overline{\SHBC^{l'+1, m'}}(\omega )\ud\omega}
%\label{eq:real_transport_expansion_unsimplified_term19}
%\\
+\left({-1}\right)^{m'}&\alpha_b\partial_z\sum_{l,m}L^{l,m}\left (\vec{x}\right)\int{Y_{\mathbb{R}}^{l,m}\overline{\SHBC^{l'+1, -m'}}(\omega )\ud\omega}
%\label{eq:real_transport_expansion_unsimplified_term20}
\end{align}

The real-valued $P_N$-equation have an intricate structure which causes many terms to cancel out. We take the first two terms (Equation~\ref{eq:real_transport_expansion_unsimplified_term1_term2}) of the $P_N$-equations and apply the following orthogonality property of SH:
\begin{align}
\int_\Omega{\SHBR^{l_1, m_1}\overline{\SHBC^{l_2, m_2}}}\ud\omega
=
\left\{
\begin{array}{lr}
\frac{i}{\sqrt{2}}
\left(
\delta_{\scaleto{\substack{l_1=l_2\\m_1=m_2}}{9pt}}
-\left({-1}\right)^{m_1}
\delta_{\scaleto{\substack{l_1=l_2\\m_1=-m_2}}{9pt}}
\right)
, & \text{for } m_1 < 0
\\
\delta_{\scaleto{\substack{l_1=l_2\\m_1=m_2}}{9pt}}, & \text{for } m_1 = 0
\\
\frac{1}{\sqrt{2}}
\left(
\delta_{\scaleto{\substack{l_1=l_2\\m_1=-m_2}}{9pt}}
+\left({-1}\right)^{m_1}
\delta_{\scaleto{\substack{l_1=l_2\\m_1=m_2}}{9pt}}
\right)
, & \text{for } m_1 > 0
\end{array}
\right.  \ .
\label{eq:real_orthogonality_property_with_complex}
\end{align}
This way we get for the first two terms:
\begin{align*}
%-i &\alpha_c\partial_y\sum_{l,m}L^{l,m}\left (\vec{x}\right)
%\int{Y_{\mathbb{R}}^{l,m}\overline{Y^{l'-1, m'-1}}(\omega )\ud\omega}
&-i\alpha_c\partial_y
\sum_{l=0}^{N}
\sum_{m=-l}^{-1}L^{l,m}\left (\vec{x}\right)
\frac{1}{\sqrt{2}}
i\delta_{\scaleto{\substack{l=l'-1\\m=m'-1}}{9pt}}
%\\
+i\alpha_c\partial_y
\sum_{l=0}^{N}
\sum_{m=-l}^{-1}L^{l,m}\left (\vec{x}\right)
\frac{1}{\sqrt{2}}
i
\left({-1}\right)^{m}
\delta_{\scaleto{\substack{l=l'-1\\m=-m'+1}}{9pt}}
\\&
-i\alpha_c\partial_y
\sum_{l=0}^{N}
L^{l,0}\left (\vec{x}\right)
\frac{1}{\sqrt{2}}
\delta_{\scaleto{\substack{l=l'-1\\0=m'-1}}{9pt}}
%\\
-i\alpha_c\partial_y
\sum_{l=0}^{N}
\sum_{m=1}^{l}L^{l,m}\left (\vec{x}\right)
\frac{1}{\sqrt{2}}
\delta_{\scaleto{\substack{l=l'-1\\m=-m'+1}}{9pt}}
\\&
-i\alpha_c\partial_y
\sum_{l=0}^{N}
\sum_{m=1}^{l}L^{l,m}\left (\vec{x}\right)
\frac{1}{\sqrt{2}}
\left({-1}\right)^{m}
\delta_{\scaleto{\substack{l=l'-1\\m=m'-1}}{9pt}}
%\\
%\int{Y_{\mathbb{R}}^{l,m}\overline{Y^{l'-1, -m'+1}}(\omega )\ud\omega}
+\left({-1}\right)^{m'}i\alpha_c\partial_y
\sum_{l=0}^{N}
\sum_{m=-l}^{-1}L^{l,m}\left (\vec{x}\right)
\frac{1}{\sqrt{2}}
i
\delta_{\scaleto{\substack{l=l'-1\\m=-m'+1}}{9pt}}
\\&
-\left({-1}\right)^{m'}i\alpha_c\partial_y
\sum_{l=0}^{N}
\sum_{m=-l}^{-1}L^{l,m}\left (\vec{x}\right)
\frac{1}{\sqrt{2}}
i
\left({-1}\right)^{m}
\delta_{\scaleto{\substack{l=l'-1\\m=m'-1}}{9pt}}
%\\
+
\left({-1}\right)^{m'}i\alpha_c\partial_y
\sum_{l=0}^{N}
L^{l,0}\left (\vec{x}\right)
\frac{1}{\sqrt{2}}
\delta_{\scaleto{\substack{l=l'-1\\0=-m'+1}}{9pt}}
\\&
+
\left({-1}\right)^{m'}i\alpha_c\partial_y
\sum_{l=0}^{N}
\sum_{m=1}^{l}L^{l,m}\left (\vec{x}\right)
\frac{1}{\sqrt{2}}
\delta_{\scaleto{\substack{l=l'-1\\m=m'-1}}{9pt}}
%\\
+
\left({-1}\right)^{m'}i\alpha_c\partial_y
\sum_{l=0}^{N}
\sum_{m=1}^{l}L^{l,m}\left (\vec{x}\right)
\frac{1}{\sqrt{2}}
\left({-1}\right)^{m}
\delta_{\scaleto{\substack{l=l'-1\\m=-m'+1}}{9pt}} \ .
\end{align*}

We apply the delta function for the sums which run over the variable $l$:
\begin{align}
\sum_{l=0}^{N}\sum_{m=a}^{b}L^{l,m}\delta_{\scaleto{\substack{l=x\\m=y}}{9pt}}=\sum_{m=a}^{b}L^{x,m}\delta_{\scaleto{m=y}{3pt}} \ .
\end{align}
We get for the first two terms of the transport term of the $P_N$-equation (Equation~\ref{eq:real_transport_expansion_unsimplified_term1_term2}):
\begin{align*}
&
%-i &\alpha_c\partial_y\sum_{l,m}L^{l,m}\left (\vec{x}\right)
%\int{Y_{\mathbb{R}}^{l,m}\overline{Y^{l'-1, m'-1}}(\omega )\ud\omega}
\mathcolor{red}{-i}
%&
\mathcolor{red}{\alpha_c\partial_y
\sum_{m=-l'+1}^{-1}L^{l'-1,m}\left (\vec{x}\right)
\frac{1}{\sqrt{2}}
i\delta_{\scaleto{m=m'-1}{4pt}}
}
%\\
% -------------------------------------------------
\mathcolor{blue}{
+i
}
%&
\mathcolor{blue}{
\alpha_c\partial_y
\sum_{m=-l'+1}^{-1}L^{l'-1,m}\left (\vec{x}\right)
\frac{1}{\sqrt{2}}
i
\left({-1}\right)^{m}
\delta_{\scaleto{m=-m'+1}{4pt}}
}
\\&
% -------------------------------------------------
\mathcolor{blue}
{
-i
}
%&
\mathcolor{blue}
{
\alpha_c\partial_y
L^{l'-1,0}\left (\vec{x}\right)
\frac{1}{\sqrt{2}}
\delta_{\scaleto{m'=1}{4pt}}
}
%\\
% -------------------------------------------------
\mathcolor{black}
{
-i
}
%&
\mathcolor{black}
{
\alpha_c\partial_y
\sum_{m=1}^{l'-1}L^{l'-1,m}\left (\vec{x}\right)
\frac{1}{\sqrt{2}}
\delta_{\scaleto{m=-m'+1}{4pt}}
}
%\\
% -------------------------------------------------
\mathcolor{blue}
{
-i
}
%&
\mathcolor{blue}
{
\alpha_c\partial_y
\sum_{m=1}^{l'-1}L^{l'-1,m}\left (\vec{x}\right)
\frac{1}{\sqrt{2}}
\left({-1}\right)^{m}
\delta_{\scaleto{m=m'-1}{4pt}}
}
\\&
% -------------------------------------------------
\mathcolor{blue}
{
+\left({-1}\right)^{m'}i
}
%&
\mathcolor{blue}{
\alpha_c\partial_y
\sum_{m=-l'+1}^{-1}L^{l'-1,m}\left (\vec{x}\right)
\frac{1}{\sqrt{2}}
i
\delta_{\scaleto{m=-m'+1}{4pt}}
}
%\\
% -------------------------------------------------
\mathcolor{red}
{
-\left({-1}\right)^{m'}i
}
%&
\mathcolor{red}
{
\alpha_c\partial_y
\sum_{m=-l'+1}^{-1}L^{l'-1,m}\left (\vec{x}\right)
\frac{1}{\sqrt{2}}
i
\left({-1}\right)^{m}
\delta_{\scaleto{m=m'-1}{4pt}}
}
% -------------------------------------------------
\\&
\mathcolor{blue}
{
+
\left({-1}\right)^{m'}i
}
%&
\mathcolor{blue}
{
\alpha_c\partial_y
L^{l'-1,0}\left (\vec{x}\right)
\frac{1}{\sqrt{2}}
\delta_{\scaleto{m'=1}{4pt}}
}
%\\
% -------------------------------------------------
\mathcolor{blue}
{
+
\left({-1}\right)^{m'}i
}
%&
\mathcolor{blue}
{
\alpha_c\partial_y
\sum_{m=1}^{l'-1}L^{l'-1,m}\left (\vec{x}\right)
\frac{1}{\sqrt{2}}
\delta_{\scaleto{m=m'-1}{4pt}}
}
\\&
% -------------------------------------------------
\mathcolor{black}
{
+
\left({-1}\right)^{m'}i
}
%&
\mathcolor{black}
{
\alpha_c\partial_y
\sum_{m=1}^{l'-1}L^{l'-1,m}\left (\vec{x}\right)
\frac{1}{\sqrt{2}}
\left({-1}\right)^{m}
\delta_{\scaleto{m=-m'+1}{4pt}}
}
\end{align*}

The variables $l'$ and $m'$ specify a particular equation within the given set of $P_N$-equations. We remember that $m'$ originated from multiplying the transport term with the real-valued SH basis function $\SHBR$ for the projection. The real-valued basis function is different for the sign of $m'$ and we derived the transport term of the $P_N$-equations under the assumption of $m'<0$ (different equations have to be derived for $m'=0$ and $m'>0$). We are able to greatly simplify the terms above when considering the parity of $m'$ and that $m'<0$.

The blue terms in the equation above all vanish, since the sums run over all negative (or positive) $m$, up to $-1$ (or l), while the Kronecker deltas in the blue terms only become non-zero for values $m>0$ (or $m<0$). This is because we derived these terms by multiplying with the real-valued SH basis function for $m'<0$.

Consider the seventh and 10th term from the equation above. Due to $\delta_{m=m'-1}$ or $\delta_{m=-m'+1}$, an even $m$ is selected if $m'$ is odd and vice versa. Therefore, we have $(-1)^m(-1)^{m'}=-1$. This causes term one and seven (red) to vanish and term four and ten (black) to collapse into one term.

Therefore, the first two terms in the expansion (Equation~\ref{eq:real_transport_expansion_unsimplified_term1_term2}), simplify to:
\begin{align*}
&-i\alpha_c\partial_y\sum_{l,m}L^{l,m}\left (\vec{x}\right)\int{Y_{\mathbb{R}}^{l,m}\overline{\SHBC^{l'-1, m'-1}}(\omega )\ud\omega}
+\left({-1}\right)^{m'}i\alpha_c\partial_y\sum_{l,m}L^{l,m}\left (\vec{x}\right)\int{Y_{\mathbb{R}}^{l,m}\overline{\SHBC^{l'-1, -m'+1}}(\omega )\ud\omega}
\\
&=-\frac{2}{\sqrt{2}}i
\alpha_c\partial_y
L^{l'-1,-m'+1}\left (\vec{x}\right)
\\
&=-\frac{2}{\sqrt{2}}i
\frac{i}{\sqrt{2}}\frac{1}{2}c^{{l'-1,m'-1}}
\partial_y
L^{l'-1,-m'+1}\left (\vec{x}\right)
\\
&=
\frac{1}{2}c^{{l'-1,m'-1}}
\partial_y
L^{l'-1,-m'+1}\left (\vec{x}\right) \ .
\end{align*}

The terms in Equation~\ref{eq:real_transport_expansion_unsimplified_term3_term4} are derived in the same way with the difference, that the signs are reversed and that we have $l'+1$ instead of $l'-1$. However, this does not affect the simplification:
\begin{align*}
&i\alpha_d\partial_y\sum_{l,m}L^{l,m}\left (\vec{x}\right)\int{Y_{\mathbb{R}}^{l,m}\overline{\SHBC^{l'+1, m'-1}}(\omega )\ud\omega}
-\left({-1}\right)^{m'}i\alpha_c\partial_y\sum_{l,m}L^{l,m}\left (\vec{x}\right)\int{Y_{\mathbb{R}}^{l,m}\overline{\SHBC^{l'+1, -m'+1}}(\omega )\ud\omega}
\\
&=-\frac{2}{\sqrt{2}}i
\alpha_d\partial_y
L^{l'-1,-m'+1}\left (\vec{x}\right)
\\
&=-\frac{2}{\sqrt{2}}i
\frac{i}{\sqrt{2}}\frac{1}{2}d^{{l'+1,m'-1}}
\partial_y
L^{l'+1,-m'+1}\left (\vec{x}\right)
\\
&=
\frac{1}{2}d^{{l'+1,m'-1}}
\partial_y
L^{l'+1,-m'+1}\left (\vec{x}\right) \ .
\end{align*}
Carrying out the same simplifications for the remaining terms, results in the following real-valued $P_N$-equations for $m'<0$:
\begin{align*}
&-\frac{1}{2}c^{{l'-1,m'-1}}
\partial_y
L^{l'-1,-m'+1}
%\\
+\frac{1}{2}d^{{l'+1,m'-1}}
\partial_y
L^{l'+1,-m'+1}
%\\
-\frac{1}{2}\beta^{m'}e^{{l'-1,m'+1}}
\partial_y
L^{l'-1,-m'-1}
\\&
+\frac{1}{2}\beta^{m'}f^{{l'+1,m'+1}}
\partial_y
L^{l'+1,-m'-1}
%\\
+\frac{1}{2}c^{{l'-1,m'-1}}
\partial_x
L^{l'-1,m'-1}
\\&
-\frac{1}{2}\delta_{\scaleto{m'\neq -1}{4pt}}e^{{l'-1,m'+1}}
\partial_x
L^{l'-1,m'+1}
%\\
+\frac{1}{2}\delta_{\scaleto{m'\neq -1}{4pt}}f^{{l'+1,m'+1}}
\partial_x
L^{l'+1,m'+1}
%\\
-\frac{1}{2}d^{{l'+1,m'-1}}
\partial_x
L^{l'+1,m'-1}
\\&
+a^{{l'-1,m'}}
\partial_z
L^{l'-1,m'}
%\\
+b^{{l'+1,m'}}
\partial_z
L^{l'+1,m'} \ .
\end{align*}
with
\begin{align}
\label{eq:real_sh_basis}
\beta^{x}=
\left\{
\begin{array}{lr}
\frac{2}{\sqrt{2}}, & \text{for } \vert x\vert = 1\\
1, & \text{for } \vert x\vert \neq 1
\end{array}
\right. \ .
\end{align}

We now carry out the same derivation for the assumption of $m'>0$. We multiply Equation~\ref{sec:complex_transport_term} with the definition of the real-valued SH basis for $m'>0$ and get:
\begin{align*}
\int{\left(\frac{1}{\sqrt{2}}\overline{Y_{\mathbb{C}}^{l', -m'}}(\omega )+\frac{1}{\sqrt{2}}\left({-1}\right)^{m'}\overline{Y_{\mathbb{C}}^{l', m'}}(\omega )\right)\left(\omega_{x}\partial_xL\left (\vec{x} ,\omega \right )+\omega_{y}\partial_yL\left (\vec{x} ,\omega \right )+\omega_{z}\partial_zL\left (\vec{x} ,\omega \right )\right)\ud\omega}
\end{align*}

We expand the integrand and split the integral. Then we apply the recursive relation from Equation~\ref{eq:recursive_relation} and get:
\begin{align*}
&
\frac{1}{\sqrt{2}}\frac{1}{2}c^{{l'-1,-m'-1}}\int{\partial_xL\left (\vec{x} ,\omega \right )\overline{Y_{\mathbb{C}}^{l'-1, -m'-1}}(\omega )\ud\omega}
-\frac{1}{\sqrt{2}}\frac{1}{2}d^{{l'+1,-m'-1}}\int{\partial_xL\left (\vec{x} ,\omega \right )\overline{Y_{\mathbb{C}}^{l'+1, -m'-1}}(\omega )\ud\omega}
\\&
-\frac{1}{\sqrt{2}}\frac{1}{2}e^{{l'-1,-m'+1}}\int{\partial_xL\left (\vec{x} ,\omega \right )\overline{Y_{\mathbb{C}}^{l'-1, -m'+1}}(\omega )\ud\omega}
+\frac{1}{\sqrt{2}}\frac{1}{2}f^{{l'+1,-m'+1}}\int{\partial_xL\left (\vec{x} ,\omega \right )\overline{Y_{\mathbb{C}}^{l'+1, -m'+1}}(\omega )\ud\omega}
\\&
-\frac{1}{\sqrt{2}}\frac{i}{2}c^{{l'-1,-m'-1}}\int{\partial_yL\left (\vec{x} ,\omega \right )\overline{Y_{\mathbb{C}}^{l'-1, -m'-1}}(\omega )\ud\omega}
+\frac{1}{\sqrt{2}}\frac{i}{2}d^{{l'+1,-m'-1}}\int{\partial_yL\left (\vec{x} ,\omega \right )\overline{Y_{\mathbb{C}}^{l'+1, -m'-1}}(\omega )\ud\omega}
\\&
-\frac{1}{\sqrt{2}}\frac{i}{2}e^{{l'-1,-m'+1}}\int{\partial_yL\left (\vec{x} ,\omega \right )\overline{Y_{\mathbb{C}}^{l'-1, -m'+1}}(\omega )\ud\omega}
+\frac{1}{\sqrt{2}}\frac{i}{2}f^{{l'+1,-m'+1}}\int{\partial_yL\left (\vec{x} ,\omega \right )\overline{Y_{\mathbb{C}}^{l'+1, -m'+1}}(\omega )\ud\omega}
\\&
+\frac{1}{\sqrt{2}}a^{{l'-1,-m'}}\int{\partial_zL\left (\vec{x} ,\omega \right )\overline{Y_{\mathbb{C}}^{l'-1, -m'}}(\omega )\ud\omega}
+\frac{1}{\sqrt{2}}b^{{l'+1,-m'}}\int{\partial_zL\left (\vec{x} ,\omega \right )\overline{Y_{\mathbb{C}}^{l'+1, -m'}}(\omega )\ud\omega}
\\&
+\frac{1}{\sqrt{2}}\left({-1}\right)^{m'}\frac{1}{2}c^{{l'-1,m'-1}}\int{\partial_xL\left (\vec{x} ,\omega \right )\overline{Y_{\mathbb{C}}^{l'-1, m'-1}}(\omega )\ud\omega}
\\&
-\frac{1}{\sqrt{2}}\left({-1}\right)^{m'}\frac{1}{2}d^{{l'+1,m'-1}}\int{\partial_xL\left (\vec{x} ,\omega \right )\overline{Y_{\mathbb{C}}^{l'+1, m'-1}}(\omega )\ud\omega}
\\&
-\frac{1}{\sqrt{2}}\left({-1}\right)^{m'}\frac{1}{2}e^{{l'-1,m'+1}}\int{\partial_xL\left (\vec{x} ,\omega \right )\overline{Y_{\mathbb{C}}^{l'-1, m'+1}}(\omega )\ud\omega}
\\&
+\frac{1}{\sqrt{2}}\left({-1}\right)^{m'}\frac{1}{2}f^{{l'+1,m'+1}}\int{\partial_xL\left (\vec{x} ,\omega \right )\overline{Y_{\mathbb{C}}^{l'+1, m'+1}}(\omega )\ud\omega}
\\&
-\frac{1}{\sqrt{2}}\left({-1}\right)^{m'}\frac{i}{2}c^{{l'-1,m'-1}}\int{\partial_yL\left (\vec{x} ,\omega \right )\overline{Y_{\mathbb{C}}^{l'-1, m'-1}}(\omega )\ud\omega}
\\&
+\frac{1}{\sqrt{2}}\left({-1}\right)^{m'}\frac{i}{2}d^{{l'+1,m'-1}}\int{\partial_yL\left (\vec{x} ,\omega \right )\overline{Y_{\mathbb{C}}^{l'+1, m'-1}}(\omega )\ud\omega}
\\&
-\frac{1}{\sqrt{2}}\left({-1}\right)^{m'}\frac{i}{2}e^{{l'-1,m'+1}}\int{\partial_yL\left (\vec{x} ,\omega \right )\overline{Y_{\mathbb{C}}^{l'-1, m'+1}}(\omega )\ud\omega}
\\&
+\frac{1}{\sqrt{2}}\left({-1}\right)^{m'}\frac{i}{2}f^{{l'+1,m'+1}}\int{\partial_yL\left (\vec{x} ,\omega \right )\overline{Y_{\mathbb{C}}^{l'+1, m'+1}}(\omega )\ud\omega}
\\&
+\frac{1}{\sqrt{2}}\left({-1}\right)^{m'}a^{{l'-1,m'}}\int{\partial_zL\left (\vec{x} ,\omega \right )\overline{Y_{\mathbb{C}}^{l'-1, m'}}(\omega )\ud\omega}
+\frac{1}{\sqrt{2}}\left({-1}\right)^{m'}b^{{l'+1,m'}}\int{\partial_zL\left (\vec{x} ,\omega \right )\overline{Y_{\mathbb{C}}^{l'+1, m'}}(\omega )\ud\omega}
\end{align*}

We simplify these using the identities from Equation~\ref{eq:recursion_identities}:
\begin{align*}
&
\alpha_c\int{\partial_xL\left (\vec{x} ,\omega \right )\overline{Y_{\mathbb{C}}^{l'-1, -m'-1}}(\omega )\ud\omega}
-\left({-1}\right)^{m'}\alpha_c\int{\partial_xL\left (\vec{x} ,\omega \right )\overline{Y_{\mathbb{C}}^{l'-1, m'+1}}(\omega )\ud\omega}
\\&
-\alpha_d\int{\partial_xL\left (\vec{x} ,\omega \right )\overline{Y_{\mathbb{C}}^{l'+1, -m'-1}}(\omega )\ud\omega}
+\left({-1}\right)^{m'}\alpha_d\int{\partial_xL\left (\vec{x} ,\omega \right )\overline{Y_{\mathbb{C}}^{l'+1, m'+1}}(\omega )\ud\omega}
\\&
-\alpha_e\int{\partial_xL\left (\vec{x} ,\omega \right )\overline{Y_{\mathbb{C}}^{l'-1, -m'+1}}(\omega )\ud\omega}
+\left({-1}\right)^{m'}\alpha_e\int{\partial_xL\left (\vec{x} ,\omega \right )\overline{Y_{\mathbb{C}}^{l'-1, m'-1}}(\omega )\ud\omega}
\\&
+\alpha_f\int{\partial_xL\left (\vec{x} ,\omega \right )\overline{Y_{\mathbb{C}}^{l'+1, -m'+1}}(\omega )\ud\omega}
-\left({-1}\right)^{m'}\alpha_f\int{\partial_xL\left (\vec{x} ,\omega \right )\overline{Y_{\mathbb{C}}^{l'+1, m'-1}}(\omega )\ud\omega}
\\&
-i \alpha_c\int{\partial_yL\left (\vec{x} ,\omega \right )\overline{Y_{\mathbb{C}}^{l'-1, -m'-1}}(\omega )\ud\omega}
-\left({-1}\right)^{m'}i \alpha_c\int{\partial_yL\left (\vec{x} ,\omega \right )\overline{Y_{\mathbb{C}}^{l'-1, m'+1}}(\omega )\ud\omega}
\\&
+i \alpha_d\int{\partial_yL\left (\vec{x} ,\omega \right )\overline{Y_{\mathbb{C}}^{l'+1, -m'-1}}(\omega )\ud\omega}
+\left({-1}\right)^{m'}i \alpha_d\int{\partial_yL\left (\vec{x} ,\omega \right )\overline{Y_{\mathbb{C}}^{l'+1, m'+1}}(\omega )\ud\omega}
\\&
-i \alpha_e\int{\partial_yL\left (\vec{x} ,\omega \right )\overline{Y_{\mathbb{C}}^{l'-1, -m'+1}}(\omega )\ud\omega}
-\left({-1}\right)^{m'}i \alpha_e\int{\partial_yL\left (\vec{x} ,\omega \right )\overline{Y_{\mathbb{C}}^{l'-1, m'-1}}(\omega )\ud\omega}
\\&
+i \alpha_f\int{\partial_yL\left (\vec{x} ,\omega \right )\overline{Y_{\mathbb{C}}^{l'+1, -m'+1}}(\omega )\ud\omega}
+\left({-1}\right)^{m'}i \alpha_f\int{\partial_yL\left (\vec{x} ,\omega \right )\overline{Y_{\mathbb{C}}^{l'+1, m'-1}}(\omega )\ud\omega}
\\&
+\alpha_a\int{\partial_zL\left (\vec{x} ,\omega \right )\overline{Y_{\mathbb{C}}^{l'-1, -m'}}(\omega )\ud\omega}
+\left({-1}\right)^{m'}\alpha_a\int{\partial_zL\left (\vec{x} ,\omega \right )\overline{Y_{\mathbb{C}}^{l'-1, m'}}(\omega )\ud\omega}
\\&
+\alpha_b\int{\partial_zL\left (\vec{x} ,\omega \right )\overline{Y_{\mathbb{C}}^{l'+1, -m'}}(\omega )\ud\omega}
+\left({-1}\right)^{m'}\alpha_b\int{\partial_zL\left (\vec{x} ,\omega \right )\overline{Y_{\mathbb{C}}^{l'+1, m'}}(\omega )\ud\omega}
\end{align*}
with
\begin{align*}
\alpha_c = \frac{1}{\sqrt{2}}\frac{1}{2}c^{{l'-1,-m'-1}}
,\qquad
%\\
\alpha_e = \frac{1}{\sqrt{2}}\frac{1}{2}e^{{l'-1,-m'+1}}
,\qquad
%\\
\alpha_d = \frac{1}{\sqrt{2}}\frac{1}{2}d^{{l'+1,-m'-1}}
\\
\alpha_f = \frac{1}{\sqrt{2}}\frac{1}{2}f^{{l'+1,-m'+1}}
,\qquad
%\\
\alpha_a = \frac{1}{\sqrt{2}}a^{{l'-1,-m'}}
,\qquad
%\\
\alpha_b = \frac{1}{\sqrt{2}}b^{{l'+1,-m'}}  \ .
\end{align*}

We substitute the radiance field function L with its spherical harmonics expansion and arrive
at the following expression after further expansions and transformations:
\begin{align*}
&
\alpha_c\partial_x\sum_{l,m}L^{l,m}\int{\SHBR^{l,m}\overline{Y_{\mathbb{C}}^{l'-1, -m'-1}}(\omega )\ud\omega}
-\left({-1}\right)^{m'}\alpha_c\partial_x\sum_{l,m}L^{l,m}\int{\SHBR^{l,m}\overline{Y_{\mathbb{C}}^{l'-1, m'+1}}(\omega )\ud\omega}
\\&
-\alpha_d\partial_x\sum_{l,m}L^{l,m}\int{\SHBR^{l,m}\overline{Y_{\mathbb{C}}^{l'+1, -m'-1}}(\omega )\ud\omega}
+\left({-1}\right)^{m'}\alpha_d\partial_x\sum_{l,m}L^{l,m}\int{\SHBR^{l,m}\overline{Y_{\mathbb{C}}^{l'+1, m'+1}}(\omega )\ud\omega}
\\&
-\alpha_e\partial_x\sum_{l,m}L^{l,m}\int{\SHBR^{l,m}\overline{Y_{\mathbb{C}}^{l'-1, -m'+1}}(\omega )\ud\omega}
+\left({-1}\right)^{m'}\alpha_e\partial_x\sum_{l,m}L^{l,m}\int{\SHBR^{l,m}\overline{Y_{\mathbb{C}}^{l'-1, m'-1}}(\omega )\ud\omega}
\\&
+\alpha_f\partial_x\sum_{l,m}L^{l,m}\int{\SHBR^{l,m}\overline{Y_{\mathbb{C}}^{l'+1, -m'+1}}(\omega )\ud\omega}
-\left({-1}\right)^{m'}\alpha_f\partial_x\sum_{l,m}L^{l,m}\int{\SHBR^{l,m}\overline{Y_{\mathbb{C}}^{l'+1, m'-1}}(\omega )\ud\omega}
\\&
-i \alpha_c\partial_y\sum_{l,m}L^{l,m}\int{\SHBR^{l,m}\overline{Y_{\mathbb{C}}^{l'-1, -m'-1}}(\omega )\ud\omega}
-\left({-1}\right)^{m'}i \alpha_c\partial_y\sum_{l,m}L^{l,m}\int{\SHBR^{l,m}\overline{Y_{\mathbb{C}}^{l'-1, m'+1}}(\omega )\ud\omega}
\\&
+i \alpha_d\partial_y\sum_{l,m}L^{l,m}\int{\SHBR^{l,m}\overline{Y_{\mathbb{C}}^{l'+1, -m'-1}}(\omega )\ud\omega}
+\left({-1}\right)^{m'}i \alpha_d\partial_y\sum_{l,m}L^{l,m}\int{\SHBR^{l,m}\overline{Y_{\mathbb{C}}^{l'+1, m'+1}}(\omega )\ud\omega}
\\&
-i \alpha_e\partial_y\sum_{l,m}L^{l,m}\int{\SHBR^{l,m}\overline{Y_{\mathbb{C}}^{l'-1, -m'+1}}(\omega )\ud\omega}
-\left({-1}\right)^{m'}i \alpha_e\partial_y\sum_{l,m}L^{l,m}\int{\SHBR^{l,m}\overline{Y_{\mathbb{C}}^{l'-1, m'-1}}(\omega )\ud\omega}
\\&
+i \alpha_f\partial_y\sum_{l,m}L^{l,m}\int{\SHBR^{l,m}\overline{Y_{\mathbb{C}}^{l'+1, -m'+1}}(\omega )\ud\omega}
+\left({-1}\right)^{m'}i \alpha_f\partial_y\sum_{l,m}L^{l,m}\int{\SHBR^{l,m}\overline{Y_{\mathbb{C}}^{l'+1, m'-1}}(\omega )\ud\omega}
\\&
+\alpha_a\partial_z\sum_{l,m}L^{l,m}\int{\SHBR^{l,m}\overline{Y_{\mathbb{C}}^{l'-1, -m'}}(\omega )\ud\omega}
+\left({-1}\right)^{m'}\alpha_a\partial_z\sum_{l,m}L^{l,m}\int{\SHBR^{l,m}\overline{Y_{\mathbb{C}}^{l'-1, m'}}(\omega )\ud\omega}
\\&
+\alpha_b\partial_z\sum_{l,m}L^{l,m}\int{\SHBR^{l,m}\overline{Y_{\mathbb{C}}^{l'+1, -m'}}(\omega )\ud\omega}
+\left({-1}\right)^{m'}\alpha_b\partial_z\sum_{l,m}L^{l,m}\int{\SHBR^{l,m}\overline{Y_{\mathbb{C}}^{l'+1, m'}}(\omega )\ud\omega}
\end{align*}
Again we apply the identity given in Equation~\ref{eq:real_orthogonality_property_with_complex}. For the first two terms we for example get:
\begin{align*}
&
%-i &\alpha_c\partial_y\sum_{l,m}L^{l,m}\left (\vec{x}\right)
%\int{Y_{\mathbb{R}}^{l,m}\overline{Y^{l'-1, m'-1}}(\omega )\ud\omega}
\mathcolor{red}{}
%&
\mathcolor{red}{\alpha_c\partial_x
\sum_{m=-l'+1}^{-1}L^{l'-1,m}\left (\vec{x}\right)
\frac{i}{\sqrt{2}}
\delta_{\scaleto{m=-m'-1}{4pt}}
}
%\\
% -------------------------------------------------
\mathcolor{blue}{
-
}
%&
\mathcolor{blue}{
\alpha_c\partial_x
\sum_{m=-l'+1}^{-1}L^{l'-1,m}\left (\vec{x}\right)
\frac{i}{\sqrt{2}}
\left({-1}\right)^{m}
\delta_{\scaleto{m=m'+1}{4pt}}
}
\\&
% -------------------------------------------------
\mathcolor{blue}
{
+
}
%&
\mathcolor{blue}
{
\alpha_c\partial_x
L^{l'-1,0}\left (\vec{x}\right)
\frac{1}{\sqrt{2}}
\delta_{\scaleto{-m'=1}{4pt}}
}
%\\
% -------------------------------------------------
\mathcolor{black}
{
+
}
%&
\mathcolor{black}
{
\alpha_c\partial_x
\sum_{m=1}^{l'-1}L^{l'-1,m}\left (\vec{x}\right)
\frac{1}{\sqrt{2}}
\delta_{\scaleto{m=m'+1}{4pt}}
}
%\\
% -------------------------------------------------
\mathcolor{blue}
{
+
}
%&
\mathcolor{blue}
{
\alpha_c\partial_x
\sum_{m=1}^{l'-1}L^{l'-1,m}\left (\vec{x}\right)
\frac{1}{\sqrt{2}}
\left({-1}\right)^{m}
\delta_{\scaleto{m=-m'-1}{4pt}}
}
\\&
% -------------------------------------------------
\mathcolor{blue}
{
-\left({-1}\right)^{-m'}
}
%&
\mathcolor{blue}{
\alpha_c\partial_x
\sum_{m=-l'+1}^{-1}L^{l'-1,m}\left (\vec{x}\right)
\frac{i}{\sqrt{2}}
\delta_{\scaleto{m=m'+1}{4pt}}
}
%\\
% -------------------------------------------------
\mathcolor{red}
{
+\left({-1}\right)^{-m'}
}
%&
\mathcolor{red}
{
\alpha_c\partial_x
\sum_{m=-l'+1}^{-1}L^{l'-1,m}\left (\vec{x}\right)
\frac{i}{\sqrt{2}}
\left({-1}\right)^{m}
\delta_{\scaleto{m=-m'-1}{4pt}}
}
% -------------------------------------------------
\\&
\mathcolor{blue}
{
-
\left({-1}\right)^{-m'}
}
%&
\mathcolor{blue}
{
\alpha_c\partial_x
L^{l'-1,0}\left (\vec{x}\right)
\frac{1}{\sqrt{2}}
\delta_{\scaleto{-m'=1}{4pt}}
}
%\\
% -------------------------------------------------
\mathcolor{blue}
{
-
\left({-1}\right)^{-m'}
}
%&
\mathcolor{blue}
{
\alpha_c\partial_x
\sum_{m=1}^{l'-1}L^{l'-1,m}\left (\vec{x}\right)
\frac{1}{\sqrt{2}}
\delta_{\scaleto{m=-m'-1}{4pt}}
}
\\&
% -------------------------------------------------
\mathcolor{black}
{
-
\left({-1}\right)^{-m'}
}
%&
\mathcolor{black}
{
\alpha_c\partial_x
\sum_{m=1}^{l'-1}L^{l'-1,m}\left (\vec{x}\right)
\frac{1}{\sqrt{2}}
\left({-1}\right)^{m}
\delta_{\scaleto{m=m'+1}{4pt}}
}
\end{align*}
As with the $m'<0$ case, the blue and red terms cancel each other out, leaving only the black terms. The first two terms of the real-valued $P_N$-equations for the transport term therefore are:
\begin{align*}
&
\alpha_c\partial_x
\sum_{m=1}^{l'-1}L^{l'-1,m}\left (\vec{x}\right)
\frac{1}{\sqrt{2}}
\delta_{\scaleto{m=m'+1}{4pt}}
-
\left({-1}\right)^{-m'}
\alpha_c\partial_x
\sum_{m=1}^{l'-1}L^{l'-1,m}\left (\vec{x}\right)
\frac{1}{\sqrt{2}}
\left({-1}\right)^{m}
\delta_{\scaleto{m=m'+1}{4pt}}
\\&
=
\frac{2}{\sqrt{2}}
\alpha_c\partial_x
L^{l'-1,m'+1}\left (\vec{x}\right)
=
\frac{2}{\sqrt{2}}
\left(\frac{1}{2\sqrt{2}}c^{l'-1, -m'-1}\right)\partial_x
L^{l'-1,m'+1}\left (\vec{x}\right)
\\&
=
\frac{1}{2}c^{l'-1,-m'-1}L^{l'-1,m'+1}\left (\vec{x}\right) \ .
\end{align*}

Following this through for the remaining terms gives us the real-valued $P_N$-equations for $m'>0$:
\begin{align}
&
\frac{1}{2}c^{l'-1,-m'-1}\partial_x L^{l'-1,m'+1}\left (\vec{x}\right)
%\\&
-\frac{1}{2}d^{l'+1,-m'-1}\partial_x L^{l'+1,m'+1}\left (\vec{x}\right)
%\\&
-\frac{1}{2}\beta^{m'}e^{l'-1,m'-1}\partial_x L^{l'-1,m'-1}\left (\vec{x}\right)
\nonumber
\\&
\frac{1}{2}\beta^{m'}f^{l'+1,-m'+1}\partial_x L^{l'+1,m'-1}\left (\vec{x}\right)
%\\&
\frac{1}{2}c^{l'-1,-m'-1}\partial_y L^{l'-1,-m'-1}\left (\vec{x}\right)
%\\&
-\frac{1}{2}d^{l'+1,-m'-1}\partial_y L^{l'+1,-m'-1}\left (\vec{x}\right)
\nonumber
\\&
\delta_{\scaleto{m'\neq 1}{4pt}}\frac{1}{2}e^{l'-1,-m'+1}\partial_y L^{l'-1,-m'+1}\left (\vec{x}\right)
%\\&
-\delta_{\scaleto{m'\neq 1}{4pt}}\frac{1}{2}f^{l'+1,-m'+1}\partial_y L^{l'+1,-m'+1}\left (\vec{x}\right)
%\\&
a^{l'-1,-m'}\partial_z L^{l'-1,m'}\left (\vec{x}\right)
\nonumber
\\&
b^{l'+1,-m'}\partial_z L^{l'+1,m'}\left (\vec{x}\right)
\nonumber
\end{align}

Finally the $m'=0$ case needs to be derived. The derivation starts very similar to the complex-valued $P_N$-equations as in this case, the real-valued SH basis function is identical to the complex-valued SH basis function. We multiply Equation~\ref{sec:complex_transport_term} with the definition of the real-valued SH basis for $m'=0$ and get:
\begin{align*}
\int{\overline{Y_{\mathbb{C}}^{l', m'}}(\omega )\left(\omega_{x}\partial_xL\left (\vec{x} ,\omega \right )+\omega_{y}\partial_yL\left (\vec{x} ,\omega \right )+\omega_{z}\partial_zL\left (\vec{x} ,\omega \right )\right)\ud\omega}
\end{align*}

Expanding the integrand and applying the recursion relation (Equation~\ref{eq:recursion_identities}) produces the following set of terms:
\begin{align*}
&
\frac{1}{2}c^{{l'-1,m'-1}}\int{\partial_xL\left (\vec{x} ,\omega \right )\overline{Y_{\mathbb{C}}^{l'-1, m'-1}}(\omega )\ud\omega}
-\frac{1}{2}e^{{l'-1,m'+1}}\int{\partial_xL\left (\vec{x} ,\omega \right )\overline{Y_{\mathbb{C}}^{l'-1, m'+1}}(\omega )\ud\omega}
\\&
-\frac{1}{2}d^{{l'+1,m'-1}}\int{\partial_xL\left (\vec{x} ,\omega \right )\overline{Y_{\mathbb{C}}^{l'+1, m'-1}}(\omega )\ud\omega}
+\frac{1}{2}f^{{l'+1,m'+1}}\int{\partial_xL\left (\vec{x} ,\omega \right )\overline{Y_{\mathbb{C}}^{l'+1, m'+1}}(\omega )\ud\omega}
\\&
-\frac{i}{2}c^{{l'-1,m'-1}}\int{\partial_yL\left (\vec{x} ,\omega \right )\overline{Y_{\mathbb{C}}^{l'-1, m'-1}}(\omega )\ud\omega}
-\frac{i}{2}e^{{l'-1,m'+1}}\int{\partial_yL\left (\vec{x} ,\omega \right )\overline{Y_{\mathbb{C}}^{l'-1, m'+1}}(\omega )\ud\omega}
\\&
+\frac{i}{2}d^{{l'+1,m'-1}}\int{\partial_yL\left (\vec{x} ,\omega \right )\overline{Y_{\mathbb{C}}^{l'+1, m'-1}}(\omega )\ud\omega}
+\frac{i}{2}f^{{l'+1,m'+1}}\int{\partial_yL\left (\vec{x} ,\omega \right )\overline{Y_{\mathbb{C}}^{l'+1, m'+1}}(\omega )\ud\omega}
\\&
+a^{{l'-1,m'}}\int{\overline{Y_{\mathbb{C}}^{l'-1, m'}}(\omega )\partial_zL\left (\vec{x} ,\omega \right )\ud\omega}
+b^{{l'+1,m'}}\int{\overline{Y_{\mathbb{C}}^{l'+1, m'}}(\omega )\partial_zL\left (\vec{x} ,\omega \right )\ud\omega}
\end{align*}

Again we will replace the radiance field $L$ with its real-valued SH projection and get:
\begin{align*}
&
\frac{1}{2}c^{{l'-1,m'-1}}\partial_x\sum_{l,m}L^{l,m}\int{\SHBR^{l,m}\overline{Y_{\mathbb{C}}^{l'-1, m'-1}}(\omega )\ud\omega}
-\frac{1}{2}e^{{l'-1,m'+1}}\partial_x\sum_{l,m}L^{l,m}\int{\SHBR^{l,m}\overline{Y_{\mathbb{C}}^{l'-1, m'+1}}(\omega )\ud\omega}
\\&
-\frac{1}{2}d^{{l'+1,m'-1}}\partial_x\sum_{l,m}L^{l,m}\int{\SHBR^{l,m}\overline{Y_{\mathbb{C}}^{l'+1, m'-1}}(\omega )\ud\omega}
+\frac{1}{2}f^{{l'+1,m'+1}}\partial_x\sum_{l,m}L^{l,m}\int{\SHBR^{l,m}\overline{Y_{\mathbb{C}}^{l'+1, m'+1}}(\omega )\ud\omega}
\\&
-\frac{i}{2}c^{{l'-1,m'-1}}\partial_y\sum_{l,m}L^{l,m}\int{\SHBR^{l,m}\overline{Y_{\mathbb{C}}^{l'-1, m'-1}}(\omega )\ud\omega}
-\frac{i}{2}e^{{l'-1,m'+1}}\partial_y\sum_{l,m}L^{l,m}\int{\SHBR^{l,m}\overline{Y_{\mathbb{C}}^{l'-1, m'+1}}(\omega )\ud\omega}
\\&
+\frac{i}{2}d^{{l'+1,m'-1}}\partial_y\sum_{l,m}L^{l,m}\int{\SHBR^{l,m}\overline{Y_{\mathbb{C}}^{l'+1, m'-1}}(\omega )\ud\omega}
+\frac{i}{2}f^{{l'+1,m'+1}}\partial_y\sum_{l,m}L^{l,m}\int{\SHBR^{l,m}\overline{Y_{\mathbb{C}}^{l'+1, m'+1}}(\omega )\ud\omega}
\\&
+a^{{l'-1,m'}}\partial_z\sum_{l,m}L^{l,m}\int{\overline{Y_{\mathbb{C}}^{l'-1, m'}}(\omega )\SHBR^{l,m}\ud\omega}
+b^{{l'+1,m'}}\partial_z\sum_{l,m}L^{l,m}\int{\overline{Y_{\mathbb{C}}^{l'+1, m'}}(\omega )\SHBR^{l,m}\ud\omega}
\end{align*}

These terms also have an intricate structure where many terms cancel out and simplify. This is seen once we apply the SH orthogonality property (Equation~\ref{eq:real_orthogonality_property_with_complex}) and further consider that $m'=0$. We show this for the first two terms, which expand to:
\begin{align*}
&
\mathcolor{red}
{
\frac{1}{2}c^{{l'-1,-1}}\frac{i}{\sqrt{2}}\partial_x\left(\sum_{m=-l'+1}^{-1}{L^{{l'-1,m}}\left (\vec{x} \right )\delta_{-1,m}}\right)
}
\mathcolor{blue}
{
-\frac{1}{2}c^{{l'-1,-1}}\frac{i}{\sqrt{2}}\partial_x\left(\sum_{m=-l'+1}^{-1}{L^{{l'-1,m}}\left (\vec{x} \right )\left({-1}\right)^{m}\delta_{-1,-m}}\right)
}
\\&
\mathcolor{blue}
{
\frac{1}{2}c^{{l'-1,-1}}\partial_x\left(L^{{l'-1,0}}\left (\vec{x} \right )\delta_{-1,0}\right)
}
\frac{1}{2}c^{{l'-1,-1}}\frac{1}{\sqrt{2}}\partial_x\left(\sum_{m=1}^{l'-1}{L^{{l'-1,m}}\left (\vec{x} \right )\delta_{-1,-m}}\right)
\\&
\mathcolor{blue}
{
\frac{1}{2}c^{{l'-1,-1}}\frac{1}{\sqrt{2}}\partial_x\left(\sum_{m=1}^{l'-1}{L^{{l'-1,m}}\left (\vec{x} \right )\left({-1}\right)^{m}\delta_{-1,m}}\right)
}
\mathcolor{blue}
{
-\frac{1}{2}e^{{l'-1,1}}\frac{i}{\sqrt{2}}\partial_x\left(\sum_{m=-l'+1}^{-1}{L^{{l'-1,m}}\left (\vec{x} \right )\delta_{1,m}}\right)
}
\\&
\mathcolor{red}
{
\frac{1}{2}e^{{l'-1,1}}\frac{i}{\sqrt{2}}\partial_x\left(\sum_{m=-l'+1}^{-1}{L^{{l'-1,m}}\left (\vec{x} \right )\left({-1}\right)^{m}\delta_{1,-m}}\right)
}
\mathcolor{blue}
{
-\frac{1}{2}e^{{l'-1,1}}\partial_x\left(L^{{l'-1,0}}\left (\vec{x} \right )\delta_{1,0}\right)
}
\\&
\mathcolor{blue}
{
-\frac{1}{2}e^{{l'-1,1}}\frac{1}{\sqrt{2}}\partial_x\left(\sum_{m=1}^{l'-1}{L^{{l'-1,m}}\left (\vec{x} \right )\delta_{1,-m}}\right)
}
-\frac{1}{2}e^{{l'-1,1}}\frac{1}{\sqrt{2}}\partial_x\left(\sum_{m=1}^{l'-1}{L^{{l'-1,m}}\left (\vec{x} \right )\left({-1}\right)^{m}\delta_{1,m}}\right)
\end{align*}

Again the blue terms vanish since the delta functions will never be non-zero under the sums. The red terms cancel each other out since $c^{l,-1}=e^{l,1}$ and $-1^m=-1$ for $m=-1$. The terms in black simplify to:
\begin{align*}
&
\frac{1}{2}c^{{l'-1,-1}}\frac{1}{\sqrt{2}}\partial_x\left(\sum_{m=1}^{l'-1}{L^{{l'-1,m}}\left (\vec{x} \right )\delta_{-1,-m}}\right)
-\frac{1}{2}e^{{l'-1,1}}\frac{1}{\sqrt{2}}\partial_x\left(\sum_{m=1}^{l'-1}{L^{{l'-1,m}}\left (\vec{x} \right )\left({-1}\right)^{m}\delta_{1,m}}\right)
\\&
=
\frac{1}{2}c^{{l'-1,-1}}\frac{1}{\sqrt{2}}\partial_xL^{{l'-1,1}}\left (\vec{x} \right )
-\frac{1}{2}c^{{l'-1,-1}}\frac{1}{\sqrt{2}}\partial_xL^{{l'-1,1}}\left (\vec{x} \right )\left({-1}\right)^{1}
\\&
=
\frac{1}{\sqrt{2}}c^{{l'-1,-1}}\partial_xL^{{l'-1,1}}\left (\vec{x} \right ) \ .
\end{align*}
Similar simplifications apply to the remaining terms of the SH expansion of the transport term for $m=0$, resulting in the final expression:
\begin{align*}
&
\frac{1}{\sqrt{2}}c^{{l'-1,-1}}\partial_x L^{{l'-1,1}}\left (\vec{x} \right )
-\frac{1}{\sqrt{2}}d^{{l'+1,-1}}\partial_x L^{{l'+1,1}}\left (\vec{x} \right )
\\&
\frac{1}{\sqrt{2}}c^{{l'-1,-1}}\partial_y L^{{l'-1,-1}}\left (\vec{x} \right )
-\frac{1}{\sqrt{2}}d^{{l'+1,-1}}\partial_y L^{{l'+1,-1}}\left (\vec{x} \right )
\\&
a^{{l'-1,0}}\partial_z L^{{l'-1,0}}\left (\vec{x} \right)
+b^{{l'+1,0}}\partial_z L^{{l'+1,0}}\left (\vec{x} \right)
\end{align*}

\subsubsection{Collision Term}

The collision term of the RTE is given as:
\begin{align*}
-\sigma_t\left(\vec{x}\right)L\left(\vec{x}, \omega\right)
\end{align*}
We first replace the radiance field $L$ with its real-valued SH expansion:
\begin{align*}
-\sigma_t\left(\vec{x}\right)
\sum_{l,m}
{
L^{l,m}\left(\vec{x}\right )\SHBR^{l,m}\left(\omega\right)
}
\end{align*}

In order to project the term into SH, we have to multiply with the real-valued SH basis function and integrate over solid angle. Since the basis function is different depending on $m'<0$, $m=0$ or $m>0$, we have to derive separate $P_N$-equations for each case.

We first derive the SH projection of the collision term for the case $m'<0$. Multiplying with the SH basis and integrating over solid angle gives, after some further transformations and application of the SH orthogonality property:
\begin{align*}
&
\mathcolor{black}
{
-\frac{i}{\sqrt{2}}\sigma_t\left (\vec{x} \right )\frac{i}{\sqrt{2}}\sum_{m=-l'}^{-1}{L^{{l',m}}\left (\vec{x} \right )\delta_{m',m}}
}
\mathcolor{blue}
{
+\frac{i}{\sqrt{2}}\sigma_t\left (\vec{x} \right )\frac{i}{\sqrt{2}}\sum_{m=-l'}^{-1}{L^{{l',m}}\left (\vec{x} \right )\left({-1}\right)^{m}\delta_{m',-m}}
}
\\&
\mathcolor{blue}
{
-\frac{i}{\sqrt{2}}\sigma_t\left (\vec{x} \right )L^{{l',0}}\left (\vec{x} \right )\delta_{m',0}
}
\mathcolor{red}
{
-\frac{i}{\sqrt{2}}\sigma_t\left (\vec{x} \right )\frac{1}{\sqrt{2}}\sum_{m=1}^{l'}{L^{{l',m}}\left (\vec{x} \right )\delta_{m',-m}}
}
\\&
\mathcolor{blue}
{
-\frac{i}{\sqrt{2}}\sigma_t\left (\vec{x} \right )\frac{1}{\sqrt{2}}\sum_{m=1}^{l'}{L^{{l',m}}\left (\vec{x} \right )\left({-1}\right)^{m}\delta_{m',m}}
}
\mathcolor{blue}
{
+\frac{i}{\sqrt{2}}\left({-1}\right)^{m'}\sigma_t\left (\vec{x} \right )\frac{i}{\sqrt{2}}\sum_{m=-l'}^{-1}{L^{{l',m}}\left (\vec{x} \right )\delta_{-m',m}}
}
\\&
\mathcolor{black}
{
-\frac{i}{\sqrt{2}}\left({-1}\right)^{m'}\sigma_t\left (\vec{x} \right )\frac{i}{\sqrt{2}}\sum_{m=-l'}^{-1}{L^{{l',m}}\left (\vec{x} \right )\left({-1}\right)^{m}\delta_{-m',-m}}
}
\mathcolor{blue}
{
+\frac{i}{\sqrt{2}}\left({-1}\right)^{m'}\sigma_t\left (\vec{x} \right )L^{{l',0}}\left (\vec{x} \right )\delta_{-m',0}
}
\\&
\mathcolor{blue}
{
+\frac{i}{\sqrt{2}}\left({-1}\right)^{m'}\sigma_t\left (\vec{x} \right )\frac{1}{\sqrt{2}}\sum_{m=1}^{l'}{L^{{l',m}}\left (\vec{x} \right )\delta_{-m',-m}}
}
\mathcolor{red}
{
+\frac{i}{\sqrt{2}}\left({-1}\right)^{m'}\sigma_t\left (\vec{x} \right )\frac{1}{\sqrt{2}}\sum_{m=1}^{l'}{L^{{l',m}}\left (\vec{x} \right )\left({-1}\right)^{m}\delta_{-m',m}}
}
\end{align*}

As for the transport term derivation, the blue terms vanish due to the delta function being always zero under the sum. The red terms cancel each other out. The remaining term (black) determines the SH projection of the collision term for $m'<0$:
\begin{align}
\sigma_t L^{l',m'}
\end{align}
The derivation for the SH projection of the collision term for $m>0$ follows the same structure and likewise results in:
\begin{align}
\sigma_t L^{l',m'}
\end{align}
The real-values SH projection of the collision term for $m=0$ also is:
\begin{align}
\sigma_t L^{l',m'}
\end{align}

\subsubsection{Scattering Term}

The scattering term is given as a convolution of the radiance field $L$ with the phase function $p$ using a rotation $R_\omega$:
\begin{align*}
&
\sigma_s(\vec{x})\int_{\Omega'}p(\vec{x}, \omega'\cdot\omega)L(\vec{x}, \omega')\ud\omega'
\\
&= \sigma_s(\vec{x})(L\circ \rho_{R(\omega)}(p))(\omega)
\end{align*}
where the convolution can be also expressed as a inner product integral:
\begin{align*}
(L\circ \rho_{R(\omega)}(p)) &= \int_{\Omega'}{L(\vec{x}, \omega')\rho_{R(\omega)}(p)(\omega')\ud\omega'} \\
&= \langle L,  \rho_{R(\omega)}(p)\rangle \ .
\end{align*}

We substitute $L$ with its real-valued SH-expansion (Equation~\ref{eq:real_sh_exp_L}) in the inner product integral and perform some further factorizations to get:
%\begin{align*}
%L\left(\vec{x}, \omega\right)=
%&
%\frac{i}{\sqrt{2}}\left(\sum_{l=0}^{N}{\sum_{m=-l}^{-1}{L^{{l,m}}\left (\vec{x} \right )\SHBC^{l, m}(\omega )}}\right)-\frac{i}{\sqrt{2}}\left(\sum_{l=0}^{N}{\sum_{m=-l}^{-1}{L^{{l,m}}\left (\vec{x} \right )\left({-1}\right)^{m}\SHBC^{l, -m}(\omega )}}\right)
%\\&
%+\sum_{l=0}^{N}{L^{{l,0}}\left (\vec{x} \right )\SHBC^{l, 0}(\omega )}+\frac{1}{\sqrt{2}}\left(\sum_{l=0}^{N}{\sum_{m=1}^{l}{L^{{l,m}}\left (\vec{x} \right )\SHBC^{l, -m}(\omega )}}\right)
%\\&
%+\frac{1}{\sqrt{2}}\left(\sum_{l=0}^{N}{\sum_{m=1}^{l}{L^{{l,m}}\left (\vec{x} \right )\left({-1}\right)^{m}\SHBC^{l, m}(\omega )}}\right)
%\end{align*}
\begin{align*}
&
\frac{i}{\sqrt{2}}
\sum_{l=0}^{N}{
\sum_{l'}^{N}{
\sum_{m=-l}^{-1}{
L^{{l,m}}\left (\vec{x} \right )
f^{l'0}
\left<
\SHBC^{l, m}(\omega )
, \rho_{R\left(\omega\right)}
\left(
\SHBC^{l'0}
\right)
\right>
}
}
}
\\
-
&
\frac{i}{\sqrt{2}}
\sum_{l=0}^{N}{
\sum_{l'}^{N}{
\sum_{m=-l}^{-1}{
\left({-1}\right)^{m}
L^{{l,m}}\left (\vec{x} \right )
f^{l'0}
\left<
\SHBC^{l, -m}(\omega )
, \rho_{R\left(\omega\right)}
\left(
\SHBC^{l'0}
\right)
\right>
}
}
}
\\
+
&
\sum_{l=0}^{N}{
\sum_{l'}^{N}{
L^{{l,0}}\left (\vec{x} \right )
f^{l'0}
\left<
\SHBC^{l, 0}(\omega )
, \rho_{R\left(\omega\right)}
\left(
\SHBC^{l'0}
\right)
\right>
}
}
\\
+
&
\frac{1}{\sqrt{2}}
\sum_{l=0}^{N}{
\sum_{l'}^{N}{
\sum_{m=1}^{l}{
L^{{l,m}}\left (\vec{x} \right )
f^{l'0}
\left<
\SHBC^{l, -m}(\omega )
, \rho_{R\left(\omega\right)}
\left(
\SHBC^{l'0}
\right)
\right>
}
}
}
\\
+
&
\frac{1}{\sqrt{2}}
\sum_{l=0}^{N}{
\sum_{l'}^{N}{
\sum_{m=1}^{l}{
\left({-1}\right)^{m}
L^{{l,m}}\left (\vec{x} \right )
f^{l'0}
\left<
\SHBC^{l, m}(\omega )
, \rho_{R\left(\omega\right)}
\left(
\SHBC^{l'0}
\right)
\right>
}
}
}
\end{align*}

The spherical harmonics basis functions $\SHBC^{lm}$ are orthogonal. We therefore have $\left < Y^{lm}, \rho_{R(\omega)}\left(Y^{l'm'}\right) \right > = 0$, for all $l\ne l'$, which further simplifies our scattering operator to
\begin{align}
&
\frac{i}{\sqrt{2}}
\sum_{l=0}^{N}{
\sum_{m=-l}^{-1}{
L^{{l,m}}\left (\vec{x} \right )
f^{l0}
\left<
\SHBC^{l, m}(\omega )
, \rho_{R\left(\omega\right)}
\left(
\SHBC^{l0}
\right)
\right>
}
}
\\
-
&
\frac{i}{\sqrt{2}}
\sum_{l=0}^{N}{
\sum_{m=-l}^{-1}{
\left({-1}\right)^{m}
L^{{l,m}}\left (\vec{x} \right )
f^{l0}
\left<
\SHBC^{l, -m}(\omega )
, \rho_{R\left(\omega\right)}
\left(
Y^{l0}
\right)
\right>
}
}
\\
+
&
\sum_{l=0}^{N}{
L^{{l,0}}\left (\vec{x} \right )
f^{l0}
\left<
\SHBC^{l, 0}(\omega )
, \rho_{R\left(\omega\right)}
\left(
\SHBC^{l0}
\right)
\right>
}
\\
+
&
\frac{1}{\sqrt{2}}
\sum_{l=0}^{N}{
\sum_{m=1}^{l}{
L^{{l,m}}\left (\vec{x} \right )
f^{l0}
\left<
\SHBC^{l, -m}(\omega )
, \rho_{R\left(\omega\right)}
\left(
\SHBC^{l0}
\right)
\right>
}
}
\\
+
&
\frac{1}{\sqrt{2}}
\sum_{l=0}^{N}{
\sum_{m=1}^{l}{
\left({-1}\right)^{m}
L^{{l,m}}\left (\vec{x} \right )
f^{l0}
\left<
\SHBC^{l, m}(\omega )
, \rho_{R\left(\omega\right)}
\left(
\SHBC^{l0}
\right)
\right>
}
}
\end{align}

What remains to be resolved are the inner products. We use the fact that the spherical harmonics basis functions $ \SHBC^{lm}$ are eigenfunctions of the inner product integral operator in the equation above, i.e.
\begin{align}
\left < \SHBC^{lm}, \rho_{R(\omega)}\left ( \SHBC^{l0} \right )\right > = \lambda_l \SHBC^{lm}
\end{align}
which results in:
\begin{align}
&
\frac{i}{\sqrt{2}}
\sum_{l=0}^{N}{
\sum_{m=-l}^{-1}{
L^{{l,m}}\left (\vec{x} \right )
f^{l0}
\lambda_l
\SHBC^{l, m}(\omega )
}
}
\\
-
&
\frac{i}{\sqrt{2}}
\sum_{l=0}^{N}{
\sum_{m=-l}^{-1}{
\left({-1}\right)^{m}
L^{{l,m}}\left (\vec{x} \right )
f^{l0}
\lambda_l
\SHBC^{l, -m}(\omega )
}
}
\\
+
&
\sum_{l=0}^{N}{
L^{{l,0}}\left (\vec{x} \right )
f^{l0}
\lambda_l
\SHBC^{l, 0}(\omega )
}
\\
+
&
\frac{1}{\sqrt{2}}
\sum_{l=0}^{N}{
\sum_{m=1}^{l}{
L^{{l,m}}\left (\vec{x} \right )
f^{l0}
\lambda_l
\SHBC^{l, -m}(\omega )
}
}
\\
+
&
\frac{1}{\sqrt{2}}
\sum_{l=0}^{N}{
\sum_{m=1}^{l}{
\left({-1}\right)^{m}
L^{{l,m}}\left (\vec{x} \right )
f^{l0}
\lambda_l
\SHBC^{l, m}(\omega )
}
} \ .
\end{align}

The next step is to project the scattering term into real-valued SH. Again we will have to use different terms for $m<0$, $m=0$ and $m>0$ due to the definition of the real-valued SH basis functions. Multiplying with the real-valued SH basis function for $m<0$ and after applying further transformations, we get:
\begin{align*}
&
\frac{i}{\sqrt{2}}\sigma_s\left (\vec{x} \right )\frac{i}{\sqrt{2}}p^{{l',0}}\left (\vec{x} \right )\lambda_{{l'}}\sum_{m=-l'}^{-1}{L^{{l',m}}\left (\vec{x} \right )\delta_{m',m}}
\mathcolor{blue}
{
-\frac{i}{\sqrt{2}}\sigma_s\left (\vec{x} \right )\frac{i}{\sqrt{2}}p^{{l',0}}\left (\vec{x} \right )\lambda_{{l'}}\sum_{m=-l'}^{-1}{\left({-1}\right)^{m}L^{{l',m}}\left (\vec{x} \right )\delta_{m',-m}}
}
\\&
\mathcolor{blue}
{
+\frac{i}{\sqrt{2}}\sigma_s\left (\vec{x} \right )L^{{l',0}}\left (\vec{x} \right )p^{{l',0}}\left (\vec{x} \right )\lambda_{{l'}}\delta_{m',0}
}
\mathcolor{red}
{
+\frac{i}{\sqrt{2}}\sigma_s\left (\vec{x} \right )\frac{1}{\sqrt{2}}p^{{l',0}}\left (\vec{x} \right )\lambda_{{l'}}\sum_{m=1}^{l'}{L^{{l',m}}\left (\vec{x} \right )\delta_{m',-m}}
}
\\&
\mathcolor{blue}
{
+\frac{i}{\sqrt{2}}\sigma_s\left (\vec{x} \right )\frac{1}{\sqrt{2}}p^{{l',0}}\left (\vec{x} \right )\lambda_{{l'}}\sum_{m=1}^{l'}{\left({-1}\right)^{m}L^{{l',m}}\left (\vec{x} \right )\delta_{m',m}}
}
\mathcolor{blue}
{
-\frac{i}{\sqrt{2}}\left({-1}\right)^{m'}\sigma_s\left (\vec{x} \right )\frac{i}{\sqrt{2}}p^{{l',0}}\left (\vec{x} \right )\lambda_{{l'}}\sum_{m=-l'}^{-1}{L^{{l',m}}\left (\vec{x} \right )\delta_{-m',m}}
}
\\&
+\frac{i}{\sqrt{2}}\left({-1}\right)^{m'}\sigma_s\left (\vec{x} \right )\frac{i}{\sqrt{2}}p^{{l',0}}\left (\vec{x} \right )\lambda_{{l'}}\sum_{m=-l'}^{-1}{\left({-1}\right)^{m}L^{{l',m}}\left (\vec{x} \right )\delta_{-m',-m}}
\mathcolor{blue}
{
-\frac{i}{\sqrt{2}}\left({-1}\right)^{m'}\sigma_s\left (\vec{x} \right )L^{{l',0}}\left (\vec{x} \right )p^{{l',0}}\left (\vec{x} \right )\lambda_{{l'}}\delta_{-m',0}
}
\\&
\mathcolor{blue}
{
-\frac{i}{\sqrt{2}}\left({-1}\right)^{m'}\sigma_s\left (\vec{x} \right )\frac{1}{\sqrt{2}}p^{{l',0}}\left (\vec{x} \right )\lambda_{{l'}}\sum_{m=1}^{l'}{L^{{l',m}}\left (\vec{x} \right )\delta_{-m',-m}}
}
\mathcolor{red}
{
-\frac{i}{\sqrt{2}}\left({-1}\right)^{m'}\sigma_s\left (\vec{x} \right )\frac{1}{\sqrt{2}}p^{{l',0}}\left (\vec{x} \right )\lambda_{{l'}}\sum_{m=1}^{l'}{\left({-1}\right)^{m}L^{{l',m}}\left (\vec{x} \right )\delta_{-m',m}}
}
\end{align*}
Again, the blue terms vanish, because the delta functions will always be zero under the sum. The red terms cancel each other out. The black terms reduce to:
\begin{align*}
-\sigma_s\left(\vec{x}\right)\lambda_{l'}p^{l',0}\left(\vec{x}\right)L^{l',m'}\left(\vec{x}\right)
\end{align*}
The same happens for the derivation for $m>0$ and $m=0$ resulting in the same term.

\subsubsection{Emission Term}

The derivation of the real-valued SH projection of the emission term is exactly the same as for the collision and scattering term.
After replacing the emission term $Q$ with its real-valued SH expansion, we multiply by the real-valued SH basis function for $m<0$ and integrate over solid angle. After some transformations we arrive at the following expression:
\begin{align*}
&
-\frac{i}{\sqrt{2}}\frac{i}{\sqrt{2}}\sum_{m=-l'}^{-1}{Q^{{l',m}}\left (\vec{x} \right )\delta_{m',m}}
\mathcolor{blue}
{
+\frac{i}{\sqrt{2}}\frac{i}{\sqrt{2}}\sum_{m=-l'}^{-1}{Q^{{l',m}}\left (\vec{x} \right )\left({-1}\right)^{m}\delta_{m',-m}}
}
\\&
\mathcolor{blue}
{
-\frac{i}{\sqrt{2}}\delta_{m',0}Q^{{l',0}}\left (\vec{x} \right )
}
\mathcolor{red}
{
-\frac{i}{\sqrt{2}}\frac{1}{\sqrt{2}}\sum_{m=1}^{l'}{Q^{{l',m}}\left (\vec{x} \right )\delta_{m',-m}}
}
\\&
\mathcolor{blue}
{
-\frac{i}{\sqrt{2}}\frac{1}{\sqrt{2}}\sum_{m=1}^{l'}{Q^{{l',m}}\left (\vec{x} \right )\left({-1}\right)^{m}\delta_{m',m}}
}
\mathcolor{blue}
{
+\frac{i}{\sqrt{2}}\left({-1}\right)^{m'}\frac{i}{\sqrt{2}}\sum_{m=-l'}^{-1}{Q^{{l',m}}\left (\vec{x} \right )\delta_{-m',m}}
}
\\&
-\frac{i}{\sqrt{2}}\left({-1}\right)^{m'}\frac{i}{\sqrt{2}}\sum_{m=-l'}^{-1}{Q^{{l',m}}\left (\vec{x} \right )\left({-1}\right)^{m}\delta_{-m',-m}}
\mathcolor{blue}
{
+\frac{i}{\sqrt{2}}\left({-1}\right)^{m'}\delta_{-m',0}Q^{{l',0}}\left (\vec{x} \right )
}
\mathcolor{blue}
{
+\frac{i}{\sqrt{2}}\left({-1}\right)^{m'}\frac{1}{\sqrt{2}}\sum_{m=1}^{l'}{Q^{{l',m}}\left (\vec{x} \right )\delta_{-m',-m}}
}
\\&
\mathcolor{red}
{
+\frac{i}{\sqrt{2}}\left({-1}\right)^{m'}\frac{1}{\sqrt{2}}\sum_{m=1}^{l'}{Q^{{l',m}}\left (\vec{x} \right )\left({-1}\right)^{m}\delta_{-m',m}}
}
\end{align*}

Again, the blue terms vanish and the red terms cancel each other out. The black terms collapses to:
\begin{align*}
Q^{l',m'}
\end{align*}

for $m<0$, $m=0$ and $m>0$.

\subsection{Final equation}

Putting all projected terms from previous subsections together, we get for $m=0$:
\begin{align}
&
\frac{1}{\sqrt{2}}c^{\scaleto{l-1,-1}{4pt}}\partial_x L^{\scaleto{l-1,1}{4pt}}
-\frac{1}{\sqrt{2}}d^{\scaleto{l+1,-1}{4pt}}\partial_x L^{\scaleto{l+1,1}{4pt}}
\frac{1}{\sqrt{2}}c^{\scaleto{l-1,-1}{4pt}}\partial_y L^{\scaleto{l-1,-1}{4pt}}
\nonumber
\\&
-\frac{1}{\sqrt{2}}d^{\scaleto{l+1,-1}{4pt}}\partial_y L^{\scaleto{l+1,-1}{4pt}}
%\\&
a^{\scaleto{l-1,0}{4pt}}\partial_z L^{\scaleto{l-1,0}{4pt}}
+b^{\scaleto{l+1,0}{4pt}}\partial_z L^{\scaleto{l+1,0}{4pt}}
\nonumber
\\&
+\sigma_t L^{\scaleto{l,m}{4pt}}
-\sigma_s\lambda_{\scaleto{l}{4pt}}p^{\scaleto{l,0}{4pt}}L^{\scaleto{l,m}{4pt}}
= Q^{\scaleto{l,m}{4pt}} \ .
\end{align}
For $m<0$:
\begin{align}
&-\frac{1}{2}c^{\scaleto{l-1,m-1}{4pt}}
\partial_y
L^{\scaleto{l-1,-m+1}{4pt}}
%\\
+\frac{1}{2}d^{\scaleto{l+1,m-1}{4pt}}
\partial_y
L^{\scaleto{l+1,-m+1}{4pt}}
%\\
-\frac{1}{2}\beta^{\scaleto{m}{4pt}}e^{\scaleto{l-1,m+1}{4pt}}
\partial_y
L^{\scaleto{l-1,-m-1}{4pt}}
\nonumber
\\&
+\frac{1}{2}\beta^{\scaleto{m}{4pt}}f^{\scaleto{l+1,m+1}{4pt}}
\partial_y
L^{\scaleto{l+1,-m-1}{4pt}}
%\\
+\frac{1}{2}c^{\scaleto{l-1,m-1}{4pt}}
\partial_x
L^{\scaleto{l-1,m-1}{4pt}}
\nonumber
\\&
-\frac{1}{2}\delta_{\scaleto{m\neq -1}{4pt}}e^{{l-1,m+1}}
\partial_x
L^{\scaleto{l-1,m+1}{4pt}}
%\\
+\frac{1}{2}\delta_{\scaleto{m\neq -1}{4pt}}f^{\scaleto{l+1,m+1}{4pt}}
\partial_x
L^{\scaleto{l+1,m+1}{4pt}}
%\\
-\frac{1}{2}d^{\scaleto{l+1,m-1}{4pt}}
\partial_x
L^{\scaleto{l+1,m-1}{4pt}}
\nonumber
\\&
+a^{\scaleto{l-1,m}{4pt}}
\partial_z
L^{\scaleto{l-1,m}{4pt}}
%\\
+b^{\scaleto{l+1,m}{4pt}}
\partial_z
L^{\scaleto{l+1,m}{4pt}}
%\\
+\sigma_t L^{\scaleto{l,m}{4pt}}
%\\
-\sigma_s\lambda_{\scaleto{l}{4pt}}p^{\scaleto{l,0}{4pt}}L^{\scaleto{l,m}{4pt}}
%\\
= Q^{\scaleto{l,m}{4pt}} \ .
\end{align}
And for $m>0$:
\begin{align}
&
\frac{1}{2}c^{\scaleto{l-1,-m-1}{4pt}}\partial_x
%\\&
-\frac{1}{2}d^{\scaleto{l+1,-m-1}{4pt}}\partial_x L^{\scaleto{l+1,m+1}{4pt}}
%\\&
-\frac{1}{2}\beta^{\scaleto{m}{4pt}}e^{\scaleto{l-1,m-1}{4pt}}\partial_x L^{\scaleto{l-1,m-1}{4pt}}
\nonumber
\\&
+\frac{1}{2}\beta^{\scaleto{m}{4pt}}f^{l+1,-m+1}\partial_x L^{\scaleto{l+1,m-1}{4pt}}
%\\&
+\frac{1}{2}c^{\scaleto{l-1,-m-1}{4pt}}\partial_y L^{\scaleto{l-1,-m-1}{4pt}}
\nonumber
\\&
-\frac{1}{2}d^{\scaleto{l+1,-m-1}{4pt}}\partial_y L^{\scaleto{l+1,-m-1}{4pt}}
%\nonumber
%\\&
+\delta_{\scaleto{m\neq 1}{4pt}}\frac{1}{2}e^{\scaleto{l-1,-m+1}{4pt}}\partial_y L^{\scaleto{l-1,-m+1}{4pt}}
\nonumber
\\&
-\delta_{\scaleto{m\neq 1}{4pt}}\frac{1}{2}f^{\scaleto{l+1,-m+1}{4pt}}\partial_y L^{\scaleto{l+1,-m+1}{4pt}}
%\\&
+a^{\scaleto{l-1,-m}{4pt}}\partial_z L^{\scaleto{l-1,m}{4pt}}
%\nonumber
%\\&
+b^{\scaleto{l+1,-m}{4pt}}\partial_z L^{\scaleto{l+1,m}{4pt}}
\nonumber
\\&
+\sigma_t L^{\scaleto{l,m}{4pt}}
-\sigma_s\lambda_{\scaleto{l}{4pt}}p^{\scaleto{l,0}{4pt}}L^{\scaleto{l,m}{4pt}}
= Q^{\scaleto{l,m}{4pt}} \ .
\end{align}
Where we defined:
\begin{align*}
\beta^{x}=
\left\{
\begin{array}{ll}
\frac{2}{\sqrt{2}}, & \text{for } \vert x\vert = 1\\
1, & \text{for } \vert x\vert \neq 1
\end{array}
\right.
,\quad
\delta_{x\neq y}=
\left\{
\begin{array}{ll}
1, & \text{for } x \neq y\\
0, & \text{for } x = y
\end{array}
\right. \ .
\end{align*}
and
\begin{align*}
&
a^{\scaleto{l,m}{4pt}}= \sqrt{\frac{\left(l-m+1\right)\left(l+m+1\right)}{\left(2l+1\right)\left(2l-1\right)}} \, , \qquad
b^{\scaleto{l,m}{4pt}}= \sqrt{\frac{\left(l-m\right)\left(l+m\right)}{\left(2l+1\right)\left(2l-1\right)}}
\\&
c^{\scaleto{l,m}{4pt}}= \sqrt{\frac{\left(l+m+1\right)\left(l+m+2\right)}{\left(2l+3\right)\left(2l+1\right)}} \, , \qquad
d^{\scaleto{l,m}{4pt}}= \sqrt{\frac{\left(l-m\right)\left(l-m-1\right)}{\left(2l+1\right)\left(2l-1\right)}}
\\&
e^{\scaleto{l,m}{4pt}}= \sqrt{\frac{\left(l-m+1\right)\left(l-m+2\right)}{\left(2l+3\right)\left(2l+1\right)}} \, , \qquad
f^{\scaleto{l,m}{4pt}}= \sqrt{\frac{\left(l+m\right)\left(l+m-1\right)}{\left(2l+1\right)\left(2l-1\right)}}
\end{align*}
\begin{align*}
\lambda_l=\sqrt{\frac{4\pi}{2l+1}} \ .
\end{align*}

These equations can be written in a more compact form by using $\pm$ and $\mp$ to write the equations for $m<0$ and $m>0$ as one. This gives the expressions of Section~\ref{sec:real_valued_pn_eq}.

%\begin{align*}
%&-\frac{1}{2}c^{\scaleto{l'-1,m'-1}{4pt}}
%\partial_y
%L^{\scaleto{l'-1,-m'+1}{4pt}}
%%\\
%+\frac{1}{2}d^{\scaleto{l'+1,m'-1}{4pt}}
%\partial_y
%L^{\scaleto{l'+1,-m'+1}{4pt}}
%%\\
%-\frac{1}{2}\beta^{\scaleto{m'}{4pt}}e^{\scaleto{l'-1,m'+1}{4pt}}
%\partial_y
%L^{\scaleto{l'-1,-m'-1}{4pt}}
%\\&
%+\frac{1}{2}\beta^{\scaleto{m'}{4pt}}f^{\scaleto{l'+1,m'+1}{4pt}}
%\partial_y
%L^{\scaleto{l'+1,-m'-1}{4pt}}
%%\\
%+\frac{1}{2}\delta_{\scaleto{m'\neq -1}{4pt}}c^{\scaleto{l'-1,m'-1}{4pt}}
%\partial_x
%L^{\scaleto{l'-1,m'-1}{4pt}}
%\\&
%-\frac{1}{2}\delta_{\scaleto{m'\neq -1}{4pt}}e^{{l'-1,m'+1}}
%\partial_x
%L^{\scaleto{l'-1,m'+1}{4pt}}
%%\\
%+\frac{1}{2}f^{\scaleto{l'+1,m'+1}{4pt}}
%\partial_x
%L^{\scaleto{l'+1,m'+1}{4pt}}
%%\\
%-\frac{1}{2}d^{\scaleto{l'+1,m'-1}{4pt}}
%\partial_x
%L^{\scaleto{l'+1,m'-1}{4pt}}
%\\&
%+a^{\scaleto{l'-1,m'}{4pt}}
%\partial_z
%L^{\scaleto{l'-1,m'}{4pt}}
%%\\
%+b^{\scaleto{l'+1,m'}{4pt}}
%\partial_z
%L^{\scaleto{l'+1,m'}{4pt}}
%\end{align*}
%\vspace{1cm}

\end{document}